# Integrated omics reveals actionable drivers of bioactive variation in US milk

Cheng-En Tan[1,2,3], Mariana Barboza[2,4], Pagkratios Tagkopoulos[2,3,5], Shanghyeon Kim[2,3], Lukas Maximilian Masopust[3], Ivor Prado[3], Muhammad Adil Salim[2,6], Fangzhou Li[1,2,3], George Berdovskiy[2], Ilias Apostolakos[5], Brandon Invergo[5], Karen M. Kalanetra[7], Danielle G. Lemay[2,3,8,9], David A. Mills[7], Armin Oloumi[10], Cheng-Yu Weng[10], Carlito B. Lebrilla[10], Xuan He[7,9], Carolyn Slupsky[7,9], Oliver Fiehn[11], Ibuki Kusumoto[7,12], Ameer Y. Taha[7,11,13], Yu Wang[7], Daniela Barile[7], Alexis Davis[14,15], David E. Olson[13,14,16,17], Justin B. Siegel[2,16,17*], and Ilias Tagkopoulos[1,2,3*]

[1]*Department of Computer Science, University of California, Davis, Davis, California, 95616 United States*

[2]*Genome Center, University of California, Davis, Davis, California, 95616 United States*

[3]*USDA/NSF AI Institute for Next Generation Food Systems (AIFS), University of California, Davis, Davis, California, 95616 United States*

[4]*Innovation Institute for Food and Health, University of California, Davis, Davis, California, 95616 United States*

[5]*Process Integration and Predictive Analytics, PIPA LLC, Davis, California, 95616 United States*

[6]*Microbiology Graduate Group, University of California, Davis, Davis, California, 95616 United States*

[7]*Department of Food Science and Technology, University of California, Davis, Davis, California, 95616 United States*

[8]*USDA ARS Western Human Nutrition Research Center, Davis, California, 95616 United States*

[9]*Department of Nutrition, University of California, Davis, Davis, California, 95616 United States*

[10]*Department of Chemistry, College of Letters and Sciences, University of California, Davis, Davis, California, 95616 United States*

[11]*West Coast Metabolomics Center, UC Davis Genome Center, University of California, Davis, Davis, California, 95616 United States*

[12]*Food Function Analysis Laboratory, Graduate School of Agricultural Science, Tohoku University, Sendai, 980-8572, Japan*

[13]*Center for Neuroscience, University of California, Davis, Davis, California, 95616 United States*

[14]*Institute for Psychedelics and Neurotherapeutics, University of California, Davis, Davis, CA 95616, USA*

[15]*Chemistry and Chemical Biology Graduate Program, University of California, Davis, Davis, CA 95616, USA*

[16]*Department of Chemistry, University of California, Davis, Davis, California, 95616 United States*

[17]*Department of Biochemistry and Molecular Medicine, University of California, Davis, Davis, California, 95616 United States*

[*]Corresponding authors: itagkopoulos@ucdavis.edu and jbsiegel@ucdavis.edu

## Abstract

Despite milk being a global dietary staple consumed by billions, the molecular basis of its health-relevant bioactivity and the factors governing fine-scale compositional variation remain poorly understood. Here, we present an integrated seven-layer omics characterization of 60 U.S. retail milk samples spanning ten geographic regions, combining genomics, transcriptomics, peptidomics, proteomics, lipidomics, metabolomics, and glycomics on the same samples. We further organized the identified and quantified milk compounds into the Dairy Molecule Database (DMD), a web-accessible resource designed to provide a molecular foundation for discovering milk-derived health effects and linking compound concentrations with SNPs, miRNAs, and product-level factors to support future milk quality optimization. In total, we identified 6,714 compounds and quantified 5,288 with absolute concentrations, including 5,220 with absolute concentration estimates that were not available in existing milk compound databases. These absolute concentration profiles enabled us to estimate bioactivity efficacy and identify factors associated with variation in milk compound concentrations. We tested 54 computationally predicted antimicrobial candidate peptides, of which 35 exhibited activity and 6 showed $IC_{50}$ values below 256 μg/mL against *A. baumannii*, one of the ESKAPE pathogens. We further identified associations of product-level factors, including purchase region, purchase ambient temperature, and packaging opacity, as well as SNPs, estimated Jersey breed proportion, and miRNA abundance, with downstream bioactive compound concentrations and estimated bioactivity efficacy. Overall, our results suggest that milk's bioactivity efficacy profiles are associated with purchase regions, selected SNPs, Jersey breed proportion, and selected miRNAs. These findings indicate that selective breeding and supply-chain optimization could help enhance the bioactive quality of commercial milk.

## Main

Milk is a complex colloidal dispersion and heterogeneous biological matrix whose molecular composition underlies its nutritional and functional relevance to human health[1–3]. In the United States, per capita milk consumption expanded in the late 19th and early 20th centuries, driven by advances in pasteurization that reduced health risks and by the growth of industrial-scale dairy production[4]. Today, dairy products remain a global dietary staple, providing energy, essential macro- and micronutrients, and high-quality protein containing all nine essential amino acids in a readily absorbed form, making milk composition directly relevant to public health[5–9].

Given these attributes, milk has long been studied to understand how its composition influences health. Milk macro- and micronutrients have been extensively characterized[10], and important links between specific milk molecules and health markers have been established. For example, peptides derived from α- and β-casein, as well as glycomacropeptide-associated glycans and peptides, have been associated with immune regulation and gut health, whereas long-chain polyunsaturated fatty acids, sphingolipids, and B vitamins have been linked to neural metabolism and development[11–13]. Milk peptides have also been reported to exhibit multiple bioactivities, including antimicrobial, antihypertensive, antioxidative, antithrombotic, and immunomodulatory effects.[14–16] However, although these studies have partially revealed connections between milk nutrient composition and human health, many questions remain. The full molecular diversity of milk, including thousands of bioactive and functional molecules, remains incompletely defined[17,18]. In addition, quantitative variation in milk components across breeds, feeding systems, geographic regions, and processing conditions, as well as the relationship of this variation to human health outcomes, remains insufficiently understood[2,18]. Because of these limitations, dietary guidelines from national and international agencies, including the World Health Organization (WHO), and food composition databases still largely rely on gross or averaged nutrient values, offering limited insight into the heterogeneous molecular composition that shapes milk's nutritional quality and biological activity[5,19,20]. These gaps hinder efforts to connect production and processing factors to detailed milk composition and, ultimately, to develop evidence-based, composition-guided nutritional strategies.

Over the last two decades, advances in analytical chemistry and computational biology, enabled by nano- and ultra-high-pressure liquid chromatography coupled to tandem or high-resolution mass spectrometry (MS), nuclear magnetic resonance (NMR), and high-throughput sequencing, together with progress in bioinformatics, have made it possible to measure thousands of metabolites, lipids, peptides, glycans, and nucleic acids across multiple molecular layers. This multi-layered measurement paradigm is collectively referred to as omics[21–23]. Integrating these data within multi-omics correlation frameworks enables a more systematic exploration of milk composition[1,17,21]. More recently, artificial intelligence (AI) and advanced machine learning (ML) methods, including deep neural networks, graph-based models, and transformer architectures, have further expanded the ability to integrate multimodal datasets, reveal molecular patterns, link molecular profiles to phenotypes, and model interactions across omics layers in ways that are difficult to achieve with traditional statistical models[3,24].

Several milk omics databases have been developed, including Bovine Milk Proteome Database (BoMiProt) for milk proteins[25,26], Milk Bioactive Peptide Database (MBPDB) for bioactive milk peptides[27,28], and Milk Composition Database (MCDB) for small molecules in milk[29]. Previous studies have also identified proteins[30], peptides[31], and small molecules[29] in milk, and integrated analyses combining metabolomics, lipidomics, and proteomics have been reported[32]. However, existing milk databases contain inconsistent compound identifications and concentration values because they are curated from different studies using different milk samples, while many experimental milk omics studies focus on a single molecular layer. In addition, prior integrated omics studies have often involved limited sample numbers and have not provided absolute concentration measurements, which are critical for estimating their bioactivity and possible health relevance.

To address these limitations, we performed a comprehensive multi-omics analysis of 60 commercial whole-milk samples collected from ten U.S. regions in 2022. In total, we detected 6,714 compounds and approximately 26 million single nucleotide polymorphisms (SNPs). After excluding 980 metabolites detected only by untargeted metabolomics, 18 oligosaccharides that could not be quantified with the available internal standards, and 428 detected miRNAs, we quantified 5,288 compounds as absolute concentrations, including 4,412 peptides, 645 proteins, 175 lipid species, 51 metabolites, and 5 oligosaccharides. This integrated omics analysis addressed several previous gaps by profiling multiple molecular layers from the same bovine milk samples, linking product-level factors and genomic information to downstream compound profiles, and providing absolute concentration measurements that support bioactivity estimation. Compared to three published databases, including MCDB[29], the MBPDB[27,28], and the BoMiProt[25,26] , 5,335 compounds detected in our study were not represented in these reference databases under our compound-matching criteria, and 5,220 of these compounds were quantified with absolute concentration estimates.

Together with curated bioactivity data from 34 assays for 284 bioactive milk compounds from the PubChem BioAssay database[33] and MBPDB[27,28], experimental bioactivity measurements identifying docosahexaenoic acid (DHA) as anti-inflammatory and six milk peptides as antibacterial against one of the six highly virulent ESKAPE pathogens[34] , and extensive metadata on purchase region, purchase date, storage temperature, nutritional labels, and packaging conditions, these omics profiles form a comprehensive compendium of U.S. retail milk.

Using this resource for hypothesis-driven analyses, we identified associations of product-level factors, including purchase region, purchase temperature, and packaging opacity, as well as SNPs, cow breed ratio, and miRNA abundance, with downstream bioactive compound concentrations and estimated bioactivity efficacy. Overall, our study aimed to systematically characterize variation in milk bioactive components across genetic backgrounds and purchase regions, with the goal of identifying factors that may inform future strategies to optimize milk bioactive compound content through selective breeding, production, and adjustments in production and processing. The integrated omics data, including SNPs, miRNAs, identified compounds, quantified compound concentrations, and associated metadata, are further organized into the Dairy Molecule Database (DMD). This database provides a molecular foundation for discovering milk-derived health effects and linking compound concentrations with SNPs, miRNAs, and product-level factors to support future milk quality optimization.

## Results

### Multi-omics profiling of 60 U.S. milk samples expands the catalog of known milk compounds

We profiled 60 U.S. milk samples collected from ten regions across seven omics layers, including genomics, transcriptomics, peptidomics, proteomics, metabolomics, lipidomics, and glycomics, to quantify SNP allele frequencies, microRNA (miRNA) counts, and compound concentrations. In parallel, we recorded sample metadata, including Nutrition Facts label values, labeled production information, purchase dates, brands, purchase ambient temperature, and package opacity (**Table 1**). Overall, we identified 25,999,881 SNPs with allele frequencies, 428 miRNAs, and 6,714 compounds across five measured molecular layers. Except for compounds identified by untargeted metabolomics, detected compounds were quantified as absolute concentrations. Peptide concentrations were estimated by normalizing relative abundances based on extracted-ion chromatogram areas to total peptide concentration, whereas other quantified compound classes were measured using internal standards. In total, 79.0% of detected compounds were absolutely quantified, comprising 5,288 compounds: 4,412 peptides, 645 proteins, 51 metabolites, 175 lipids, and 5 oligosaccharides (**Figure 1a**).

By integrating multi-omics profiles and absolutely quantifying most identified compounds, we generated a dairy database with broader omics-layer coverage and a larger number of quantified compounds than existing resources focused on a single molecular type, including the Milk Composition Database (MCDB)[29], the Milk Bioactive Peptide Database (MBPDB)[27,28], and the Bovine Milk Proteome Database (BoMiProt)[25,26] (**Table 2**). Compared with the compounds recorded in these three databases, which together contain 1,649 quantified and 12,184 identified compounds, 5,335 compounds detected in our study were not represented in these reference databases, and 5,220 of these were quantified with absolute concentration estimates (**Supplementary Figure 1**). Interestingly, 15 proteins that were not represented in the reference milk databases ranked among the top 100 quantified compounds by average concentration. These included sequence variants of well-known milk proteins, such as α-S1-casein, as well as less-characterized proteins including a tyrosine-specific protein kinase, an Ig-like domain-containing protein, and a fatty acid synthase (**Supplementary Table 1**). These findings suggest that our approach can resolve multiple sequence variants of the same protein in milk samples, providing additional molecular detail beyond conventional protein-level annotations. Relative to the largest single-experiment datasets available for proteomics[30], peptidomics[31], and small-molecule profiling[29], our study provides absolute quantification for 645 proteins and 4,412 peptides, additionally quantified 186 small molecules, and detected 3,473 compounds not reported in those datasets (**Supplementary Figure 2**).

### Integrated omics profiling for milk composition

The milk samples exhibited a typical composition profile in which proteins; metabolites, especially lactose; and lipids comprised the largest measured fractions of total milk mass other than water, contributing approximately 4.77%, 4.55%, and 1.91% of total milk weight, respectively (**Figure 2a**). A small subset of high-abundance compounds accounted for most of the quantified milk solids. Specifically, the top 10 compounds (six proteins, three fatty acids, and lactose) represented approximately 91.3% of the sum of quantified compounds (**Figure 2a, Supplementary Figure 3**). In addition, low relative standard deviations were observed for these key compounds except κ-casein (**Supplementary Figure 4**).

We compared measured concentrations of major nutritional components with the values reported on Nutrition Facts labels. The summed concentration of 645 quantified proteins was significantly higher than labeled total protein values (mean difference: 45.8%; range: 14% to 78%; $p < 2.2*10^{-16}$), whereas the summed concentrations of 175 lipid species and cholesterol concentrations were significantly lower than labeled total fat and cholesterol values, respectively (mean difference for total fat: -42.4%; range: -56% to -33%; mean difference for cholesterol: -55.8%; range: -76% to -26%; $p < 2.2*10^{-16}$ for both comparisons) (**Figure 2b, Supplementary Figure 5**). These differences likely reflect methodological differences between the current analytical workflow and conventional approaches used for Nutrition Facts labeling. In this study, fatty acids were measured following Folch extraction instead of AOAC or similar methods optimized for total lipids. Additionally, the fatty acid GC method was not optimized for shorter chain fatty acids (C4 to C12) or trans fatty acid isomers, resulting in lower total fat estimates. For proteins, the digestion-based LC-MS workflow enabled broader protein detection and sequence-level specificity, but differences in quantification strategy may contribute to discrepancies relative to Nutrition Facts label values based on conventional total-protein methods such as the Kjeldahl method[35] and to reported concentrations of specific milk proteins measured by intact-protein MS approaches.[36,37]

**Bioactivity profiling expands the known bioactive compound landscape of U.S. milk**

We profiled milk bioactivity using curated data and wet-lab experiments. We found 20 small molecules (13 lipids, 7 metabolites) in milk and 264 bioactive peptides from 9 source milk proteins with bioactivity records in the PubChem BioAssay database and MBPDB, respectively. These compounds and peptides were associated with 34 bioactivities (**Supplementary Table 2**). We then evaluated bioactivity efficacy for each compound–bioactivity pair based on quantified milk concentrations and curated Hill-curve parameters[38]. Using the highest single agent (HSA) approach[39,40], which assigns each bioactivity to the compound with the highest estimated efficacy, we identified 18 compounds, including 7 lipids, 6 peptides, and 5 metabolites, as the main contributors to the estimated in vitro efficacy of 34 bioactivities in five categories (**Table 3**; **Figure 3a, b; Supplementary Figure 6a**).

To approximate in vivo efficacy, we estimated postprandial serum concentrations after milk ingestion using reported milk-to-serum conversion ratios for a representative small molecule (lactose[41]) and peptide (the tripeptide IPP[35]). These ratios, approximately 1.47 $\times 10^{-4}$ for small molecules and $6 \times 10^{-6}$ for proteins and peptides, illustrate the order-of-magnitude attenuation from milk concentration to serum concentration. Applying these conversion factors to all small molecules and proteins/peptides in our data, we identified three compounds – cytidine, linoleic acid (octadecadienoic acid-(9Z,12Z)), and stearic acid (octadecanoic acid) – that were predicted to reach at least 5% estimated in vivo efficacy in five anticancer bioassays, one antioxidant bioassay (15-hLO-1 enzyme inhibition), and one antiviral bioassay targeting Chikungunya virus, respectively (**Figure 3c; Supplementary Figure 6b**).

In addition to database-derived bioactivity estimates, we experimentally evaluated selected milk compounds in three bioactivity assays focused on neurite outgrowth, anti-inflammatory activity, and antibacterial activity. None of the tested compounds showed significant neurite outgrowth activity (**Supplementary Figure 7a–d**). In the neuroinflammation assay, docosahexaenoic acid (DHA) exhibited anti-inflammatory bioactivity at 50 μM, significantly reducing the lipopolysaccharide-induced increase in microglial area without reducing neuronal cell count (**Supplementary Figure 7e–h**). In addition, six milk-derived peptides exhibited $IC_{50}$ values below 256 μg/mL against *A. baumannii*, while showing no inhibitory activity against the probiotic species *L. plantarum* or *L. rhamnosus* within the tested concentration range. Notably, three peptides showed substantially higher selectivity indices (>40.0), calculated as the ratio of the $IC_{50}$ value against the probiotic species to that against *A. baumannii*, compared with melittin (8.8) (**Supplementary Table 3**). Furthermore, 35 peptides significantly inhibited the growth of *A. baumannii* at 1,024 μg/mL, supporting their antibacterial activity (**Supplementary Table 4**). Together, these results suggest that selected milk-derived compounds and peptides may exert bioactivities relevant to neuroinflammation and pathogen inhibition, including antibacterial effects that spare probiotic bacteria.

By integrating 284 curated bioactive milk compounds associated with 34 bioactivities and 64 experimentally tested compounds associated with antibacterial or anti-inflammatory assays, we identified bioactivities not previously reported. To our knowledge, the antibacterial activities of these 35 endogenous milk-derived peptides have not previously been reported. In addition, we experimentally validated DHA, a known anti-inflammatory omega-3 fatty acid detected in our milk samples, as active in our LPS-induced neuroinflammation assay. In total, our analysis captured 34 previously reported bioactivities and 284 known bioactive milk-associated compounds or peptides. However, 29 MBPDB functional categories and 356 known bioactive peptides were not represented in our data under exact endogenous-peptide matching criteria. This is likely because our study focused on endogenous peptides naturally present in retail milk, whereas many MBPDB entries correspond to peptides generated by enzymatic hydrolysis, gastrointestinal digestion, fermentation, or other processing conditions. In addition, bioactivities lacking sufficient quantitative parameters for efficacy estimation, such as $EC_{50}$ values, were excluded from the efficacy analysis.

**Metadata analysis links product-level factors to compound profiles and estimated bioactivity efficacy**

We identified several product-level factors associated with milk compound profiles and estimated bioactivity efficacy. Variance decomposition analysis showed that purchase region was associated with five bioactive compounds, explaining at least 35% of the variance in their concentrations (**Supplementary Figure 8**). Among these compounds, two showed significant regional differences in concentration: κ-casein and docosahexaenoic acid (DHA). After adjustment for purchase time and brand, κ-casein was significantly more abundant in the West and South Central regions and significantly less abundant in the Southeast and Great Lakes regions, compared with the estimated marginal mean across all regions. DHA was significantly more abundant in the Southeast region (Figure 4a). Regional variation in DHA concentration may have further contributed to variation in the estimated efficacy of two bioactivities, including one anticancer activity and one antimicrobial activity, with regional differences ranging from approximately −20% to 20% relative to the average (**Figure 4b, c**).

We also found that purchase ambient temperature, package opacity, and their interaction were associated with the concentrations of selected milk compounds. Eight compounds were significantly correlated with purchase temperature (**Supplementary Table 5; Supplementary Figure 9a, b**), and aspartic acid was significantly more abundant in clear packages than in opaque packages (**Supplementary Figure 9c**). In addition, the interaction between purchase temperature and packaging was significantly associated with total lipid concentration: total lipid concentration was negatively correlated with temperature in clear packaging but not in opaque packaging (**Supplementary Figure 9d, e**). These results suggest that product-level and retail-condition factors, including purchase-region characteristics, purchase ambient temperature, and packaging opacity, may influence selected milk compound concentrations and could potentially be considered in strategies to optimize milk composition.

**Genomic and transcriptomic profiling links SNPs and miRNAs to milk compound profiles and bioactivities**

Genomic profiling provided insight into breed composition and genetic influences on milk composition. Overall, the variation of milk concentration was partly predictable from a small number of SNPs. Centered milk compound profiles, obtained after subtracting the across-sample mean for each compound, were reconstructed using only 61 selected SNPs, achieving a leave-one-region-out cross-validation Pearson correlation coefficient of 0.64.

We then analyzed how SNPs were associated with dairy breeds, specific compounds, and downstream bioactivity estimates. Using breed-informative SNPs, we estimated the proportional contributions of two major dairy breeds, Holstein and Jersey, to each milk sample (**Supplementary Figure 10**). Milk from the West region showed a significantly higher estimated Jersey proportion than milk from other regions (**Supplementary Figure 11a**). Estimated Jersey proportion was also significantly correlated with concentrations of several abundant and/or bioactive compounds; for example, Jersey proportion was positively correlated with kappa-casein concentration (**Supplementary Table 6; Supplementary Figure 11b**). After the breed composition evaluation, we found several associations between the SNPs and bioactive compounds, including 24 SNP-bioactive compound pairs that passed a stricter correlation threshold of 0.55 (for both Pearson correlation and Spearman correlation) together with their corresponding bioactivities (**Figure 5; Supplementary Table 7**). In addition to SNPs, we also found 7 miRNA-bioactive compound pairs that significantly correlated with correlations greater than 0.35 (**Supplementary Table 8; Supplementary Figure 12**). These results suggest that specific milk compounds and associated bioactivities may be modulated through breed selection and through SNP-level genetic selection. Finally, we summarized the associations among product-level factors such as regions, estimated Jersey breed ratio, SNPs, bioactive compounds, and bioactivity. These summarized associations suggest that bioactivity efficacy estimates are mainly associated with purchase region, Jersey breed proportion, and SNPs (**Figure 5**), highlighting the need for further investigation into regional variation in bioactive compound concentrations and the potential for breeding-based bioactivity optimization.

## Discussion

To our knowledge, this study provides the first retail milk compendium that combines seven omics layers-genomics, transcriptomics, peptidomics, proteomics, lipidomics, metabolomics, and glycomics-with metadata describing purchase and production factors. This integrated database highlights milk-associated compounds and estimated bioactivity profiles that are not fully captured by previous milk omics studies and databases, and enabled us to associate milk-compound profiles with product-level factors, connect omics layers to bioactivity estimates and explore strategies for improving milk quality by increasing selected bioactive compounds through breeding, storage, or packaging interventions.

Composition analysis showed that the top 10 compounds by average concentration in whole milk – lactose; three fatty acids (palmitic acid, oleic acid, and stearic acid); and six proteins (α-S1-casein, α-S2-casein, β-casein, κ-casein, β-lactoglobulin, and α-lactalbumin) – collectively accounted for approximately 90% of the measured dry milk mass. Thus, a small set of molecules dominated the overall composition. We also observed differences between the summed concentrations of 645 quantified proteins and 175 lipid species and the corresponding Nutrition Facts label values. These differences are mainly attributable to analytical-method differences or differences between label definitions and the molecular species quantified here, and they motivate further evaluation of how label values are determined for key macronutrients.

By combining compound profiles with curated and experimentally measured bioactivity data, we further characterized the bioactivity potential of milk and identified compounds likely to contribute to those activities. The samples contained compounds associated with five main bioactivity classes-anticancer, disease therapy (including type-2 diabetes, Alzheimer's, and hypertension), antioxidant, antimicrobial, and antiviral activity-covering 34 bioactivities primarily contributed by 18 bioactive compounds with estimated in vitro efficacy. We also estimated in vivo efficacy using limited published data on postprandial serum concentrations after milk consumption. Because these conversion factors are approximate, we evaluated sensitivity by multiplying the estimated serum concentrations by variation factors (0.01, 0.1, 10, and 100), thereby illustrating the possible range of in vivo efficacy under substantial uncertainty in serum exposure.

The bioactive compounds identified here can be linked to product and purchase metadata to develop future strategies for optimizing compound concentrations. Several bioactive compounds, especially docosahexaenoic acid, showed significant regional differences in average concentration; these differences affected the corresponding estimated bioactivities, including G9a enzyme inhibition, related to anticancer activity and anti-malarial activity. We also observed associations between selected compound concentrations and other product-level factors, particularly purchase-region temperature and package opacity. Together, these results suggest that production and purchase factors are associated with milk compound concentrations and corresponding bioactivity estimates, highlighting packaging and other supply-chain variables as potential levers for optimization.

Finally, we connected genomic and transcriptomic profiles to compound variation. We showed that overall compound variation can be mainly explained by a few SNPs. Using 61 SNPs selected from more than 8 million detected SNPs, we reconstructed centered compound profiles with a leave-one-region-out cross-validation Pearson correlation coefficient of 0.64. We then discovered the associations among the SNPs, Jersey breed ratio, miRNA, bioactivity compounds, and bioactivity efficacy. We found that 24 SNPs, 7 miRNAs, and estimated Jersey breed proportion were associated with bioactive compound concentrations and estimated bioactivity efficacies. The summary associations among product-level factors, especially purchase region, Jersey breed ratio, and SNPs suggest that bioactive compound concentrations may be optimized through breed selection and, in some cases, SNP-level selection within breeds. Further investigation of the factors underlying regional variation in bioactive compound concentrations may reveal additional opportunities for optimization.

Overall, this study presents an integrated multi-omics profile of retail milk that links genotype, molecular composition, and bioactivity. We catalogued milk's molecular composition, estimated bioactive efficacies across a broad set of health-related assays, and identified multiple factors-including genetic markers, inferred breed composition, miRNA abundance, production region, and handling conditions-associated with variation in milk composition and potential health-relevant properties. These results highlight the value of integrating multi-omics measurements with bioactivity information and detailed metadata to achieve a systems-level understanding of food composition. As additional datasets become available with larger sample sizes, broader seasonal coverage, and more detailed farming-practice information, more nuanced relationships are likely to emerge. Such knowledge could support precision nutrition strategies and enable dairy production and processing to be tailored toward beneficial milk components. Beyond biological interpretation, the absolute concentration measurements quantified here may also support future evaluations of the commercial feasibility of recovering or producing milk-derived compounds, given information on extraction costs, recovery rates, and, for peptides, hydrolysis efficiency and yield from source proteins. Future efforts may also leverage AI and machine learning methods to predict and optimize milk quality from multimodal datasets. In conclusion, our work lays a foundation for evidence-based improvements in milk quality and may help guide breeding programs and production practices aimed at maximizing the nutritional and bioactive benefits of milk.

## Methods

### Milk sample collection

Commercial milk samples were collected from ten metropolitan regions across the United States. These regions were based on the IRI/Circana retail market-region framework[42], with California further subdivided into three city-level purchase regions. For each region, two brands of pasteurized whole milk were purchased at three time points (June, July, and August 2022), except for the pilot samples collected in the California-2 region, which were collected approximately one month earlier. In total, 60 samples were obtained, including 58 whole-milk samples and 2 reduced-fat milk samples from the South Central region. For each sample, we recorded metadata related to production and sale, including Nutrition Facts label values, purchase ambient temperature, purchase date/time, expiration date, and packaging appearance (opaque vs. translucent container).

Upon purchase, samples were immediately placed in insulated shipping containers designed to maintain temperatures between 2°C and 8°C for up to 96 h (FedEx). Samples were shipped overnight to the University of California, Davis (UC Davis). Milk sample temperature was continuously monitored throughout transportation using data loggers placed within each shipping container to verify maintenance of the cold chain. Upon arrival at UC Davis, samples were immediately aliquoted into 1.5 mL, 20 mL, and 50 mL polypropylene tubes according to the requirements of the downstream analyses. Aliquots were frozen and stored at −20°C until distribution to the respective omics laboratories for analysis.

## Omics analysis

**Genomic analysis for SNP identification.** Genomic DNA was extracted from milk somatic cells and sequenced to identify SNPs. DNA extraction and library preparation followed a whole-genome sequencing (WGS) protocol (**Supplementary Materials, section 3.1**). Briefly, cow somatic cells were isolated from milk using centrifugation and washing steps to remove fat and other components, and genomic DNA was extracted using a cell lysis and precipitation method. DNA quality was assessed using absorbance ratios and gel electrophoresis, and a polymerase chain reaction (PCR) test with cow-specific primers confirmed amplifiable cow DNA (**Supplementary Materials, section 3.1; Supplementary Table 9**).

Sequencing libraries were prepared using the KAPA HyperPlus kit (Roche) with approximately 100 ng genomic DNA input. DNA was fragmented to approximately 400–750 bp, adapter-ligated, PCR-amplified, and size-selected to remove short fragments. Whole-genome sequencing was performed on an Illumina NovaSeq S4 platform (paired-end 150 bp reads) at the DNA Technologies and Expression Analysis Core at the UC Davis Genome Center.

Raw sequencing reads were quality-trimmed using Trimmomatic v0.32[43] to remove adapter sequences and low-quality bases. SNP calling and genotyping were performed using the bcbio-nextgen pipeline (v1.2.93)[44], which incorporates BWA-MEM37 for alignment to the bovine reference genome (UMD 3.1.1[45]) and GATK HaplotypeCaller[46] for variant detection, with joint genotyping enabled across all samples. Default parameters were used throughout, with the analysis type set to “variant” and all 60 samples processed together as a single batch to facilitate joint variant calling. SNPs were annotated with SnpEff[47] to evaluate functional consequences.

**Transcriptomic analysis for microRNA.** Transcriptomic analysis focused on exosomal microRNA (miRNA) isolated and sequenced from milk samples. Milk RNA exists in multiple compartments, including somatic cells, cytoplasmic crescents of milk fat globules, and extracellular vesicles. Because RNA in cytoplasmic crescents and milk cells can be rapidly degraded during processing and digestion, we targeted exosome-associated miRNA, which has been reported to survive milk processing and to be absorbed by intestinal epithelial cells, where it may influence consumer gene expression.

To obtain sufficient exosomal miRNA for sequencing, large milk volumes were processed (200 mL per sample). Milk was subjected to sequential centrifugation steps to remove fat, debris, and somatic cells. The TRIzol LS method from Sanoji et al. 2021 was used to extract miRNA from milk exosomes[48]. Purified miRNA was quantified using a Qubit microRNA Assay Kit prior to library preparation and sequencing. Purified miRNA was sequenced on an Illumina NextSeq500 in a single-read 75 bp format. De-multiplexed and adapter-trimmed reads were mapped to miRNA transcripts of the cattle genome, version ARS-UCD1.3, using the Spliced Transcripts Alignment to a Reference (STAR) aligner[49] with some key adjustments. miRNAs were retained for downstream analysis if they had annotated miRBase[50] IDs and nonzero counts in at least one sample. Details of miRNA extraction, sequencing, and quantification are provided in **Supplementary Materials, section 3.2**.

**Proteomics and peptidomics analysis.** Milk proteins and peptides were analyzed using a multi-step workflow combining protein digestion, peptide extraction, and high-resolution mass spectrometry (MS), adapted from established methods[51–55]. Milk samples were denatured with dithiothreitol, alkylated with iodoacetamide, and digested overnight with sequencing-grade trypsin[51]. Free peptides were extracted from 200 µL milk by Folch partitioning[52,53]. The aqueous phase was purified using C18 solid-phase extraction (SPE) and eluted sequentially with 40% and 80% acetonitrile containing 0.1% TFA.

Samples were analyzed using an UltiMate WPS-3000RS nanoLC system coupled to an Orbitrap Fusion Lumos MS (Thermo Fisher). Data were acquired in positive ionization mode over 375–1500 m/z at 120,000 resolution, with HCD fragmentation[54]. Proteins were identified using Byonic against the UniProt bovine database with a 1% false discovery rate (FDR). Relative abundances were quantified in Byologic using extracted ion chromatogram (XIC) peak areas. Absolute quantification of six major milk proteins (α-, β-, and κ-casein; β-lactoglobulin; α-lactalbumin; and lactotransferrin) was performed using calibration curves generated from commercial protein standards[55]. Detailed extraction procedures, LC gradients, and data-processing parameters are provided in **Supplementary Materials, section 3.3**.

**Targeted metabolite quantification (NMR).** Polar milk metabolites were quantified using targeted 1H NMR spectroscopy[56,57]. Samples were adjusted to pH 6.7 ± 0.1 and analyzed on a 600 MHz Bruker Avance system using a NOESY presaturation pulse sequence. Spectra were processed and quantified using Chenomx NMR Suite v8.6. Fifty-one metabolites were identified and quantified against internal standards (DSS-d6)[58]. Detailed extraction procedures and NMR acquisition settings are provided in **Supplementary Materials, section 3.4**.

**Untargeted metabolomics analysis.** Complementary untargeted metabolomics analyses were performed to expand molecular coverage and sensitivity. Primary metabolites were analyzed by automated liner exchange cold injection GC-TOF MS (ALEX-CIS). Raw data were processed in ChromaTOF and BinBase for peak detection, deconvolution, and identification. Quantification was based on peak heights normalized by metabolite total ion current (mTIC)[59].

Small polar metabolites and biogenic amines were analyzed by hydrophilic interaction liquid chromatography QTOF MS/MS (HILIC-ESI). Separation was performed on a Waters BEH Amide column. Features were annotated in MS-DIAL using in-house mzRT libraries and NIST/MoNA spectral matching, followed by manual curation. Data normalization used internal standard total ion current (iTIC) or SERRF correction for large datasets[60]. Detailed extraction procedures and MS acquisition and processing parameters are provided in **Supplementary Materials, section 3.5**.

**Lipidomics analysis.** Total and free oxylipins, total fatty acids, and total cholesterol were quantified following established protocols[61–64]. Milk total lipid was extracted by Folch partitioning using methanol / chloroform (1:2 v/v), and disodium EDTA/KCl aqueous buffer.

An aliquot of total lipid extract was subjected to hydrolysis, followed by solid phase extraction using Oasis HLB SPE columns to obtain total oxylipins. Free oxylipins were extracted separately by protein precipitation, followed by the same SPE purification steps used for total oxylipins post-hydrolysis. Deuterated surrogate standards were added before hydrolysis or protein precipitation to account for extraction recovery. Oxylipins were analyzed on an Agilent 1290 Infinity UPLC system coupled to an Agilent 6460 Triple Quadrupole mass-spectrometer (Agilent Technologies, Santa Clara, CA, USA), using a ZORBAX RRHD Eclipse Plus C18 column (2.1 × 150 mm, 1.8 μm, Agilent Technologies) and previously reported binary gradients with slight modifications[62,63]. Quantification used calibration curves from 77 oxylipin standards, each adjusted for their response factor and corresponding surrogate[64]. An aliquot of total lipid extract obtained above was subjected to transesterification to derivatize fatty acids and release cholesterol. Fatty acids and cholesterol were analyzed by gas chromatography with flame ionization detection (GC-FID) on a Perkin Elmer Clarus 500Gas Chromatograph system (Perkin Elmer, Shelton, CT, USA) using a DB-FFAP capillary column (30 m × 0.25 mm inner diameter, 0.25 μm film thickness, Agilent Technologies). Tridecanoate (C19:0-TAG) and 5α-cholestane were used as internal standards for fatty acids and cholesterol[62,63]. Detailed extraction methods and MS acquisition parameters are provided in **Supplementary Materials, section 3.6, Supplementary Tables 10 and 11**.

**Glycomics analysis.** Oligosaccharides were extracted from 200 μL milk following the protocol developed by Aldredge et al. (2013)[65]. After defatting and protein precipitation, the aqueous fraction was purified using C18 and porous graphitized carbon (PGC) SPE microplates.

Absolute quantification was performed by LC-QQQ (MRM mode) using an Agilent AdvanceBio Glycan Mapping column with ammonium acetate–acetonitrile gradients. Relative profiling was performed using an Agilent 6520 Nano LC Chip Q-TOF MS system with graphitized carbon enrichment and analytical columns. Data were processed using MassHunter and GlycoNote libraries to verify structure-specific fragment ions. Detailed extraction and acquisition parameters are provided in **Supplementary Materials, section 3.7**.

### Data processing and analysis

**Data preprocessing and compound composition analysis.** All omics data and metadata were integrated into a comprehensive milk composition compendium that included SNP allele frequencies, identified and quantified compound concentrations, compound annotations, product information such as brand and region, purchase conditions, Nutrition Facts label values, and storage and packaging data. For all quantified compounds, concentrations were initially recorded as mass per serving (μg per 240 mL serving) and converted to weight/weight percentage (% w/w) assuming a milk density of 1.03 g/mL, based on reported milk density values[66,67]. Compound identities were standardized across platforms using unique identifiers: PubChem Compound IDs for small molecules[68], UniProt accessions for proteins[69], and InChI Keys for chemical structures[70]. Compounds were also mapped to a common ontology, including resources such as ChEBI[71], to assign consistent molecular categories, such as lipid, peptide, and saccharide.

**Bioactivity assay data collection and efficacy estimation.** We curated bioactivity data for each compound from the PubChem BioAssay database[33,68] and the Milk Bioactive

Peptide Database (MBPDB)[27,28] for small molecules, including metabolites and lipids, and large molecules, including proteins and peptides, respectively. PubChem Compound IDs for small milk molecules were used to search PubChem BioAssay entries. For each entry, we curated the bioassay identifier (AID), assay purpose, positive and negative controls defining 100% and 0% efficacy, the bioactive compound, four fitted Hill-curve parameters (slope, minimum efficacy at zero concentration, maximum efficacy at infinite concentration, and $EC_{50}$), and the normalized maximum efficacy relative to the positive control. When no specific negative control was annotated, we assumed the negative control corresponded to zero compound concentration; entries missing other required attributes were excluded. We also retained only entries labeled as active in the database. In total, we curated 114 entries from 28 bioactivities (48 AIDs) contributed by 13 lipids and 7 metabolites.

For peptide and protein bioactivity curation, we searched the identified milk peptide sequences in MBPDB and curated bioactive peptide sequences derived from identified milk proteins. We identified six bioactivities from 73 studies, including angiotensin-converting enzyme (ACE) inhibition, dipeptidyl peptidase IV (DPP-IV) inhibition, prolyl endopeptidase inhibition, and three assays measuring antioxidant activity. Each entry contained the study, peptide sequence, protein of origin, and corresponding $EC_{50}$ value. Because the peptide bioactivity data lacked full Hill curves, which are required for efficacy estimation, we constructed a representative Hill-curve model by averaging normalized responses across the 114 curated small-molecule Hill curves (**Supplementary Figure 13a**). We then used this representative curve to estimate the in vitro efficacy contributed by peptides. To assess estimation accuracy, we performed bootstrap validation with 100 iterations. In each iteration, 10%, 50%, or 90% of the Hill-curve dataset was randomly selected to build the representative model, and the remaining curves were used to evaluate prediction error (**Supplementary Figure 13b**).

To estimate in vivo efficacy, we translated milk concentrations to approximate post-consumption serum concentrations using published human data. For small molecules, we used data for lactose from Pimentel et al.[41] (14 healthy men; postprandial serum concentration of approximately 0.002 mM at 120 min after consuming 800 g milk). For proteins and peptides, we used data for the tripeptide IPP from Foltz et al.[72] (6 healthy adults; postprandial serum concentration of approximately 0.0006 µM at approximately 45 min after consuming a 250 mL beverage containing 100 µM peptide added to yogurt). These data were used to estimate approximate conversion factors from ingested dose to peak serum concentration for small molecules and proteins/peptides, respectively (**Supplementary Figure 13c**). Because digestion, absorption, metabolism, and compound-specific bioavailability can strongly affect this conversion[73], these in vivo efficacy estimates should be interpreted as upper-bound approximations of which compounds might reach active levels in circulation. All calculated in vitro and in vivo efficacy values for each milk sample and assay are compiled in **Table 3** and **Supplementary Table 12** and **13**.

**Variance decomposition analysis.** For variance decomposition, we used the R package lme4[74] to fit linear mixed-effects models for each compound. Because each region contained two sampled milk products, we modeled brand/product identity as nested within region rather than as a global brand effect. Specifically, compound concentration was modeled as a function of random effects for region, purchase time, and brand-within-region. The variance components estimated from this model were used to calculate the proportion of total variance attributable to each factor.

$$Conc_i = \mu + \mu_{region\,(i)} + \mu_{time\,(i)} + \mu_{brand-region\,(i)} + \epsilon_i$$

$$\mu_{region} \sim N\left(0, \sigma^2_{region}\right), \mu_{time} \sim N(0, \sigma^2_{time})$$

$$\mu_{brand-region} \sim N\left(0, \sigma^2_{brand-region}\right), \epsilon \sim N(0, \sigma^2_{\epsilon})$$

After identifying compounds for which a substantial proportion of variance was explained by region, we further evaluated the direction and statistical significance of regional effects using a second mixed-effects model in which region and purchase time were treated as fixed effects and brand-within-region was retained as a random intercept.

$$Conc_i = \mu + \beta_{region\,(i)} + \beta_{time\,(i)} + \mu_{brand-region\,(i)} + \epsilon_i$$

$$\mu_{brand-region} \sim N\left(0, \sigma^2_{brand-region}\right), \epsilon \sim N(0, \sigma^2_{\epsilon})$$

We then estimated regional marginal means using the fitted model and compared each region against the overall marginal mean to identify regions with significantly higher or lower compound concentrations.

**Correlation analysis linking compound concentrations with SNPs, miRNAs, and Jersey breed proportion.** To identify factors potentially associated with milk compound concentrations, including SNPs, miRNAs, and Jersey breed proportion estimated from SNP allele frequencies, we computed both Pearson and Spearman correlation coefficients[75,76] between compound concentrations and SNP allele frequencies, miRNA molecule counts, and estimated Jersey breed proportions. To estimate Jersey breed proportion, we assumed that each milk sample was derived primarily from Holstein and Jersey cows, which account for approximately 80% and 8% of the U.S. dairy herd, respectively[77]. We then used a non-negative least squares (NNLS) approach[78] to estimate the breed-mixture proportions from the allele-frequency matrix of 8,179,104 detected SNPs annotated in the Bovine Genome Variation Database (BGVD) across the 60 milk samples, using the corresponding reference allele-frequency profiles of Holstein and Jersey cattle in BGVD. Statistical significance was evaluated using two-sided p-values, followed by multiple-testing correction using the Benjamini–Hochberg procedure[79]. A SNP–compound or miRNA–compound association was considered significant when the corresponding false discovery rate-adjusted q-value was < 0.1.

**The mediation model for determining the contribution factors of SNP-compound association.** We applied latent variable analysis (implemented as a R package lavaan[80]) to evaluate whether SNP–compound associations were mediated by estimated Jersey breed proportion using a structural equation model (SEM).

$$Jersey_i = \alpha_0 + a \cdot Freq_i + \varepsilon_{Jersey_i}$$

$$Conc_i = \beta_0 + c' \cdot Freq_i + b \cdot Jersey_i + \varepsilon_{Conc_i}$$

In this framework, the SNP allele frequency was treated as the predictor, Jersey proportion as the mediator, and compound concentration as the outcome. The indirect effect was defined as the product of the SNP-to-Jersey path coefficient and the Jersey-to-compound path coefficient (a*b). SNP–compound pairs were classified as mediated by Jersey proportion when the indirect effect (a*b) was significantly different from zero and the direct SNP effect (c’) was not significantly different from zero. They were classified as direct-only when the direct SNP effect was significantly different from zero, but the indirect effect was not. Pairs were classified as partially mediated when both the indirect and direct effects were significantly different from zero. Statistical significance was assessed using 2,000 bootstrap resamples, and an effect was considered non-significant when its 95% bootstrap confidence interval included zero.

**Bioactivity measurement assays**

**Neurite outgrowth activity measurement of selected milk compounds.** Primary cortical cultures were prepared from embryonic day 18 Sprague Dawley rats and treated at DIV4 with selected milk-derived compounds or corresponding vehicle controls. After 72 h, cells were fixed, immunostained for MAP2, counterstained with Hoechst, and imaged using a CellInsight CX7 high-content screening system. Primary neurite number was quantified using HCS Studio software by an experimenter blinded to treatment conditions. Tested compounds included fatty acids, amino acids, oligosaccharide-related metabolites, lactoferrin, and other milk-associated bioactive candidates across compound-specific concentration ranges. Detailed methods are provided in **Supplementary Materials, section 4.1**.

**Neuron anti-inflammatory activity measurement of selected milk compounds.** Cortical tricultures were prepared from postnatal day 0 Sprague Dawley rats and treated with LPS at DIV7 (after 7 days in vitro) to induce neuroinflammation. At DIV9, cells were treated with selected milk-derived compounds or corresponding vehicle controls for 24 h. Cells were then fixed, immunostained for MAP2, GFAP, and Iba1, counterstained with Hoechst, and imaged using a CellInsight CX7 high-content screening system. Glial cell count and area were quantified using HCS Studio software by an experimenter blinded to treatment conditions. Detailed methods are provided in **Supplementary Materials, section 4.2**.

**Antibacterial activity measurement of selected milk peptides.** An AI-based platform was used to select candidate antimicrobial peptides from the 4,412 quantified milk peptides. Briefly, the platform applied a QSAR-based approach to predict peptide bioactivity from protein language model embeddings of sequences with known active or inactive labels. The balanced training set consisted of 8,440 cases curated from the literature and public databases. To account for model stochasticity, five independently initialized models were trained. For each peptide, the mean predicted probability of activity and the 95% confidence interval of the mean were calculated. Peptides were predicted to be active when the confidence interval for the mean predicted probability was entirely above 0.5.

The 54 selected candidates were screened against six ESKAPE pathogens, which are recognized as major public-health priorities for antimicrobial development[34]. *Enterococcus faecalis* (ATCC 29212), *Staphylococcus aureus* (ATCC 25923), *Klebsiella pneumoniae* (ATCC 700603), *Acinetobacter baumannii* (ATCC 19606), *Pseudomonas aeruginosa* (ATCC 27853), and *Escherichia coli* (ATCC 700728). Among them, six peptides exhibited IC50 values below 256 µg/mL against *A. baumannii*, whereas none of the peptides showed IC50 values below 256 µg/mL against the other five ESKAPE species. These six most potent peptides were further evaluated against *E. coli* and two probiotic species, *Lactiplantibacillus plantarum* (ATCC 14917) and *Lactobacillus rhamnosus* GG (ATCC 53103), using a higher maximum tested concentration of 1,024 µg/mL. Melittin, an antimicrobial peptide reported to be in Phase III clinical trials, was used as a positive control, and selectivity indices were calculated as the ratio of the IC50 value against the probiotic species to that against the pathogen (**Supplementary Table 3**). In addition, all 54 peptides were tested against *A. baumannii* at 1,024 μg/mL, and OD600 values were compared with those of the negative control without peptide treatment. Overall, 35 peptides exhibited antibacterial activity against *A. baumannii* at 1,024 μg/mL, as indicated by significantly lower average $OD_{600}$ values than the negative control without peptide treatment (**Supplementary Table 4**). Detailed methods are provided in **Supplementary Materials, section 4.3**.

## Data Availability

The Dairy Molecule Database (DMD), including the integrated milk compound information generated in this study, is available at dairymolecules.com. Access to the database requires account registration; requests for access can be made through the website or by contacting the corresponding authors.

## Code Availability

Custom code used to generate the figures, tables, and key computational analyses is available in the GitHub repository https://github.com/IBPA/DMI. Because access to some underlying study data is restricted, the repository does not include all raw or processed input data, and some figures or tables may not be fully reproducible from the public repository alone. Documentation is provided to describe the expected input formats and analysis workflow. Access to additional code or data needed for review can be provided to editors and reviewers upon reasonable request, subject to applicable data-use restrictions.

## Acknowledgements

We thank Nikhil Joshi from the UCD Bioinformatics Core for pre-processing of the miRNA-Seq data.

## Funding Statement

This genomic analysis was performed at the DNA Technologies and Expression Analysis Core at the UC Davis Genome Center and supported by NIH Shared Instrumentation Grant 1S10OD010786-01. The support for the 600 MHz NMR for targeted metabolomics analysis was provided by the National Institutes of Health grant 1S10RR011973-01.

## Competing interests

This study received research funding from Dairy Management Inc. (DMI), an organization with interests in dairy research and the dairy industry. The authors declare no other competing interests.

# Tables

## Table 1. Summary of the milk product metadata

| Factors | | Values |
|---|---|---|
| Product Information* | Brand | Brand A, Brand B<br>(*Two sampled brands per region.*) |
| | Product Type | Whole milk (N = 58), Reduced fat milk (N = 2) |
| | Packaging Opaqueness | Transparent (N = 42), Opaque (N = 18) |
| | Expiration Date | 2022/06/07 – 2022/8/28<br>(5 ~ 34 days after the purchase) |
| Purchase Information | Purchase Region | Northeast, Mid-South, West, South Central, Plains, California-1, Southeast, Great Lakes, California-2, California-3 |
| | Purchase Time | Time 1: 2022 Jun (May for Pilot Samples (California-2))<br>Time 2: 2022 Jul (Jun for Pilot Samples (California-2))<br>Time 3: 2022 Aug (Jul for Pilot Samples (California-2)) |
| | Temperature (°F)<br>(Purchase region) | 53 – 89 |
| | Store Crowdedness | Empty (N = 41), Middle (N = 18), Crowded (N = 1) |
| | Dairy Case Crowdedness | Empty (N = 43), Middle (N = 12), Crowded (N = 5) |
| | Dairy Case Type | Closed (N = 54), Open (N = 6) |
| Nutritional Fact (Per Serving) ** | Calories (kcal) | 120 (Reduced fat milk), 150 – 160 (Whole milk) |
| | Calcium<br>(% of the Daily value (DV)) | 20 – 30 |
| | Saturated Fat (g) | 3 (Reduced fat milk), 4.5 – 5 (Whole milk) |
| | Total Fat (g) | 5 (Reduced fat milk), 8 – 9 (Whole milk) |
| | Cholesterol (mg) | 20 – 35 |
| | Sodium (mg) | 85 – 135 |
| | Total sugars (g) | 10 – 13 |
| | Total carbohydrates (g) | 11 – 13 |
| | Vitamin A (% of the DV) | 0 (N = 25), 6 (N = 6), 10 (N = 27), 15 (N = 2) |
| | Vitamin C (% of the DV) | 0 (N = 57), 4 (N = 3) |
| | Vitamin D (% of the DV) | 10 – 25 |
| | Vitamin E (% of the DV) | 0 (N = 57), 35 (N = 3) |
| | Potassium (% of the DV) | 6 – 10 |
| | Serving Size (mL) | 236 (N = 3), 240 (N = 57) |
| | Servings Per Container | 4 (Quart, N = 15), 8 (Half gallon, N = 40), 16 (One gallon, N = 5) |

* High-Temperature Short-Time (HTST) was applied to all 60 products for 16 seconds.

** No fiber, iron, or trans fats were reported in these milk products.

**Table 2.** Comparison of the number of detected compounds and SNPs in this study versus prior milk omics resources and studies.

| | DMD (our study) | MCDB[29] | MBPDB[27,28] | BoMiProt[25,26] |
|---|---|---|---|---|
| Data Source | Direct Measurement (60 US Milk Samples) | Direct Measurement (4 Canadian Milk Samples) | Literature Curation | Literature Curation |
| Peptides | 4,412 (Quantified) | N/A | 490 (Not Quantified) | N/A |
| Proteins | 645 (Quantified) | N/A | N/A | 9,336 (Not Quantified) |
| Small Molecules (Metabolites, Lipids, Oligosaccharides) | 231 (Quantified)<br>998 (Not Quantified) | 1,649 (Quantified)<br>709 (Not Quantified) | N/A | N/A |
| miRNAs | 428 (Not Quantified) | N/A | N/A | N/A |
| DNA SNPs | 25,999,881 SNPs | N/A | N/A | N/A |

[*] MCDB, MBPDB, and BoMiProt were accessed on May 26, 2026.

**Table 3.** Summary of the bioactivity of the U.S. milk.

| Bioactivity Category | Main Compound Contributed to the Bioactivity (Average Concentration) | Bioactivity | Estimated Average Efficacy (%) | |
|---|---|---|---|---|
| | | | In vitro | In vivo |
| Anticancer | Cytidine **(13.8 μM)** | Leukemia (AML) cell viability reduction (AID: 1963878) | 80.9 | 18.0 |
| | | Ewing sarcoma cell viability reduction (AID: 1347089) | 72.6 | 9.7 |
| | | Carcinoma (MCC) cell viability reduction (AID: 1296009) | 69.5 | 8.3 |
| | | Neuroblastoma cell viability reduction (AID: 1347100) | 65.6 | 13.1 |
| | | AT/RT cell viability reduction (AID: 1347094) | 53.4 | 9.9 |
| | | Osteosarcoma cell viability reduction (AID:1347103) | 36.4 | 0.9 |
| | | Myeloma cell viability reduction (AID:1918981) | 28.9 | 0.0 |
| | Linolenic acid (Octadecatrienoic acid-(9Z,12Z,15Z)) **(288.4 μM)** | Polymerase Eta inhibition (AID:588591) | 96.1 | 0.0 |
| | | Tdp1 inhibition (AID:485290) | 89.8 | 0.8 |
| | | Polymerase Iota inhibition (AID:720496) | 83.8 | 0.0 |
| | | WRN inhibition (AID:651768) | 78.5 | 0.0 |
| | Citric acid **(8585.4 μM)** | Jurkat Cell toxicity (AID:655) | 99.4 | 0.0 |
| | Palmitoleic acid (9Z-hexadecenoic acid) **(1336.0 μM)** | GPX4 inhibition (AID:1845192) | 99.3 | 0.8 |
| | Arachidonic acid (Icosatetraenoic acid-(5Z,8Z,11Z,14Z)) **(76.9 μM)** | RECQ1 inhibition (AID:2549) | 63.5 | 0.1 |
| | DHA (Docosahexaenoic acid-(4Z,7Z,10Z,13Z,16Z,19Z)) **(7.4 μM)** | G9a inhibition (AID:504332) | 12.7 | 0.1 |
| Disease Therapy | Oleic acid (9-octadecenoic acid) **(18879.3 μM)** | GCN5L2 inhibition - multiple diseases therapy (AID:504327) | 107.5 | 0.5 |
| | | USP1 inhibition - multiple diseases therapy (AID:504865) | 94.0 | 0.0 |
| | Peptide: MAP from β-casein **(269.7 μM)***  | ACE inhibition - hypertension therapy (Norris et al. 2014) | 100.0* | 0.0* |
| | Peptide: IPI from κ-casein **(207.0 μM)*** | DPP-IV inhibition - T2D therapy (Nongonierma et al. 2013) | 99.8* | 0.0* |
| | Peptide: VYPFPGPIP from β-casein **(269.7 μM)*** | PREP inhibition - multiple diseases therapy (Asano et al. 1992) | 93.8* | 0.0* |
| | Linolenic acid (Octadecatrienoic acid-(9Z,12Z,15Z)) **(288.4 μM)** | VDR inhibition - hyperparathyroidism therapy (AID:504847) | 100.9 | 0.0 |
| | | Tau aggregation inhibition - Alzheimer's therapy (AID:1460) | 95.0 | 0.0 |
| | Citric acid **(8585.4 μM)** | ALDH1A1 inhibition - metabolic disease therapy (AID:1030) | 100.0 | 0.5 |
| Anti-oxidation | Peptide: VYPFPGPIPN (from β-casein) **(269.7 μM)*** | ABTS reduction (Eisele et al. 2013) | 99.8* | 0.0* |
| | Peptide: EL (from α-S1-casein) **(1984.4 μM)**** | XTT reduction (Suetsuna et al. 2000) | 99.6* | 0.0* |
| | Peptide: YFYPEL (from α-S1-casein) **(661.5 μM)*** | Radical scavenging - ORAC (Contreras et al. 2013) | 96.9* | 0.0* |
| | Linoleic acid (Octadecadienoic acid-(9Z,12Z)) **(2777.0 μM)** | 15-hLO-1 inhibition (AID: 887) | 85.8 | 73.4 |
| | Linolenic acid (Octadecatrienoic acid-(9Z,12Z,15Z)) **(288.4 μM)** | 15-hLO-2 inhibition (AID:881) | 73.6 | 0.0 |
| | Tyrosine **(4.6 μM)** | 12hLO inhibition (AID:1671190) | 70.8 | 0.0 |
| Antimicrobial | Linolenic acid (Octadecatrienoic acid-(9Z,12Z,15Z)) **(288.4 μM)** | *S. mansoni* TGR inhibition (AID:485364) | 90.0 | 0.0 |
| | Glutamic acid **(366.2 μM)** | *B. subtilis* PPTases inhibition (AID:1490) | 74.9 | 0.0 |
| | DHA (Docosahexaenoic acid-(4Z,7Z,10Z,13Z,16Z,19Z)) **(7.4 μM)** | Malaria parasite viability reduction (AID:504749) | 9.3 | 0.0 |
| Antivirus | Stearic acid (Octadecanoic acid) **(9998.3 μM)** | Chikungunya nsP2 inhibition (AID: 1964101) | 96.6 | 25.8 |
| | Uridine **(34.1 μM)** | Influenza A NS1 inhibition (AID:2326) | 61.4 | 0.0 |

* The bioactivity efficacies of peptides were evaluated based on the following formula which takes the concentration of the protein, the number of intervals that contain the peptide sequences (which is equal to 1 for the most cases except few peptides), and the concentration of the free peptide (which is negligible compared to the concentration of the source proteins):

$$Conc_{Final} = Conc_{Protein} \times n_{Interval} + Conc_{Free\ Peptide}$$

**There are three intervals that contain the peptide sequences (EL) in α-S1-casein so the final concentration is three times the concentration of α-S1-casein.

## Figure Captions

**Figure 1. Overview of multi-omics profiling and bioactivity-efficacy estimation for U.S. milk samples. a,** Study design and analytical workflow. Sixty commercial whole-milk samples were collected from ten U.S. regions, with two brands per region sampled at three purchase time points. SNP profiles, compound concentrations, and bioactivity efficacies were obtained or estimated using genomic profiling, multi-omics measurements, and integrative computational analyses. **b,** The associations among the product-level factors and SNPs, bioactive compounds, and bioactivities.

**Figure 2. Composition of milk compounds across samples. a.** Composition of major molecular classes within dry milk content (total quantified solids). **b.** Relative differences between quantified total sugar, total protein, total fat, and cholesterol values and Nutrition Facts label values. The means (with the p-value of two-sided t-test) and the ranges of the relative differences in the parentheses are also shown. **c.** Comparison of the labeled total sugar, total fat, and total protein and the corresponding measured concentrations.

**Figure 3. Estimated bioactivity efficacy of milk compounds. a,** Average estimated in vitro and in vivo bioactivity efficacy across regions. In vivo efficacy was estimated from predicted postprandial serum concentrations, with additional sensitivity analyses performed by scaling the estimated serum concentrations by 0.01×, 0.1×, 10×, and 100×. **b,** Compound–in vitro bioactivity associations. The bioactivities and the corresponding compounds with estimated in vivo efficacy greater than 5% were highlighted by bold text and underlined. For each bioactivity, only the compound contributing the highest estimated efficacy is shown.

**Figure 4. Purchase regions drive bioactivity variation. a.** Concentration variation of selected bioactive compounds and **b.** corresponding estimated in vitro bioactivity efficacy variation, shown as differences from the overall mean. **c.** Regional variation in selected compounds and corresponding bioactivity efficacies.

**Figure 5. Product, regional, and genetic factors associated with milk bioactivity variability.** Integrated network linking milk compounds, estimated bioactivities, purchase region, Jersey breed ratio, and SNPs. Edges indicate significant compound–bioactivity, compound–metadata, compound–Jersey proportion, and SNP–compound associations.

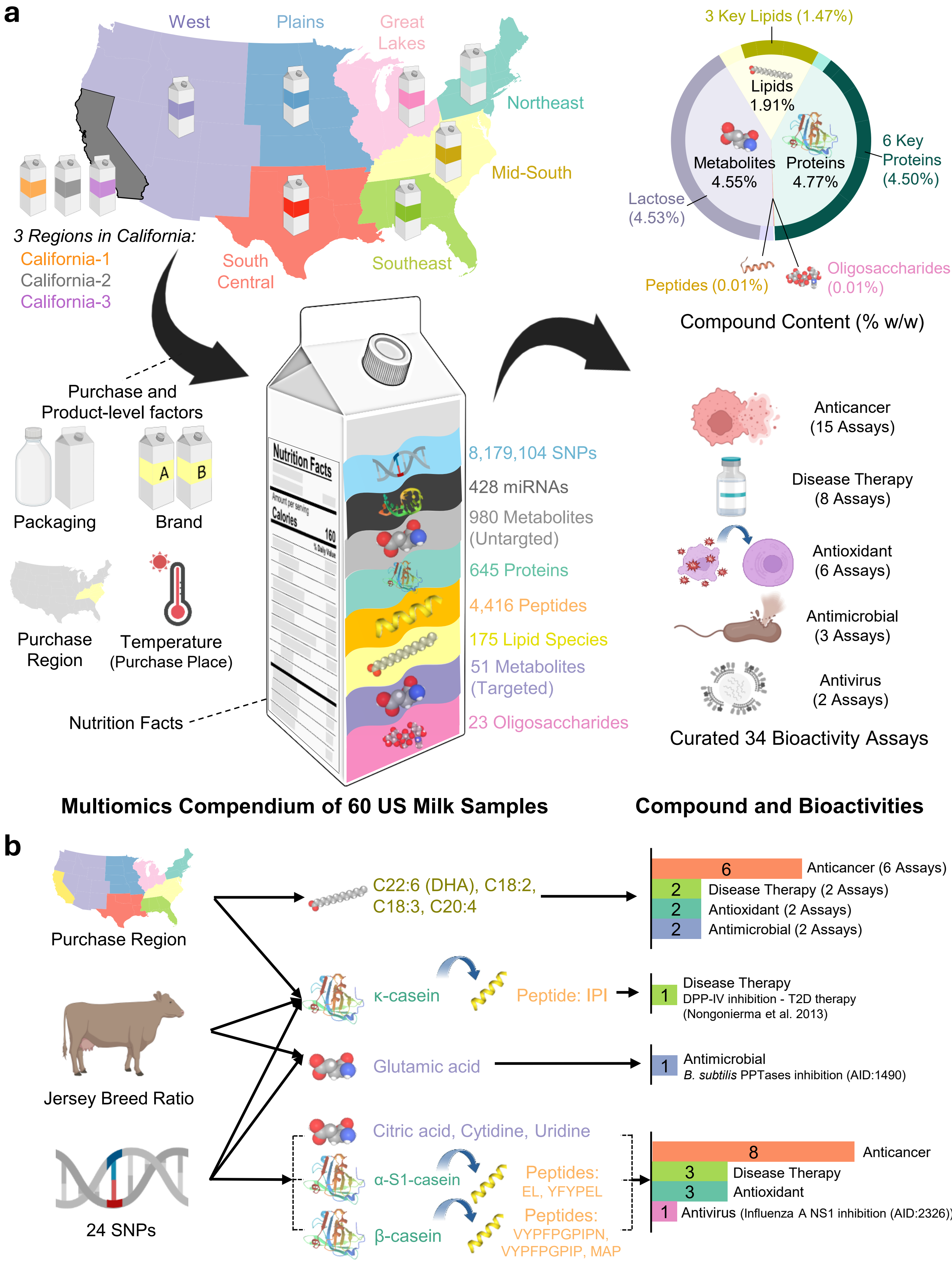

a
West
Plains
Great Lakes
Northeast
Mid-South
Southeast
South Central
3 Regions in California:
California-1
California-2
California-3
3 Key Lipids (1.47%)
Lipids 1.91%
Metabolites 4.55%
Proteins 4.77%
6 Key Proteins (4.50%)
Lactose (4.53%)
Peptides (0.01%)
Oligosaccharides (0.01%)
Compound Content (% w/w)
Purchase and Product-level factors
Packaging
Brand
A
B
Purchase Region
Temperature (Purchase Place)
Nutrition Facts
8,179,104 SNPs
428 miRNAs
980 Metabolites (Untargted)
645 Proteins
4,416 Peptides
175 Lipid Species
51 Metabolites (Targeted)
23 Oligosaccharides
Anticancer (15 Assays)
Disease Therapy (8 Assays)
Antioxidant (6 Assays)
Antimicrobial (3 Assays)
Antivirus (2 Assays)
Curated 34 Bioactivity Assays
Multiomics Compendium of 60 US Milk Samples
Compound and Bioactivities
b
Purchase Region
Jersey Breed Ratio
24 SNPs
C22:6 (DHA), C18:2, C18:3, C20:4
6 Anticancer (6 Assays)
2 Disease Therapy (2 Assays)
2 Antioxidant (2 Assays)
2 Antimicrobial (2 Assays)
κ-casein
Peptide: IPI
1 Disease Therapy
DPP-IV inhibition - T2D therapy
(Nongonierma et al. 2013)
Glutamic acid
1 Antimicrobial
B. subtilis PPTases inhibition (AID:1490)
Citric acid, Cytidine, Uridine
α-S1-casein
Peptides: EL, YFYPEL
β-casein
Peptides: VYPFPGPIPN, VYPFPGPIP, MAP
8 Anticancer
3 Disease Therapy
3 Antioxidant
1 Antivirus (Influenza A NS1 inhibition (AID:2326))

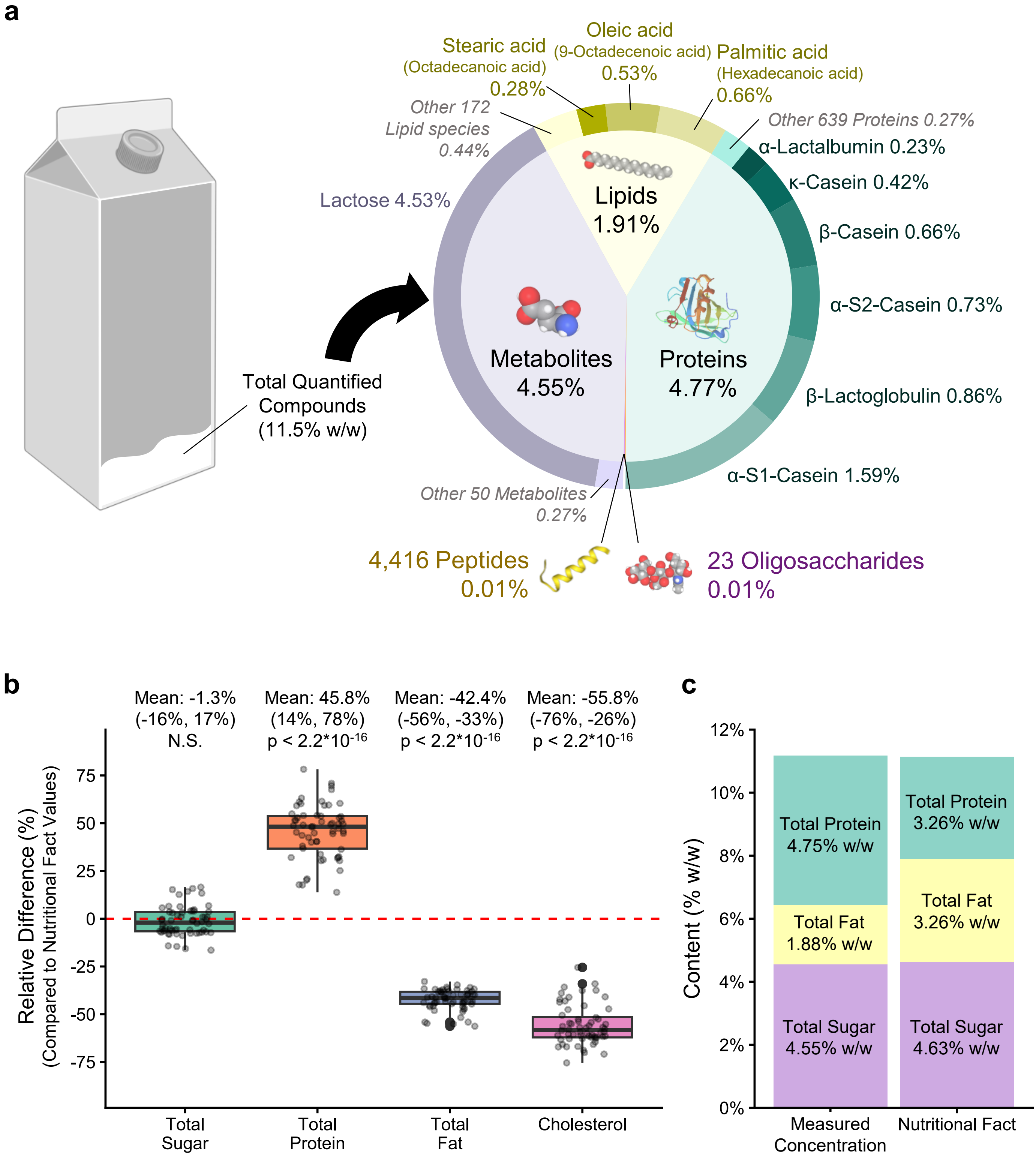
a
Stearic acid
(Octadecanoic acid)
0.28%
Oleic acid
(9-Octadecenoic acid)
0.53%
Palmitic acid
(Hexadecanoic acid)
0.66%
Other 172 Lipid species 0.44%
Other 639 Proteins 0.27%
α-Lactalbumin 0.23%
κ-Casein 0.42%
β-Casein 0.66%
α-S2-Casein 0.73%
β-Lactoglobulin 0.86%
α-S1-Casein 1.59%
Lactose 4.53%
Lipids 1.91%
Metabolites 4.55%
Proteins 4.77%
Total Quantified Compounds (11.5% w/w)
Other 50 Metabolites 0.27%
4,416 Peptides 0.01%
23 Oligosaccharides 0.01%
b
Mean: -1.3% (-16%, 17%) N.S.
Mean: 45.8% (14%, 78%) p < 2.2*10^-16
Mean: -42.4% (-56%, -33%) p < 2.2*10^-16
Mean: -55.8% (-76%, -26%) p < 2.2*10^-16
Relative Difference (%)
(Compared to Nutritional Fact Values)
75
50
25
0
-25
-50
-75
Total Sugar
Total Protein
Total Fat
Cholesterol
c
Content (% w/w)
12%
10%
8%
6%
4%
2%
0%
Total Protein 4.75% w/w
Total Fat 1.88% w/w
Total Sugar 4.55% w/w
Total Protein 3.26% w/w
Total Fat 3.26% w/w
Total Sugar 4.63% w/w
Measured Concentration
Nutritional Fact

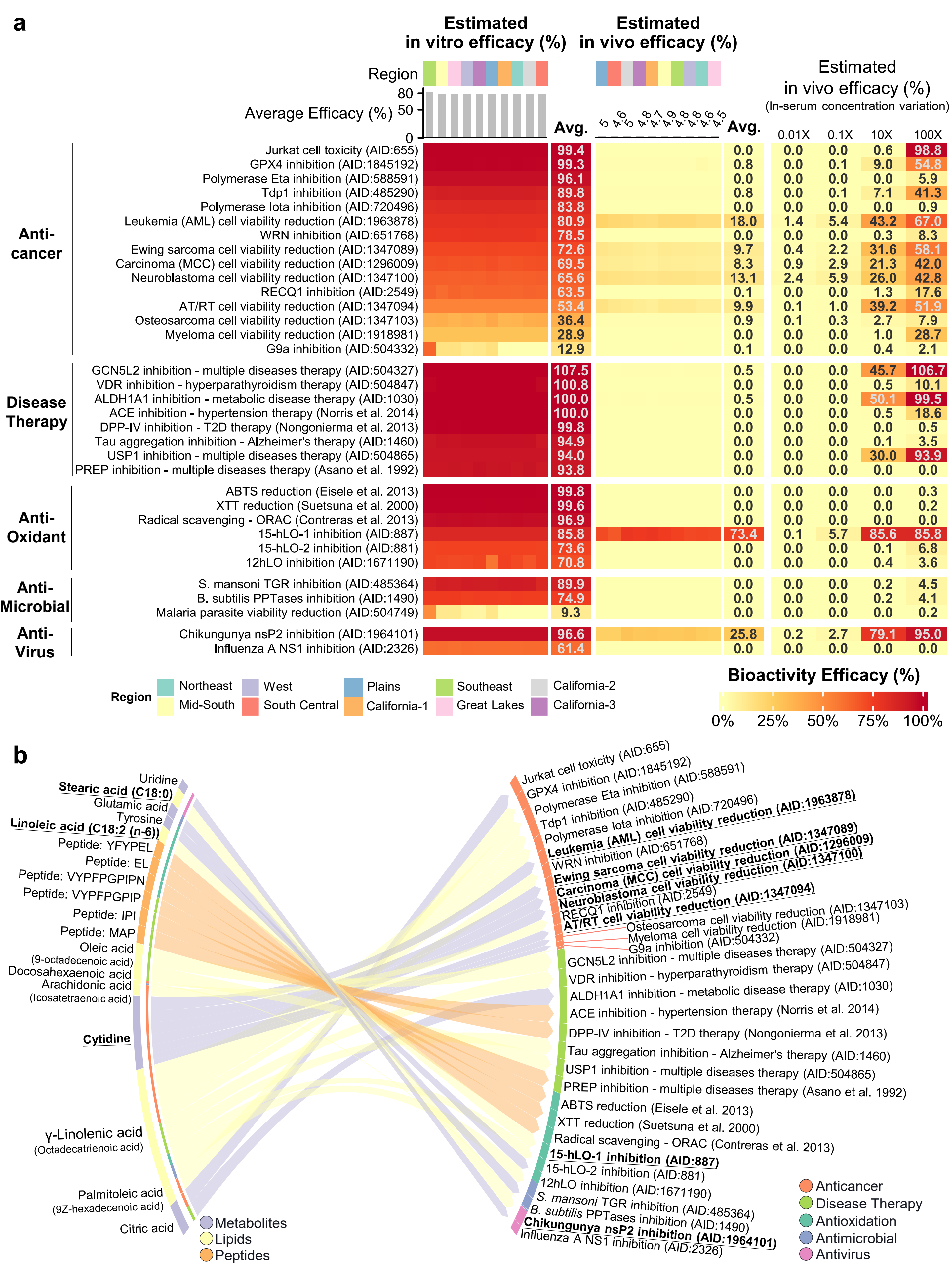

a
Estimated in vitro efficacy (%)
Estimated in vivo efficacy (%)
Region
Average Efficacy (%)
80
50
0
Avg.
Avg.
Estimated in vivo efficacy (%) (In-serum concentration variation)
0.01X
0.1X
10X
100X
Anti-cancer
Jurkat cell toxicity (AID:655) 99.4 0.0 0.0 0.0 0.6 98.8
GPX4 inhibition (AID:1845192) 99.3 0.8 0.0 0.1 9.0 54.8
Polymerase Eta inhibition (AID:588591) 96.1 0.0 0.0 0.0 0.0 5.9
Tdp1 inhibition (AID:485290) 89.8 0.8 0.0 0.1 7.1 41.3
Polymerase Iota inhibition (AID:720496) 83.8 0.0 0.0 0.0 0.0 0.9
Leukemia (AML) cell viability reduction (AID:1963878) 80.9 18.0 1.4 5.4 43.2 67.0
WRN inhibition (AID:651768) 78.5 0.0 0.0 0.0 0.3 8.3
Ewing sarcoma cell viability reduction (AID:1347089) 72.6 9.7 0.4 2.2 31.6 58.1
Carcinoma (MCC) cell viability reduction (AID:1296009) 69.5 8.3 0.9 2.9 21.3 42.0
Neuroblastoma cell viability reduction (AID:1347100) 65.6 13.1 2.4 5.9 26.0 42.8
RECQ1 inhibition (AID:2549) 63.5 0.1 0.0 0.0 1.3 17.6
AT/RT cell viability reduction (AID:1347094) 53.4 9.9 0.1 1.0 39.2 51.9
Osteosarcoma cell viability reduction (AID:1347103) 36.4 0.9 0.1 0.3 2.7 7.9
Myeloma cell viability reduction (AID:1918981) 28.9 0.0 0.0 0.0 1.0 28.7
G9a inhibition (AID:504332) 12.9 0.1 0.0 0.0 0.4 2.1
Disease Therapy
GCN5L2 inhibition - multiple diseases therapy (AID:504327) 107.5 0.5 0.0 0.0 45.7 106.7
VDR inhibition - hyperparathyroidism therapy (AID:504847) 100.8 0.0 0.0 0.0 0.5 10.1
ALDH1A1 inhibition - metabolic disease therapy (AID:1030) 100.0 0.5 0.0 0.0 50.1 99.5
ACE inhibition - hypertension therapy (Norris et al. 2014) 100.0 0.0 0.0 0.0 0.5 18.6
DPP-IV inhibition - T2D therapy (Nongonierma et al. 2013) 99.8 0.0 0.0 0.0 0.0 0.5
Tau aggregation inhibition - Alzheimer's therapy (AID:1460) 94.9 0.0 0.0 0.0 0.1 3.5
USP1 inhibition - multiple diseases therapy (AID:504865) 94.0 0.0 0.0 0.0 30.0 93.9
PREP inhibition - multiple diseases therapy (Asano et al. 1992) 93.8 0.0 0.0 0.0 0.0 0.0
Anti-Oxidant
ABTS reduction (Eisele et al. 2013) 99.8 0.0 0.0 0.0 0.0 0.3
XTT reduction (Suetsuna et al. 2000) 99.6 0.0 0.0 0.0 0.0 0.2
Radical scavenging - ORAC (Contreras et al. 2013) 96.9 0.0 0.0 0.0 0.0 0.0
15-hLO-1 inhibition (AID:887) 85.8 73.4 0.1 5.7 85.6 85.8
15-hLO-2 inhibition (AID:881) 73.6 0.0 0.0 0.0 0.1 6.8
12hLO inhibition (AID:1671190) 70.8 0.0 0.0 0.0 0.4 3.6
Anti-Microbial
S. mansoni TGR inhibition (AID:485364) 89.9 0.0 0.0 0.0 0.2 4.5
B. subtilis PPTases inhibition (AID:1490) 74.9 0.0 0.0 0.0 0.2 4.1
Malaria parasite viability reduction (AID:504749) 9.3 0.0 0.0 0.0 0.0 0.2
Anti-Virus
Chikungunya nsP2 inhibition (AID:1964101) 96.6 25.8 0.2 2.7 79.1 95.0
Influenza A NS1 inhibition (AID:2326) 61.4 0.0 0.0 0.0 0.0 0.0
Region
Northeast
Mid-South
West
South Central
Plains
California-1
Southeast
Great Lakes
California-2
California-3
Bioactivity Efficacy (%)
0%
25%
50%
75%
100%
b
Uridine
Stearic acid (C18:0)
Glutamic acid
Tyrosine
Linoleic acid (C18:2 (n-6))
Peptide: YFYPEL
Peptide: EL
Peptide: VYPFPGPIPN
Peptide: VYPFPGPIP
Peptide: IPI
Peptide: MAP
Oleic acid (9-octadecenoic acid)
Docosahexaenoic acid
Arachidonic acid (Icosatetraenoic acid)
Cytidine
γ-Linolenic acid (Octadecatrienoic acid)
Palmitoleic acid (9Z-hexadecenoic acid)
Citric acid
Metabolites
Lipids
Peptides
Anticancer
Disease Therapy
Antioxidation
Antimicrobial
Antivirus

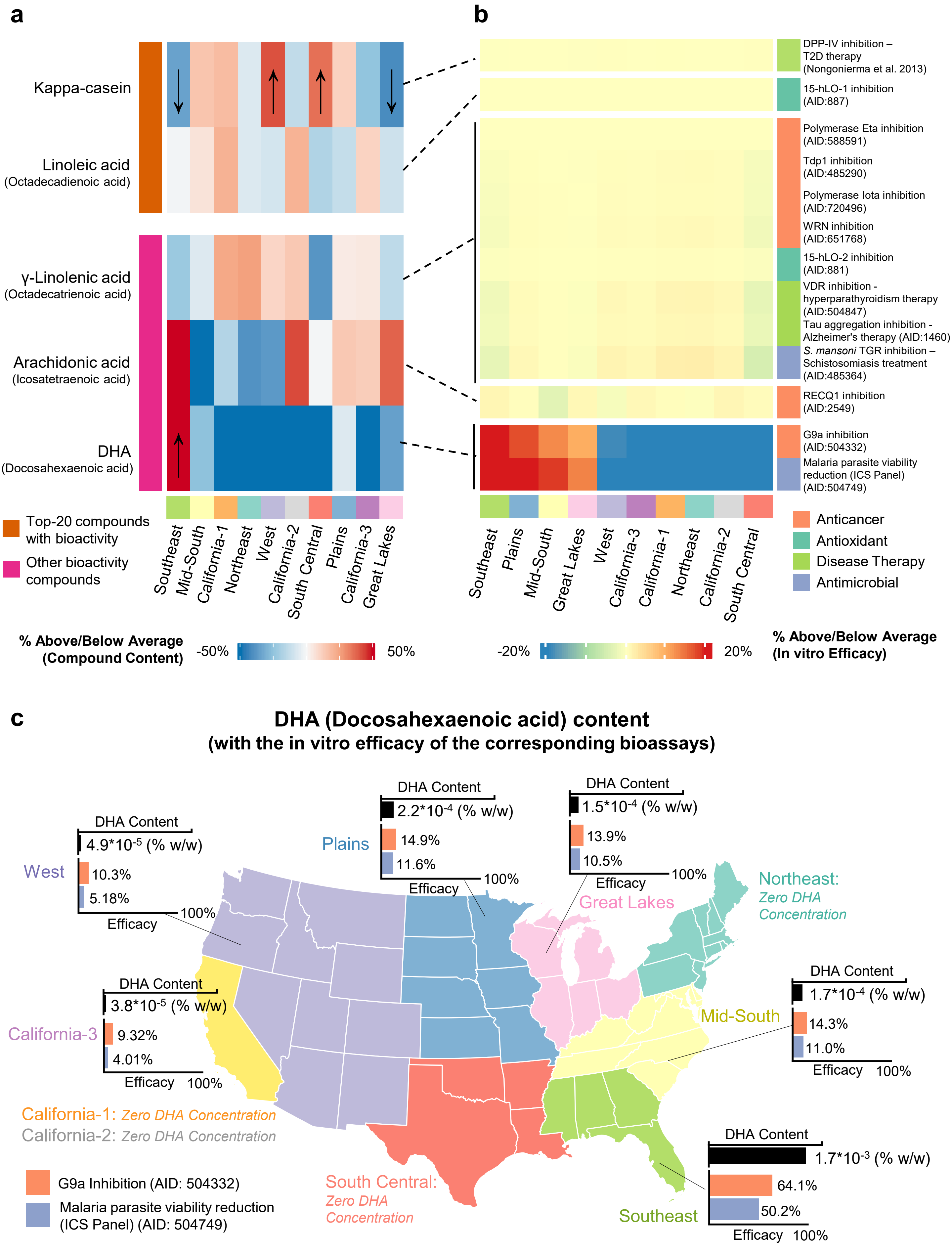

a
Kappa-casein
Linoleic acid
(Octadecadienoic acid)
γ-Linolenic acid
(Octadecatrienoic acid)
Arachidonic acid
(Icosatetraenoic acid)
DHA
(Docosahexaenoic acid)
Top-20 compounds with bioactivity
Other bioactivity compounds
Southeast
Mid-South
California-1
Northeast
West
California-2
South Central
Plains
California-3
Great Lakes
% Above/Below Average (Compound Content)
-50%
50%
b
DPP-IV inhibition – T2D therapy (Nongonierma et al. 2013)
15-hLO-1 inhibition (AID:887)
Polymerase Eta inhibition (AID:588591)
Tdp1 inhibition (AID:485290)
Polymerase Iota inhibition (AID:720496)
WRN inhibition (AID:651768)
15-hLO-2 inhibition (AID:881)
VDR inhibition - hyperparathyroidism therapy (AID:504847)
Tau aggregation inhibition - Alzheimer's therapy (AID:1460)
S. mansoni TGR inhibition – Schistosomiasis treatment (AID:485364)
RECQ1 inhibition (AID:2549)
G9a inhibition (AID:504332)
Malaria parasite viability reduction (ICS Panel) (AID:504749)
Southeast
Plains
Mid-South
Great Lakes
West
California-3
California-1
Northeast
California-2
South Central
Anticancer
Antioxidant
Disease Therapy
Antimicrobial
-20%
20%
% Above/Below Average (In vitro Efficacy)
c
DHA (Docosahexaenoic acid) content
(with the in vitro efficacy of the corresponding bioassays)
West
DHA Content
4.9*10-5 (% w/w)
10.3%
5.18%
Efficacy
100%
Plains
DHA Content
2.2*10-4 (% w/w)
14.9%
11.6%
Efficacy
100%
Great Lakes
DHA Content
1.5*10-4 (% w/w)
13.9%
10.5%
Efficacy
100%
Northeast: Zero DHA Concentration
California-3
DHA Content
3.8*10-5 (% w/w)
9.32%
4.01%
Efficacy
100%
Mid-South
DHA Content
1.7*10-4 (% w/w)
14.3%
11.0%
Efficacy
100%
California-1: Zero DHA Concentration
California-2: Zero DHA Concentration
South Central: Zero DHA Concentration
Southeast
DHA Content
1.7*10-3 (% w/w)
64.1%
50.2%
Efficacy
100%
G9a Inhibition (AID: 504332)
Malaria parasite viability reduction (ICS Panel) (AID: 504749)

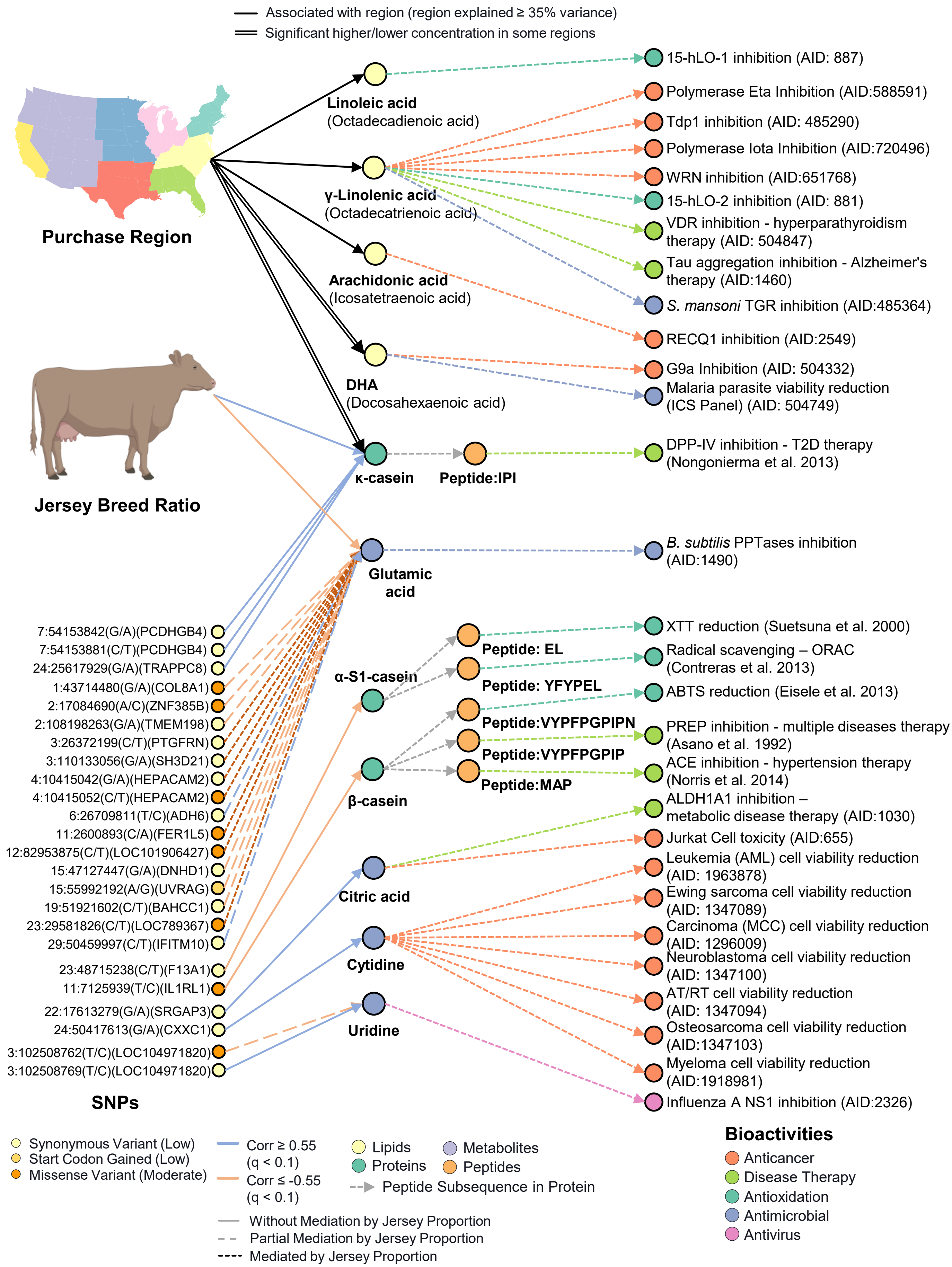

Associated with region (region explained ≥ 35% variance)
Significant higher/lower concentration in some regions
Purchase Region
Jersey Breed Ratio
Linoleic acid
(Octadecadienoic acid)
γ-Linolenic acid
(Octadecatrienoic acid)
Arachidonic acid
(Icosatetraenoic acid)
DHA
(Docosahexaenoic acid)
κ-casein
Peptide:IPI
Glutamic acid
α-S1-casein
β-casein
Peptide: EL
Peptide: YFYPEL
Peptide:VYPFPGPIPN
Peptide:VYPFPGPIP
Peptide:MAP
Citric acid
Cytidine
Uridine
15-hLO-1 inhibition (AID: 887)
Polymerase Eta Inhibition (AID:588591)
Tdp1 inhibition (AID: 485290)
Polymerase Iota Inhibition (AID:720496)
WRN inhibition (AID:651768)
15-hLO-2 inhibition (AID: 881)
VDR inhibition - hyperparathyroidism therapy (AID: 504847)
Tau aggregation inhibition - Alzheimer's therapy (AID:1460)
S. mansoni TGR inhibition (AID:485364)
RECQ1 inhibition (AID:2549)
G9a Inhibition (AID: 504332)
Malaria parasite viability reduction (ICS Panel) (AID: 504749)
DPP-IV inhibition - T2D therapy (Nongonierma et al. 2013)
B. subtilis PPTases inhibition (AID:1490)
XTT reduction (Suetsuna et al. 2000)
Radical scavenging – ORAC (Contreras et al. 2013)
ABTS reduction (Eisele et al. 2013)
PREP inhibition - multiple diseases therapy (Asano et al. 1992)
ACE inhibition - hypertension therapy (Norris et al. 2014)
ALDH1A1 inhibition – metabolic disease therapy (AID:1030)
Jurkat Cell toxicity (AID:655)
Leukemia (AML) cell viability reduction (AID: 1963878)
Ewing sarcoma cell viability reduction (AID: 1347089)
Carcinoma (MCC) cell viability reduction (AID: 1296009)
Neuroblastoma cell viability reduction (AID: 1347100)
AT/RT cell viability reduction (AID: 1347094)
Osteosarcoma cell viability reduction (AID:1347103)
Myeloma cell viability reduction (AID:1918981)
Influenza A NS1 inhibition (AID:2326)
7:54153842(G/A)(PCDHGB4)
7:54153881(C/T)(PCDHGB4)
24:25617929(G/A)(TRAPPC8)
1:43714480(G/A)(COL8A1)
2:17084690(A/C)(ZNF385B)
2:108198263(G/A)(TMEM198)
3:26372199(C/T)(PTGFRN)
3:110133056(G/A)(SH3D21)
4:10415042(G/A)(HEPACAM2)
4:10415052(C/T)(HEPACAM2)
6:26709811(T/C)(ADH6)
11:2600893(C/A)(FER1L5)
12:82953875(C/T)(LOC101906427)
15:47127447(G/A)(DNHD1)
15:55992192(A/G)(UVRAG)
19:51921602(C/T)(BAHCC1)
23:29581826(C/T)(LOC789367)
29:50459997(C/T)(IFITM10)
23:48715238(C/T)(F13A1)
11:7125939(T/C)(IL1RL1)
22:17613279(G/A)(SRGAP3)
24:50417613(G/A)(CXXC1)
3:102508762(T/C)(LOC104971820)
3:102508769(T/C)(LOC104971820)
SNPs
Synonymous Variant (Low)
Start Codon Gained (Low)
Missense Variant (Moderate)
Corr ≥ 0.55 (q < 0.1)
Corr ≤ -0.55 (q < 0.1)
Lipids
Proteins
Metabolites
Peptides
Peptide Subsequence in Protein
Without Mediation by Jersey Proportion
Partial Mediation by Jersey Proportion
Mediated by Jersey Proportion
Bioactivities
Anticancer
Disease Therapy
Antioxidation
Antimicrobial
Antivirus

# Supplementary Information

## 1. Compendium Description and Details of Analysis Methods

### 1.1 Genomic data statistics

We compared detected 25,999,881 SNPs with the published cow SNP database BGVD[1]: 8,179,104 variants (31.4%) were recorded in BGVD, and 75,346 variants (0.29%) were annotated by SnpEff[2] as having high-, moderate-, or low-impact consequences (**Supplementary Figure 14a, b**). SNP allele-frequency profiles were highly consistent across samples: 91% of BGVD-recorded SNPs were observed in at least 90% of samples, and more than 95% were observed in at least 50% of samples at a minimum sequencing depth of 5 (**Supplementary Figure 14c, d**).

### 1.2 Milk bulk composition

A small subset of high-abundance compounds, approximately the top 10-20 by concentration, accounted for most of the quantified milk solids and explained most of the variance observed by principal component analysis (PCA) (**Supplementary Figure 15**). Specifically, the top 10 compounds (six proteins, three lipids, and lactose), top 20 compounds, and top 50 compounds by concentration represented approximately 91.3%, 97.1%, and 99.2% of the total dry milk mass, respectively, where dry milk mass was defined as the sum of quantified compounds (**Supplementary Figure 3a, b**). These results indicate that a small number of major components dominate the measured compositional profile. Consistent with this observation, all 60 samples showed highly similar concentration profiles, with minimum pairwise Pearson correlation values of 0.97 and 0.98 when calculated using the top 10 highest-concentration compounds and all compounds, respectively (**Supplementary Figure 16**).

### 1.3 Milk bioactivity profiles

We profiled milk bioactivity by curating data from the PubChem BioAssay Database[3] and the Milk Bioactive Peptide Database (MBPDB)[4,5] (**Supplementary Figure 13**). The milk samples contained 20 small bioactive molecules (13 lipids, 7 metabolites) and 264 bioactive peptides from 9 source milk proteins associated with 34 curated bioactivities, which we classified into five categories: anticancer, disease therapy, antioxidant, antimicrobial, and antiviral activity (**Supplementary Table 2**). Of these, 28 bioactivities were curated from the PubChem BioAssay Database[3] and included 114 Hill-curve[6] parameter sets describing bioactivity-compound pairs with maximum efficacy relative to the positive control. Using the Hill curve, maximum efficacy, and measured bioactive compound concentration in milk, we estimated the in vitro efficacy of these 28 bioactivities contributed by 8 lipids and 7 metabolites (**Supplementary Table 12**). For the remaining six peptide-associated bioactivities curated from MBPDB[4,5] that reported IC50 values but not full Hill-curve parameters, we compiled the 114 available Hill curves from the 28 small-molecule bioactivities (**Supplementary Figure 17a**) and used them to build a representative response model. Bootstrap evaluation (100 iterations) showed that, across different fractions of Hill-curve data used for model construction, the 95th percentile of absolute estimation error and the mean absolute estimation error were approximately 25% and 10%, respectively (**Supplementary Figure 17b**). Using this framework, we estimated in vitro efficacy for six peptide-associated bioactivities based on the measured concentrations of the corresponding bioactive peptides (**Supplementary Table 13**).

In addition to database-derived bioactivity estimates, we experimentally evaluated milk compounds in three bioactivity assays, focusing on neurite outgrowth, anti-inflammatory activity, and antibacterial activity. For neurite outgrowth and anti-inflammatory activity measurement, we selected 10 compounds (Docosahexaenoic acid, arachidonic acid, lactoferrin, oleic acid, beta-hydroxybutyrate, N-acetylneuraminic acid, choline, fucose, tryptophan, and butyrate) based on the abundance in milk, ability to cross the blood brain barrier, and the likelihood of impacting a cell signaling pathway relevant to neuroplasticity or neuroinflammation. For antibacterial activity, we selected 54 peptides from the 4,416 quantified peptides using a proprietary AI platform (**Supplementary Table 14**). Six peptides inhibited *A. baumannii* with $IC_{50}$ values below 256 μg/mL, and one of these six peptides showed weak activity against *E. coli*, with an $IC_{50}$ value below 1,024 μg/mL. None of the six peptides inhibited the growth of the probiotic species *L. plantarum* or *L. rhamnosus* within the tested concentration range. Four peptides showed higher selectivity indices than melittin, with the selectivity index defined as the ratio of the $IC_{50}$ value against the probiotic species to that against the pathogen (**Supplementary Table 3**). In addition, 35 peptides exhibited antibacterial activity against *A. baumannii* at 1,024 μg/mL, as indicated by significantly lower average $OD_{600}$ values than the negative control without peptide treatment (**Supplementary Table 4**). These results indicate that selected milk-derived peptides can inhibit pathogenic bacteria while sparing probiotic bacteria.

### 1.4 Correlation analysis linking compound concentrations with packaging opacity and temperature

To evaluate whether packaging opacity, purchase ambient temperature, or their interaction was associated with compound concentration, we fit a series of linear models.

$$Conc_i \sim Opacity_i \times Temperature_i$$

$$Conc_i \sim Opacity_i + Temperature_i$$

$$Conc_i \sim Opacity_i$$

$$Conc_i \sim Temperature_i$$

Where the × operator includes opacity, temperature, and the interaction between them. The full model included packaging opacity, temperature, and their interaction. The interaction term was evaluated by comparing the full interaction model with the corresponding additive model containing only opacity and temperature, and the effects of opacity and temperature were also evaluated by comparing the single factor model with full model. These model comparisons were performed using ANOVA-based F-tests

## 2. Details of Analysis Results

### 2.1 Multi-omics factor analysis (MOFA) and variance decomposition analysis results

Multi-omics factor analysis (MOFA)[7] and variance decomposition analysis revealed associations between omics profiles and production or purchase factors. First, we performed MOFA and selected the first three factors based on their explained variance ratios (**Supplementary Figure 18 and 19**). We used the R package MOFA[7] to extract latent factors from seven omics layers: genomics (29,356 SNPs with moderate or high predicted impact), transcriptomics (428 miRNAs), proteomics (645 proteins), lipidomics (175 lipids), targeted metabolomics (51 metabolites), glycomics (5 oligosaccharides), and peptidomics (4,412 peptides). Raw allele frequencies, miRNA counts, and compound contents (% w/w) were used without additional normalization. Default parameters were used, except that double precision and slow convergence mode were enabled. Although Factor 1 showed no clear association with production or purchase factors (**Supplementary Figure 20 and 21**), Factor 2 indicated that some samples from the South Central and West regions had higher allele frequencies for selected SNPs and higher concentrations of selected proteins, as reflected by the top 10 loadings for each omics layer (**Supplementary Figure 22 and 23**).

Second, we used variance decomposition to quantify the proportion of compound variance attributable to production and purchase factors, including brand, purchase region, and purchase time. Purchase region explained at least 35% of variance for 511 compounds, including 15 bioactive compounds, and was the dominant variance component for 284 compounds, including 12 bioactive compounds. Brand was the dominant variance component for 186 compounds, including 7 bioactive compounds (**Supplementary Figure 24a, b**). Among the top 100 highest-concentration compounds, 17%-26% had variance dominated by region and 20%-28% had at least 35% of variance explained by region. In contrast, the corresponding proportions for compounds outside the top 100 were lower (5.5% and 9.9%, respectively) (**Supplementary Figure 24c, d**), suggesting that region and brand explain a larger fraction of variance for higher-abundance compounds than for low-abundance compounds. We then focused on the selected molecules, including total molecular classes, the top 20 highest-concentration compounds, and the compounds that mainly contributed to the 34 bioactivities, and found that the variance of total oligosaccharides, total peptides, and eight individual compounds was dominated by purchase region (**Supplementary Figure 8**). Among these eight compounds, κ-casein was significantly higher in two regions and lower in two regions, and docosahexaenoic acid was significantly higher in one region, compared with the estimated marginal mean across all regions after adjustment for purchase time and brand.

### 2.2 Milk compound profile reconstruction using selected SNPs

We selected 61 SNPs that correlated with the first seven principal components of the milk compound profiles, using thresholds of Pearson and Spearman correlations greater than 0.45. Linear regression models based on these SNPs predicted the first seven principal components (**Supplementary Figure 25 and 26**), enabling reconstruction of centered milk compound profiles after subtracting the across-sample mean for each compound. In leave-one-region-out cross-validation, reconstructed profiles achieved an average Pearson correlation coefficient of 0.64 between observed and predicted values (**Supplementary Figure 27a**). Reconstruction using only five predicted PCs and 44 SNPs still achieved a Pearson correlation coefficient above 0.60 (**Supplementary Figure 27b**). Prediction performance also varied across held-out purchase regions (**Supplementary Figure 27c**). Although Spearman correlation was low when reconstructing the entire compound profile, the average leave-one-region-out Spearman correlation exceeded 0.50 for the top 10 or 20 highest-concentration compounds, which contributed most of the dry milk mass (**Supplementary Figure 28**).

### 2.3 SNP/miRNA – compound association analysis

We next assessed associations between compound concentrations and individual SNP allele frequencies or miRNA counts across a range of Pearson and Spearman correlation thresholds, using a fixed FDR-adjusted q-value threshold of 0.1 for both coefficients and restricting the analysis to compounds detected in at least 10 samples. At correlation coefficient thresholds of 0.4 for SNPs and 0.35 for miRNAs, we identified 343 SNP-bioactive compound pairs and 7 miRNA-bioactive compound pairs (**Supplementary Figure 29**). We then applied latent variable analysis (implemented as a R package lavaan[8]) to evaluate whether SNP–compound associations were mediated by estimated Jersey breed proportion using a structural equation model (SEM) to the SNP-compound pairs to distinguish associations mediated through estimated Jersey breed proportion from breed-independent associations. This analysis classified 128 pairs as mediated through Jersey proportion, 108 as not mediated through Jersey proportion, 100 as partially mediated, and 7 as neither significant in the SEM model (**Supplementary Figure 30a**). SNP-bioactive compound pairs mediated through Jersey proportion were enriched for SNPs with significantly different allele frequencies between Holstein and Jersey and for bioactive compounds associated with Jersey proportion (**Supplementary Figure 30b**). Finally, we visualized the 24 SNP-bioactive compound pairs that passed a stricter correlation threshold of 0.55 together with their corresponding bioactivities (**Figure 5**) and 7 miRNA-bioactive compound pairs that passed a correlation threshold of 0.35 together with their corresponding bioactivities (**Supplementary Figure 12**). These results suggest that specific milk compounds and associated bioactivities may be modulated through breed selection and through SNP-level genetic selection.

## 3. Omics Analysis

Multiple omics approaches were performed to identify and quantify SNPs, microRNAs, and molecular compounds in the milk samples. Genomics and transcriptomics analyses identified SNPs and microRNAs, respectively. Proteomics, peptidomics, lipidomics, and glycomics quantified the proteins, peptides, lipids, and oligosaccharides in the samples. Additionally, metabolites were identified and quantified using both targeted and untargeted metabolomics analyses. Below, we provide detailed methods for each omics analysis and additional results supporting the main manuscript.

### 3.1 Genomic DNA extraction and sample preparation

**Extraction of genomic DNA.** Cow milk samples were stored at –20 °C upon arrival. The protocol for isolating somatic cells and extracting genomic DNA was adapted from Pokorska et al. (2016)[9] with minor modifications. Samples were thawed overnight at 4 °C and processed the next day. Each sample (50 mL tube) was diluted with an equal volume of 1× phosphate-buffered saline (PBS: 137 mM NaCl, 12 mM phosphate, 2.7 mM KCl, pH 7.4) and incubated at 40 °C for 10 minutes. The samples were centrifuged at 7,000 × *g* for 30 minutes at 4 °C. After centrifugation, the top fat layer and supernatant were carefully removed, leaving the pellet containing somatic cells and precipitated protein. To remove residual fat, the pellet was washed with PBS: if a high fat content remained, the pellet was transferred to a 15 mL tube and washed with 5–10 mL PBS at 4 °C for 10 minutes (centrifuging at 4,500 × *g*); lower-fat pellets were washed in 1–2 mL PBS in a 2 mL tube for 3 minutes (4,500 × *g*). Repeated washing continued until the supernatant was clear of fat. Next, 1 mL of cell lysis buffer (pH 8.8; containing 6% SDS, 3 mM $MgCl_2$, 15 mM Tris-HCl, 0.5% DMSO, 6% acetone) was added to the pellet and incubated at 65 °C for 60–90 minutes. The mixture was cooled to room temperature, then an equal volume of protein precipitation buffer (2.35 M $NH_4Cl$, 1.15 M NaCl, 38% ethanol, pH 5.0) was added and the tube was vortexed briefly. The sample was centrifuged at 16,000 × *g* for 8 minutes at 4 °C to precipitate proteins. The supernatant (containing DNA) was transferred to a new 2 mL tube, and 600 µL of ice-cold 100% isopropanol was added to precipitate the DNA. The tube was gently inverted to mix and then centrifuged for 8 minutes at 10,000 × *g*. The supernatant was discarded, and the DNA pellet was washed twice with 70% ethanol (centrifuging at 10,000 × *g* for 1 minute each wash). The pellet was air-dried for 10–15 minutes, then dissolved in 100 µL of TE buffer (10 mM Tris, 1 mM EDTA, pH 8.0) or nuclease-free water. DNA quality ($A_{260}/A_{280}$ and $A_{260}/A_{230}$ ratios) was measured with a Synergy HTX microplate reader (BioTek). If samples showed impurities, a Zymo Genomic DNA Clean & Concentrator kit (Cat. D4010) was used for further purification. DNA concentration was determined using the Qubit dsDNA HS Assay Kit (Life Technologies, Cat. Q32851). The integrity of genomic DNA was also checked on a 1% agarose gel. To confirm the presence of amplifiable cow DNA, PCR was performed with cow-specific primers targeting different regions of the bovine mitochondrial DNA (see **Supplementary Table 9** for primer sequences).

**Library preparation for whole-genome sequencing.** Sequencing libraries were prepared using the KAPA HyperPlus kit (Roche), an Illumina-compatible library prep kit[10]. Manufacturer protocols were followed (KAPA HyperPlus Kit Technical Data Sheet KR1145 v4.17) with minor adjustments. Approximately 100 ng of high-quality genomic DNA was used as input for each library. DNA was fragmented enzymatically at 37 °C for 15 minutes to achieve an average insert size in the range of 400–750 bp. Fragmented DNA underwent end-repair and A-tailing, followed by adapter ligation. The ligated libraries were PCR-amplified (5 cycles). A single 0.7× size selection (using AMPure beads) was performed to remove short fragments and enrich for the target library size range. Final library size distributions (expected ~500–700 bp fragments including adapters) were verified using an Agilent 2100 Bioanalyzer. Libraries were pooled and sequenced on an Illumina NovaSeq 6000 S4 flow cell with 150 bp paired-end reads. Sequencing was carried out at the DNA Technologies and Expression Analysis Core at UC Davis Genome Center, supported by NIH Shared Instrumentation Grant 1S10OD010786-01.

### 3.2 Transcriptomic analysis for microRNA

**Exosomal miRNA extraction.** To obtain sufficient exosomal miRNA for sequencing, large milk volumes were processed (200 mL per sample). Milk was subjected to sequential centrifugation steps to remove fat, debris, and somatic cells; pelleted somatic cells were retained for genomic DNA analysis. To extract milk exosomes, supernatant was centrifuged three times at 21,500 x g at 4°C to pellet casein. The resulting supernatant was then filtered sequentially through 0.8, 0.45 and 0.22 um filters and the preprocessed supernatant stored at -80 until milk exosome extraction. To pellet the milk exosomes the preprocessed samples were thawed overnight at at 4°C and centrifuged at 100,000 x g for 90 minutes in an ultra-centrifuge. Exosome pellets were resuspended in phosphate-buffered saline (PBS) and stored at −80 °C until miRNA extraction.

The TRizol LS method from Sanoji et al. 2021 was used to extract miRNA from milk exosomes[11]. Briefly, milk exosomes were resuspended in 3 volumes of cold TRIzol LS and homogenized by repeated pipetting and aspiration. Chloroform (200 µl) was added, and samples were shaken for 30 s, briefly vortexed, and incubated at room temperature for 10 min. Phase separation was carried out by centrifugation (12,000 × g, 15 min, 4 °C). The aqueous phase was collected and precipitated with 3 M sodium acetate (pH 5.5; 10% v/v), 4 µl glycogen, and 2.5 volumes of molecular-grade ethanol, followed by overnight incubation at −80 °C. RNA was pelleted by centrifugation (16,000 × g, 30 min, 4 °C), washed with 500 µl cold 70% ethanol, and centrifuged again (16,000 × g, 5 min, 4 °C). Residual ethanol was removed by a brief high-speed spin, and pellets were air-dried for 10 min at room temperature before resuspension in 42 µl RNase-free water. RNA concentration was measured on a NanoDrop ND-1000 spectrophotometer.

The Zymo Clean and Concentrator – 5 kit (Zymo Research) was used to separate and purify small RNA. DNase treatment was carried out following the kit's instructions prior to purification. Briefly, to the 40 ul RNA samples 5 ul DNase I and 5 ul DNA digestion buffer was added and the samples incubated for 15 min at RT. The Zymo protocol for purification for small RNAs was followed according to instructions and purified small RNA was eluted in 22 ul DNase/RNase-free water. Purified miRNA was quantified using a Qubit microRNA Assay Kit prior to library preparation and sequencing.

**miRNA sequencing and counting.** Purified miRNA was quantified using a Qubit microRNA Assay Kit prior to library preparation and sequencing. Purified miRNA was sequenced on an Illumina NextSeq500 in a single-read 75 bp format. De-multiplexed and adapter-trimmed reads were mapped to miRNA transcripts of the cattle genome, version ARS-UCD1.3, using the Spliced Transcripts Alignment to a Reference (STAR) aligner[12] with some key adjustments. STAR 2.7.10b was used with parameters changed from the defaults to enable miRNA mapping: alignIntronMax = 1, alignSJDBoverhangMin 1000, outFilterMatchNmin 16, outFilterMatchNminOverLread = 0, outFilterMatchNminOverLread = 0, outFilterMultimapNmax = 10, outFilterMultimapScoreRange = 0, and outFilterMismatchNmax = 1. Aligned reads were quantified as counts at the individual miRNA level using RSEM version 1.3.3[13] .

### 3.3 Proteomics and peptidomics analysis

**Bovine milk protein digestion.** The protein digestion protocol was adapted from Goonatilleke et al. (2019)[14]. In summary, each milk sample (7 µL) was diluted into 193 µL of 50 mM ammonium bicarbonate (to reduce milk matrix complexity and provide optimal pH for trypsin). The sample was reduced by adding 2 µL of 550 mM DTT and incubating at 60 °C for 60 min (to break disulfide bonds), then alkylated by adding 4 µL of 450 mM IAA at room temperature in the dark for 30 min (to cap cysteines). Trypsin was added at an enzyme:substrate ratio of approximately 1:50 (2 µg trypsin) and digestion proceeded at 37 °C for 18 h. The digestion was stopped by chilling to 4 °C.

**Tryptic digest clean-up.** The digest clean-up followed a C18 SPE protocol (Huang et al., 2017)[15] with a few minor changes in solution volumes and centrifugation speeds. A 25 mg bed C18 96-well plate was used. Each well was conditioned with 3 × 200 µL of 100% acetonitrile (0.1% TFA) and equilibrated with 4 × 200 µL of water (0.1% TFA). The tryptic digest (~200 µL) was loaded per well and pulled through by centrifugation (Eppendorf 5810R, ~1300 rpm). Wells were washed 3× with 200 µL water (0.1% TFA). Peptides were eluted with 2 × 200 µL of 40% acetonitrile (0.1% TFA) followed by 2 × 200 µL of 80% acetonitrile (0.1% TFA). Combined eluates were dried in a vacuum concentrator and stored at –20 °C. Samples were reconstituted in 560 µL water before LC-MS.

**Protein standards digestion (absolute quantification).** Six major milk proteins were quantified using external standards. α-casein, β-casein, κ-casein, α-lactalbumin, β-lactoglobulin, and lactotransferrin standards (purity ≥ 95%) were obtained commercially. Stock solutions of 10 µg/µL were prepared in 50 mM ammonium bicarbonate. An aliquot of each standard was mixed in a defined ratio (α-casein 45%, β-casein 19%, κ-casein 15%, β-lactoglobulin 15%, α-lactalbumin 5%, lactotransferrin 1% by mass, totaling 420 μg protein) to mimic the approximate composition of milk. This mixture was subjected to the same reduction, alkylation, digestion, and C18 clean-up procedures as the milk samples. After digestion and clean-up, the mixture was split and diluted to a series of concentrations to generate calibration points. These standard digest solutions were analyzed by LC-MS in parallel with the milk samples. The peak areas of signature peptides unique to each of the six proteins were extracted, and calibration curves (signal vs. amount loaded) were constructed. The absolute content of each protein in milk (μg per μL milk, later converted to % w/w) was determined by comparing the peptide signals in the milk sample digests to the calibration curves. All other milk proteins (beyond those six) were quantified by their relative MS signal and then scaled to total protein content measured by BCA assay, as described.

**Free peptides extraction and purification.** The workflow for endogenous peptide extraction was based on published milk peptidomics protocols[16,17]. For endogenous peptide extraction, 200 μL of milk was transferred to a 1.5 mL Eppendorf tube and centrifuged at 16,000 rcf at 4 °C for 5 min to remove the cream layer. 50 μL of skimmed milk was taken out and transferred into a new tube and 50 μL of nanopure water was added to dilute the sample. 400 μL of Folch solution (MeOH : CHCl3 = 1 : 2) was then added to perform Folch extraction, removing non-polar lipids and precipitating proteins. After 20 seconds of vortexing, the sample was left on ice for 5 minutes and then centrifuged at 16000 rcf and 4 °C for another 10 minutes. The upper layer (aqueous phase) was transferred into the new tube, dry down completely in concentrator, and reconstituted in 100 µL nanopure water.

Isolated peptides were purified from salts and sugars by solid phase extraction (SPE) using C18 96-well plate cartridges (glygen 40-µL). The solutions used for SPE include A, 0.1% TFA in acetonitrile; B, 0.1% TFA in water; C, 0.1% TFA in 40% acetonitrile and D, 0.1% TFA in 80% acetonitrile. The cartridges were first activated by adding 200 µL of solution A 3 times and equilibrated by adding 200 µL of solution B three times. 200 µL of samples was then loaded into the cartridges and washed by 200 µl of solution B for three times. After changing to the collection plates, free peptides were eluted out by adding 200 µL of solution C twice and solution D twice. After eluting out all the peptides, the collected sample was put in the concentrator to remove the solvent. 40 µL of 2% ACN and 0.1% TFA in nanopure water was added to reconstitute the sample. 10 µL of reconstituted sample was transferred to a 9mm plastic screw thread autosampler vial for the instrumental analysis.

**LC-MS/MS analysis.** The LC-MS/MS analysis method was adapted from a previous publication (Alvarez, et al. 2022[18]) with minor changes in LC and MS parameters. The cleaned tryptic peptide samples were analyzed on an UltiMate™ WPS-3000RS nanoLC system coupled to an Orbitrap Fusion Lumos MS system (ThermoFisher Scientific). 1 µL of each sample was injected, and the peptides were separated on an Acclaim™ PepMap™ 100 C18 LC Column (75 µm x 150 mm, particle size 2 µm; ThermoFisher Scientific) at a flow rate of 300 nL/min. Water in 0.1% formic acid, and 80% acetonitrile in 0.1% formic acid were used as solvents A and B, respectively. The gradient for separation started with 4% B and followed by 0-3 min (4-9% B), 3-6 min (9-12% B), 6-80 min (12-24% B), 80-95 min (24-50 % B), 95-99 min (50-99% B), 99-103 min (99% B) and 103-103.5 min (99-2% B). MS spectra were collected with a mass range of 375–1500 m/z, resolution of 120000, and scan rate of 1.5 sec/spectrum in positive ionization mode. The filtered precursor ions in each MS spectrum were subjected to fragmentation via 30% higher-energy C-trap dissociation (HCD) using nitrogen gas.

**Protein identification and quantitation.** The proteomics data analysis was adapted from a previous publication (Li, et al 2020[19]). Mass spectrometry data were analyzed using the Byos workflow (Protein Metrics). Proteins were identified on the Byonic software (Protein Metrics) based on the Uniprot Bovine proteome database (Swiss-Prot and TrEMBL combined) with a precursor mass tolerance of 20 ppm and fragment mass tolerance of 20 ppm. Data analysis parameters used included C-terminal cleavage by trypsin (K and R cleavage sites) with up to three missed cleavages, and the following peptide modifications: carbamidomethyl at the cysteine, oxidation at the methionine, deamidation at asparagine and glutamine, acetylation at protein N-terminal, glutamine to pyro-glutamate at N-terminal, glutamate to pyro-glutamate at N-terminal. Protein IDs were filtered at 1% FDR. Each protein was quantified using the Byologic software (Protein Metrics). Protein relative abundances were calculated based on the XIC (extracted ion chromatogram) area sum of all peptides from each protein, relative to XIC area sum of all peptides from all identified proteins. Additional protein information (Molecular Weight, GO (ID), etc.) were extracted from the Uniprot database.

### 3.4 Targeted analysis for quantifying metabolites

**Metabolite extraction.** Cow milk samples were centrifuged at 16 K RCF at 4°C for 20 minutes to separate into layers. The aqueous layer was carefully collected and sequentially filtered through a syringe filter with 0.22 µm pore size (Millex-GP syringe filter, Millipore, Billerica, MA) followed by an ultrafiltration filter unit with 3k Da molecular weight cut-off (Amicon ultra centrifugal filter, Millipore, Billerica, MA)) that had been washed 3 times with MiliQ water (Milipore, Billerica, MI).

The volume of each filtrate was carefully measured and the appropriate amount of MiliQ water was added to each filtrate to ensure a final volume of 207 µL. 23 µL of internal standard (5 mM 3-(trimethylsilyl)$^{-1}$-propanesulfonic acid-d6) containing 0.2% $NaN_3$ in 99.8% $D_2O$ was added to inhibit microbial growth and ensure NMR locking respectively. The pH of each sample was adjusted to 6.7 ± 0.1 by adding small amounts of NaOH and HCl to minimize pH-based peak movement. 180 µL aliquots were subsequently transferred to 3 mm Bruker NMR tubes (Bruker, Brillerica, MA) and stored at 4 °C until spectral acquisition.

A blank sample, where miliQ water was filtered as described for the milk samples, was included in each batch.

**NMR data acquisition, data processing and quantification.** $^1$H NMR spectra were acquired at 298K using NOESY $^1$H presaturation experiment ('noesypr1d') on a Bruker Avance 600-MHz NMR equipped with a SampleJet autosampler (Bruker BioSpin, Germany). Water saturation was achieved during the prescan delay (2.5 s) and mixing time (100 ms). Spectra was acquired with 8 dummy scans and 128 transients over a spectral width of 12 ppm with a total acquisition time of 2.5 s.

The resulting spectra were Fourier transformed with zero filling to 128 k data points and the Free Induction Decays (FIDs) were transformed with an exponential apodization function corresponding to a line-broadening of 0.5 Hz.

Detection and quantification of metabolites was done by experienced personnel in the lab using Chenomx NMR Suite software. Spectra were manually processed and profiled using Chenomx NMR Suite v8.6 (Chenomx Inc, Edmonton, Alberta, Canada) based on the established method of targeted profiling[20].

### 3.5 Untargeted analysis for identifying metabolites

**Sample preparation and extraction.** Milk samples were processed using solvent systems optimized for broad metabolite coverage and compatibility with downstream analytical platforms. Platform-specific extraction procedures were employed to enable comprehensive detection of chemically diverse metabolite classes.

For lipidomics, samples were extracted using a methyl-tert-butyl ether (MTBE)–based biphasic extraction adapted from Matyash et al., which separates non-polar lipids into the organic phase while retaining polar metabolites in the methanol–water phase. Internal standards were added during extraction to support quality control, retention-time alignment, and normalization. Pooled quality-control (QC) samples were generated and injected at regular intervals throughout analytical sequences to monitor instrument stability, carryover, and batch effects.

**Primary metabolism analysis by GC–TOF MS.** Primary metabolites were analyzed by gas chromatography–time-of-flight mass spectrometry (GC–TOF MS) following chemical derivatization. Chromatographic separation was performed on an Rtx-5Sil MS capillary column (30 m × 0.25 mm internal diameter, 0.25 μm film thickness) using helium as the carrier gas at a flow rate of 1 mL $min^{-1}$. Samples were injected (0.5 μL) in splitless mode using an automated liner exchange system, with the injection temperature ramped from 50 °C to 250 °C at 12 °C $s^{-1}$.

The oven temperature program was 50 °C for 1 min, ramped at 20 °C $min^{-1}$ to 330 °C, and held for 5 min. Mass spectra were acquired under electron ionization at −70 eV using a Leco Pegasus IV TOF mass spectrometer, scanning from m/z 80–500 at an acquisition rate of 17 spectra $s^{-1}$. The transfer line and ion source temperatures were set to 230 °C and 250 °C, respectively.

Raw data were processed using ChromaTOF software with baseline subtraction, peak deconvolution, and signal-to-noise filtering. Metabolite annotation and curation were performed using the BinBase database based on retention indices, mass-spectral similarity, and reference library matching. Quantification was reported as peak heights of unique ions. Data were normalized using metabolite-sum–based (mTIC) normalization, with additional QC-based corrections applied where appropriate.

**Biogenic amines and polar metabolites by HILIC–ESI–QTOF MS/MS.** Biogenic amines and other polar metabolites were analyzed by hydrophilic interaction liquid chromatography coupled to electrospray ionization quadrupole time-of-flight mass spectrometry (HILIC–ESI–QTOF MS/MS). Chromatographic separation was performed on a Waters Acquity Premier BEH amide column (1.7 µm, 2.1 mm × 50 mm) maintained at 45 °C.

Mobile phase A consisted of ultrapure water with 10 mM ammonium formate and 0.125% formic acid (pH 3), and mobile phase B consisted of 95:5 (v/v) acetonitrile:water with the same additives. The flow rate was 0.8 mL $min^{-1}$, with injection volumes of 1–5 µL. The gradient was as follows: 0–0.5 min, 100% B; 1.95 min, 70% B; 2.55 min, 30% B; 3.15–3.8 min, 100% B.

Data were acquired in both positive and negative ionization modes with high-resolution MS and MS/MS fragmentation. Raw data underwent quality-control checks based on internal-standard consistency and were processed using MS-DIAL for peak detection, alignment, and annotation. Metabolite identities were assigned using an in-house m/z–retention-time library and MS/MS spectral matching against NIST and MoNA databases, followed by manual curation. Quantification was reported as normalized peak heights using internal-standard total ion current (iTIC) normalization or alternative approaches depending on batch size.

**Lipidomics by ESI–QTOF MS/MS.** Complex lipids were analyzed using reversed-phase ultra-high-performance liquid chromatography coupled to ESI–QTOF MS/MS. Lipid extracts were separated on a Waters Acquity Premier BEH C18 column (50 mm × 2.1 mm, 1.7 µm) maintained at 65 °C. For positive-ion mode, mobile phase A was 60:40 (v/v) acetonitrile:water with 10 mM ammonium formate and 0.1% formic acid, and mobile phase B was 90:10 (v/v) isopropanol:acetonitrile with the same additives. For negative-ion mode, ammonium acetate was used in place of ammonium formate.

The flow rate was 0.8 mL min$^{-1}$. Injection volumes were 1.67 µL for positive mode and 5 µL for negative mode, with the autosampler maintained at 4 °C. The gradient progressed from 15% B to 99% B over 4.13 min, followed by re-equilibration to initial conditions. Data were acquired in both ionization modes using MS and MS/MS scans across m/z 60–1200, with collision energies of 25 eV.

Lipid identification was performed by MS/MS spectral matching against the LipidBlast library and curated in-house databases. Data processing included peak detection, blank subtraction, isotope and duplicate removal, adduct consolidation, and manual spectral verification. Lipid abundances were reported as peak heights and normalized using class-specific internal standards or systematic error removal using random forest (SERRF) normalization for large datasets.

**Data reporting and normalization.** Across all platforms, metabolomic data were reported with associated metadata, including retention time, m/z, sample identifiers, and QC information. Quantitative values represent relative, semi-quantitative normalized peak heights. Platform-specific normalization strategies were applied to reduce analytical drift and batch effects, with QC-based evaluation used to guide additional correction where necessary.

### 3.6 Lipidomics analysis

**Total lipid extraction.** For each sample, 150 µL of milk was mixed with methanol containing 0.006% butylated hydroxytoluene (BHT), chloroform, and aqueous solution containing 1 mM EDTA and 0.9% KCl, to achieve the final methanol/chloroform/aqueous phase ratio of 4:8:3. Samples were vortexed and centrifuged at 2,100 rpm for 15 min at 0 °C.

The lower organic phase was transferred to a new glass tube. The remaining aqueous phase was re-extracted with an additional chloroform, vortexed, and centrifuged under identical conditions. Organic phases were combined, evaporated under nitrogen, and reconstituted in 8 mL chloroform:isopropanol (2:1, v/v). Extracts were stored at −80 °C until further processing.

**Hydrolysis and solid-phase extraction.** Half of the total lipid extract was transferred to a clean glass tube and dried under nitrogen. Antioxidant mix (EDTA, triphenylphosphine, and BHT) and deuterated oxylipin surrogate standards were added to each sample. Samples were reconstituted in methanol containing 0.1% acetic acid and 0.1% BHT, followed by addition of 0.4 M NaOH (300 µL). Alkaline hydrolysis was performed at 60 °C for 30 min to release esterified oxylipins.

After cooling, samples were neutralized with acetic acid and diluted with ultrapure water to reduce methanol content to approximately 15%. Oxylipins were purified using 60 mg Oasis HLB solid-phase extraction cartridges. Columns were preconditioned with ethyl acetate, methanol, and SPE buffer (5% methanol, 0.1% acetic acid in water). Samples were loaded by gravity, washed with SPE buffer, and dried under low vacuum. Oxylipins were eluted sequentially with methanol and ethyl acetate, combined, and dried under nitrogen. Eluates were reconstituted in 100 µL LC–MS–grade methanol and filtered through 0.1 µm PVDF centrifugal filters. Filtered extracts were transferred to amber LC–MS vials with glass inserts and stored at −80 °C until analysis.

**Free oxylipin extraction.** Free oxylipins were extracted by protein precipitation. Milk samples (150 µL) were mixed with ice-cold methanol containing 0.1% acetic acid and 0.1% BHT (200 µL), antioxidant mix, and surrogate standards. Samples were vortexed and centrifuged at 15,871 × g for 10 min at 0 °C. Supernatants were diluted with SPE buffer to reduce methanol content below 20% and purified using Oasis HLB SPE cartridges following the same conditioning, washing, and elution steps described above.

**LC–MS/MS analysis.** Oxylipins were analyzed on an Agilent 1290 Infinity UPLC system coupled to an Agilent 6460 Triple Quadrupole mass-spectrometer (Agilent Technologies, Santa Clara, CA, USA) using a ZORBAX RRHD Eclipse Plus C18 column (2.1 × 150 mm, 1.8 µm, Agilent Technologies) under a gradient elution method with mobile phase A: ultrapure water with 0.1% acetic acid and mobile phase B: acetonitrile/methanol 85:15 with 0.1% acetic acid. The gradient profile was as follows: 35 to 85% B (0-2 min at 0.25 mL/min; 2-12 min at 0.3 mL/min); 12 – 15 min, 85% B at 0.25 mL/min; 85 to 100% B (15.1 – 17 min at 0.4 mL/min); 100 to 35% B (17.1 – 19 min at 0.4 mL/min); 35% B at 0.3 mL/min (19 – 20 min). Detailed ESI source parameters and dynamic MRM transition parameters for oxylipin analysis are provided in **Supplementary Tables 10 and 11**, respectively. Data acquisition was performed in multiple reaction monitoring mode. Extraction blanks were included with each extraction batch to monitor contamination and recovery. Oxylipins were quantified based on the external calibration curve adjusted for the response factor for corresponding surrogate standard.

**Fatty acid derivatization and gas chromatography with flame ionization detection (GC–FID) analysis.** For the analyses of total fatty acid and total cholesterol, an aliquot (1/4) of total lipid extract was derivatized via acid-catalyzed methylation.Tridecanoate (19:0-TAG, 0.0201 mg) and 5α-cholestane (0.0625 mg) were added to the total lipid extract and dried under nitrogen. Dried lipid residues were reconstituted in 400 µL toluene, followed by addition of 3 mL methanol and 600 µL freshly prepared 8% HCl in methanol. Samples were vortexed, sealed, and incubated at 90 °C for 1 h to derivatize fatty acids and release cholesterol. After cooling to room temperature for 5 min, 1 mL hexane and 1 mL distilled water were added sequentially and vortexed. Samples were allowed to stand for 15 min to promote phase separation.

The upper hexane layer was transferred to a clean microcentrifuge tube containing 450 µL distilled water, vortexed, and centrifuged at 13,000 rpm for 2 min. The hexane phase was recovered, dried under nitrogen, and reconstituted in hexane prior to analysis. Final extracts were transferred to glass inserts in GC vials for injection.

Fatty acids and cholesterol were analyzed by GC–FID on a Perkin Elmer Clarus 500Gas Chromatograph system (Perkin Elmer, Shelton, CT, USA) using a DB-FFAP capillary column (30 m × 0.25 mm inner diameter, 0.25 µm film thickness, Agilent Technologies) with helium as the carrier gas. The injector temperature was set to 285 °C and the detector temperature to 300 °C. Samples were injected at a volume of 1 µL. The oven temperature program started at 80 °C with a 2 min hold, followed by a ramp of 10 °C/min to 185 °C, then 6 °C/min to 249 °C, with a final hold of 42 min. Data acquisition and instrument control were performed using the TotalChrom software platform (Parkin Elmer). Method blanks and hexane washes were included throughout the sequence to monitor carryover and background signal. Fatty acids were quantified based on the relative peak area of each analyte to nonadecanoic acid (C19:0) derived from 19:0-TAG. Cholesterol quantification was based on the external calibration curve correcting for the response factor to 5α-cholestane.

### 3.7 Glycomics analysis

**Oligosaccharides extraction.** Oligosaccharides (OS) were isolated and purified according to previously described methods[21]. Each milk sample was extracted twice to obtain duplicate measurements. Briefly, 200 µL of each sample was mixed with an equal volume of nanopure water and centrifuged at 4000 g in a microfuge for 30 min at 4°C to remove lipids. The top fat layer was removed, and 4 volumes of chloroform/methanol (2:1, vol/vol) were added, and vigorously mixed. The mixture was centrifuged at 4000 g for 30 min at 4°C. The upper layer containing OS was carefully transferred to a tube, and 2 volumes of cold ethanol were added. The OS solution was left for 1 h at -30 °C, followed by centrifugation for 30 min at 4000 g at 4°C to remove denatured protein. The supernatant (OS-rich fraction) was freeze-dried using a Genevac evaporator.

Samples were then reconstituted in 200 µL nanopure water and purified by solid phase extraction (SPE) using a HyperSep™ C18 96-well microplate (ThermoFisher Scientific). Prior to use, each C18 microplate well was activated with 3 column volumes of acetonitrile (ACN), and centrifuged at 1300 rpm at 20°C for 1 min after each addition. Wells were then conditioned with 3 column volumes of nanopure water, and centrifuged at 1300 rpm at 20°C for 1 min after each addition. Then 200 µL of the OS-rich solution was loaded onto the microplate and centrifuged at 1300 rpm at 20°C for 2 min. The solution containing OS was eluted with 3 column volumes of nanopure water, centrifuging at 1500 rpm at 20°C for 2 min after each addition. Samples were dried in a Genevac evaporator at 30°C and then reconstituted in 200 µL nanopure water. After that, calibrated volumetric pipettes were used to transfer 20 µL reconstitued milk, 40 µL 10 mg/L internal standard xylosyl cellobiose, and 340 µL nanopure water were used to dilute the sample 20-fold before LC-MS analysis using Agilent 6470A Triple Quad mass spectrometer.

To profile the OS by nano LC chip Quadrupole-Time-of-Flight mass spectrometry, milk samples cleaned up by C18 SPE were further purified by a porous graphitized carbon (PGC) 96-well microplate (Glygen). Prior to use, each PGC microplate well was activated with 3 column volumes of 80% ACN, 0.1% trifluoroacetic acid (TFA) (v/v), centrifuged at 1300 rpm at 20°C for 1 min after each addition. Wells were then equilibrated with 3 column volumes of nanopure water, centrifuged at 1300 rpm at 20°C for 1 min after each addition. The OS-rich solution was loaded onto the cartridge and centrifuged at 1500 rpm at 20°C for 2 min. Excess lactose and salts were removed by washing with 4 column volumes of nanopure water, centrifuged at 1500 rpm at 20°C for 2 min after each addition. The OS solution was then eluted with 3 column volumes of 40% ACN, 0.1% TFA (v/v), centrifuged at 1000 rpm at 20°C for 3 min after each addition, and dried in Genevac™ miVac centrifugal evaporator at room temperature.

**Absolute quantification of oligosaccharides by liquid chromatography triple quadrupole mass spectrometry.** Samples were analyzed on an Agilent 6470A Triple Quad mass spectrometer equipped with an Agilent 1260 Infinity II liquid chromatography system (Agilent Technologies, Santa Clara, CA). Chromatographic separation was performed on an Agilent AdvanceBio Glycan Mapping LC column (2.1 × 150 mm, 2.7 µm particle size) attached to a guard column (2.1 × 5 mm, 2.7 µm particle size). The column compartment was held at a constant temperature of 35 °C. Mobile phase solvents consisted of (A) 10-mM ammonium acetate in 3% ACN and (B)10-mM ammonium acetate in 95% ACN. Samples were loaded onto the column by a binary pump operating at a flow rate of 0.3 mL/min. The gradient was held constant at 87% B from 0.0-4.0 min, then lowered from 87-80% B from 4.0-5.0 min, further lowered from 80-72% B from 5.0-9.0 min, and 72-57% B from 9.0-11.0 min. The gradient was then held constant at 57% B from 11.0-12.0 min, decreased from 57-35% B from 12.0-14.0 min, held constant at 35% B from 14.0-15.0 min, and then ramped back up from 35 to 87% B from 15.0-16.5 min. The column was flushed and equilibrated with a post-run method at 87% B for 6 min prior to the initiation of subsequent injections.

Upon eluting from the column, OS were analyzed in positive ionization mode with multiple reaction monitoring (MRM) method, with a drying gas flow rate of 10 L/min at 150 °C, sheath gas flow rate of 7 L/min at 350°C, capillary voltage of 2200 V, and nebulizer pressure of 45 psi.

**Profiling of oligosaccharides by nano LC chip Quadrupole-Time-of-Flight mass spectrometry.** Prior to MS analysis, reconstituted OS samples were diluted 2-fold and spiked with 0.2 mg/L xylosyl cellobiose as the internal standard. Profiling analysis was performed with an Agilent 6520 accurate-mass Quadrupole-Time-of-Flight mass spectrometer with a microfluidic nano-electrospray chip (Agilent Technologies, Santa Clara, CA) as previously described[22]. The chip employed contained enrichment and analytical columns, both packed with graphitized carbon. Chromatographic elution was performed with a binary gradient of (A)3% ACN/0.1% formic acid in water and (B) 90% ACN/0.1% formic acid in water. The column was initially equilibrated and eluted with a flow rate of 0.3 μL/min for the nano pump and 4 μL/min for the capillary pump. The 65-min gradient was programmed as follows: 0-2.5 min, 0% B; 2.5–20 min, 0–16% B; 20–30 min, 16–44% B; 30–35 min, 44–100% B; 35–45 min, 100% B; and 45–65 min, 0% B. Data were acquired in the positive ionization mode with a 450–2500 mass/charge (m/z) range. The electrospray capillary voltage was between 1800–1950 V to maintain a steady spray. The acquisition rate was 1 spectrum/s for both MS and MS/MS modes. Automated precursor selection was employed based on abundance, with up to 6 MS/MS per MS. The precursor isolation window was at 4 m/z. Fragmentation energy was set at 1.3 V × m/z ÷ 100 with an offset of −3.5 V. Internal calibration was performed using m/z 922.009 and 1221.991 as the reference masses (ESI-TOF Tuning Mix G1969–85000, Agilent Technologies).

## 4. Bioactivity Measurement

### 4.1 Neurite Outgrowth Experiment

Cortical cultures were prepared from tissue from embryonic day 18 (E18) Sprague Dawley rats. CellVis 96-well glass plates were coated with 0.1 mg/mL poly-D-lysine. Cells were plated at approximately 12,000 cells/well in rat plating medium consisting of 10% heat-inactivated fetal bovine serum (ThermoFisher), 1% penicillin-streptomycin (ThermoFisher), and 0.5 mM glutamine (ThermoFisher) in neurobasal (ThermoFisher). After 24 h, the plating medium was removed completely and exchanged for replacement medium consisting of 2% B27 supplement (ThermoFisher), 1% penicillin-streptomycin (ThermoFisher), 0.5 mM glutamine (ThermoFisher), and 12.5 μM glutamate (ThermoFisher) in neurobasal. After 4 days in vitro (DIV4), cells were treated by diluting either 10 mM DMSO or $H_2O$ stock solutions in neurobasal (1:100 dilution), followed by a 1:10 dilution into each well for a final concentration of 0.1% vehicle. Each treatment was compared to its vehicle control (i.e., DMSO or $H_2O$). After 72 h, cells were fixed by removing 80% of media and replacing it with 100 µL (i.e., 50% of the working volume of the well) with 4% aqueous paraformaldehyde (ThermoFisher). The cells were allowed to incubated for 15 min at room temperature and then washed with DPBS (2 x 200 µL per well) (ThermoFisher). Cells were permeabilized using 0.2% Triton x-100 (ThermoFisher) in DPBS for 15 min at room temperature. Plates were then blocked for 24 h by gently shaking at 4°C with 2% bovine serum albumin (BSA) (Fischer Scientific) in DPBS. The blocking solution was then removed and replaced with chicken anti-MAP2 antibody (1:10,000; EnCor Biotechnology) in 1% BSA and allowed to gently shake at 4°C for 24 h. The plates were then washed with DPBS (2 x 200 µL per well) and one time with 1% BSA (200 µL per well) and then allowed to incubate for 1 h at room temperature with an anti-chicken IgG secondary antibody conjugated to AlexaFluor 680 (1:1,000, ThermoFisher). The plates were then washed with DPBS (3 x 200 µL per well) and were allowed to incubate for 5 min shaking at room temperature with Hoechst (1:10,000, ThermoFisher). Plates were washed with DPBS (3 x 200 µL per well), leaving 100µL of DPBS in each well. Images were acquired using a CellInsight CX7 LZR Pro High Content Screening (HCS) Reader with a 20x objective. Images were analyzed using HCS Studio Cell Analysis Software to quantify number of primary neurites extending from the cell body. All images were taken and analyzed by an experimenter blinded to treatment conditions.

Concentration ranges (log10-fold changes) and vehicles for each treatment were as follows: Arachidonic Acid: 31.7 nM–100 µM (DMSO); *β*-Hydroxybutyric Acid: 3.18 µM–10 mM ($H_2O$); Butyrate: 3.18 µM–10 mM ($H_2O$): Choline: 6.36 µM–20 mM ($H_2O$); Docosahexaenoic Acid: 15.8 nM–50 µM (DMSO); Fucose: 6.36 µM–20 mM ($H_2O$); Lactoferrin: 127 ng/mL–400 µg/mL ($H_2O$); N-Acetylneuraminic Acid: 3.18 µM–10 mM ($H_2O$); Oleic Acid: 31.7 nM–100 µM (DMSO); Tryptophan: 158.9 nM–500 µM ($H_2O$).

**4.2 Anti-inflammatory activity measurement**

Cortical tricultures were prepared from postnatal day 0 (P0) Sprague Dawley rats. CellVis 96-well glass plates were coated with 0.1mg/mL poly-D-lysine. Cells were plated at approximately 18,000 cells/well in rat plating medium consisting of 10% heat-inactivated fetal bovine serum (ThermoFisher), 1% penicillin-streptomycin (ThermoFisher), and 0.5 mM glutamax (ThermoFisher) in neurobasal (ThermoFisher). After 3-4h, the plating medium was removed completely and exchanged for maintenance medium consisting of 2% B27 supplement (ThermoFisher), 1% penicillin-streptomycin (ThermoFisher), 0.5 mM glutamax (ThermoFisher), 10 µg/mL IL-34 (R&D), 200 ng/mL TGF-β1 (PeproTech), 1.5 mg/mL cholesterol (Sigma), and neurobasal. After 7 days in vitro (DIV7), cells were treated with 1µg/mL lipopolysaccharide (LPS) (Sigma) in neurobasal and allowed to incubate for 48 h. On DIV9, cells were treated by diluting either 10 mM DMSO or $H_2O$ stock solutions in maintenance medium (1:100 dilution), followed by a 1:10 dilution into each well for a final concentration of 0.1% vehicle. Each treatment was compared to its vehicle control (i.e., DMSO or $H_2O$). After 24 h, cells were fixed by removing 80% of media and replacing it with 100 µL (i.e., 50% of the working volume of the well) 4% aqueous paraformaldehyde (ThermoFisher). The cells were allowed to incubate for 15 min at room temperature, followed by washing with DPBS (2 x 200 µL per well) (ThermoFisher). Cells then were permeabilized using 0.2% Triton x-100 (ThermoFisher) in DPBS for 15 min at room temperature. Plates were then blocked for 24 h by gently shaking at 4°C with 2% bovine serum albumin (BSA) (Fischer Scientific) in DPBS. The blocking solution was then removed and replaced with chicken anti-MAP2 antibody (1:10,000; EnCor Biotechnology), mouse anti-GFAP antibody (1:5,000; ThermoFisher), and rabbit anti-Iba1 (1:1,000; Fujifilm) in 1% BSA and allowed to gently shake at 4°C for 24 h. The plates were then washed with DPBS (2 x 200 µL per well) and one time with 1% BSA (200 µL per well) and then allowed to incubate for 1 h at room temperature with an anti-chicken Alexa Fluor 680 (1:1,000, ThermoFisher), anti-mouse Alex Fluor 594 (1:1,000; ThermoFisher), and anti-rabbit Alexa Fluor 488 (1:1,000; ThermoFisher). The plates were then washed with DPBS (3 x 200 µL per well) and were allowed to incubate for 5 min shaking at room temperature with Hoechst (1:10,000, ThermoFisher). Plates were then washed with DPBS (3 x 200 µL per well), leaving 100 µL of DPBS in each well. Images were acquired using a CellInsight CX7 LZR Pro High Content Screening (HCS) Reader with a 20x objective. Images were analyzed using HCS Studio Cell Analysis Software to quantify glia cell count and area. All images were taken and analyzed by an experimenter blinded to treatment conditions.

Concentration ranges (log10-fold changes) and vehicles for each treatment were as follows: Arachidonic Acid: 31.7 nM–100 µM (DMSO); β-Hydroxybutyric Acid: 3.18 µM–10 mM ($H_2O$); Butyrate: 3.18 µM–10 mM ($H_2O$): Choline: 6.36 µM–20 mM ($H_2O$); Docosahexaenoic Acid: 15.8 nM–50 µM (DMSO); Fucose: 6.36 µM–20 mM ($H_2O$);

Lactoferrin: 127 ng/mL–400 µg/mL ($H_2O$); N-Acetylneuraminic Acid: 3.18 µM–10 mM ($H_2O$); Oleic Acid: 31.7 nM–100 µM (DMSO); Tryptophan: 158.9 nM–500 µM ($H_2O$).

### 4.3 Antibacterial activity measurement targeting ESKAPE species

Initial screening against six ESKAPE species was performed for all 54 peptides (**Supplementary Table 14**). Six peptides exhibited IC50 values below 256 µg/mL against *A. baumannii*, whereas none of the peptides showed IC50 values below 256 µg/mL against the other five ESKAPE species. Follow-up validation was then conducted for these six peptides, including $IC_{50}$ determination against *E. coli* and two probiotic species, *L. plantarum* and *L. rhamnosus*, using a higher maximum tested concentration of 1,024 µg/mL. In addition, all 54 peptides were tested against *A. baumannii* at a higher concentration of 1,024 µg/mL, and $OD_{600}$ values were compared with those of the negative control without peptide treatment. We intentionally used MHB medium without divalent cation supplementation ($Ca^{2+}$ and $Mg^{2+}$)[23] to increase assay sensitivity, because these cations can reduce the apparent activity of cationic peptides by weakening electrostatic interactions with the bacterial membrane[24].

Before inoculation, bacterial suspensions were adjusted to $5 \times 10^5$ CFU/mL for all assays. Assay volumes were tailored to organism-specific aeration requirements. For aerobic ESKAPE pathogens, 100 µL of peptide solution was mixed with 100 µL of bacterial inoculum (200 µL total volume). For the facultative anaerobic probiotic strains (Lactobacillus spp.), volumes were increased to 150 µL of peptide solution and 150 µL of inoculum (300 µL total volume) to minimize headspace and reduce oxygen exposure.

To prevent evaporation and cross-contamination, plates were sealed with plate-sealing film (Albert Bioscience Inc., Laguna Hills, USA). Initial screening was performed in duplicate (n = 2) by measuring endpoint optical density at 600 nm ($OD_{600}$) after 20 h incubation at 37 °C. For this screening phase, the maximum peptide concentration was set to 256 µg/mL, consistent with established high-throughput protocols in which MIC values above this threshold are typically interpreted as indicating no substantial antimicrobial activity[25].

For validation of the top five candidates, inhibitory activity was characterized in greater detail. These assays were performed in quadruplicate (n = 4) by monitoring bacterial growth kinetics over 20 h at 37 °C across an extended concentration range with a maximum concentration of 1,024 µg/mL. This upper limit was selected to encompass the theoretical physiological abundance of these peptides (approximately 900 µg/mL) that could be achieved under complete casein hydrolysis, thereby capturing any activity within a biologically relevant range.

Bacterial growth was monitored by measuring absorbance at 600 nm ($OD_{600}$) using a Synergy HTX Multi-Mode Microplate Reader (Agilent, Santa Clara, USA). Half-maximal inhibitory concentrations ($IC_{50}$) were calculated from the growth kinetics to quantify potency. Differences in bacterial growth between peptide-treated groups and controls were evaluated using two-tailed Welch's t-tests, with $p < 0.05$ considered statistically significant.

## Supplementary Tables

**Supplementary Table 1. Compounds not represented in the reference milk databases among the top 100 most abundant quantified compounds.** Fifteen compounds not represented in the reference milk databases ranked within the top 100 by average concentration. For proteins with multiple sequence variants, the variant with the highest average concentration was labeled as the majority variant.

| Compounds (Category) | Rank of the compound concentration | Average Concentration (mg/L) | (% w/w) | Concentration Range (mg/L) | (% w/w) |
|---|---|---|---|---|---|
| Tyrosine-protein kinase **(The majority variant)** (UniProt ID: A0A3Q1MYA3) | 18 | 336.98 | 0.033 | 151.80 ~ 430.41 | 0.015 ~ 0.042 |
| Ig-like domain-containing protein **(The majority variant)** (UniProt ID: A0A3Q1M3L6) | 30 | 104.09 | 0.010 | 35.57 ~ 147.71 | 0.003 ~ 0.014 |
| Ig-like domain-containing protein (The minority variant) (UniProt ID: G5E513) | 37 | 64.98 | 0.006 | 23.59 ~ 196.21 | 0.002 ~ 0.019 |
| Ig-like domain-containing protein (The minority variant) (UniProt ID: F1N160) | 39 | 56.13 | 0.005 | 26.77 ~ 80.69 | 0.003 ~ 0.008 |
| Progestagen-associated endometrial protein (UniProt ID: A0A3Q1LYE8) | 43 | 37.99 | 0.004 | 37.35 ~ 38.08 | 0.004 ~ 0.004 |
| Sulfhydryl oxidase (UniProt ID: A0A3Q1N9Y5) | 56 | 18.93 | 0.002 | 1.62 ~ 576.27 | $1.57*10^{-4}$ ~ 0.056 |
| Cathepsin D (UniProt ID: F1MMR6) | 61 | 16.96 | 0.002 | 0.27 ~ 207.05 | $2.62*10^{-5}$ ~ 0.020 |
| Ig-like domain-containing protein (The minority variant) (UniProt ID: A0A3Q1N2B9) | 63 | 15.88 | 0.002 | 3.48 ~ 45.37 | $3.38*10^{-4}$ ~ 0.004 |
| Ig-like domain-containing protein (The minority variant) (UniProt ID: G3N0V0) | 65 | 14.28 | 0.001 | 4.01 ~ 28.44 | $3.90*10^{-4}$ ~ 0.003 |
| Alpha-S1-casein (The minority variant) (UniProt ID: A0A3Q1NG86) | 84 | 7.39 | 0.001 | 3.03 ~ 10.04 | $2.94*10^{-4}$ ~ 0.001 |
| Nucleobindin 1 | 86 | 7.22 | 0.001 | 2.05 ~ 50.71 | $1.99*10^{-4}$ ~ 0.005 |
| (UniProt ID: A0A452DHZ5) | | | | | |
| Fatty acid synthase **(The majority variant)** (UniProt ID: F1N647) | 88 | 6.71 | 0.001 | 0.12 ~ 14.70 | $1.17*10^{-5}$ ~ 0.001 |
| E1A binding protein p400 (UniProt ID: F1MLB1) | 90 | 6.27 | 0.001 | 0.00 ~ 58.95 | 0 ~ 0.006 |
| Ig-like domain-containing protein (The minority variant) (UniProt ID: A0A3Q1MSF6) | 92 | 6.06 | 0.001 | 0.14 ~ 20.91 | $1.36*10^{-5}$ ~ 0.002 |
| Ig-like domain-containing protein (The minority variant) (UniProt ID: F1MZ96) | 97 | 5.38 | 0.001 | 0.06 ~ 20.70 | $6.11*10^{-6}$ ~ 0.002 |

**Supplementary Table 2. The bioactivities curated from the databases and found in milk, with the definition of the negative control (0% efficacy) and the positive control (100% efficacy).** We assumed that no compound was applied if the negative control was not mentioned in the description section of the bioassay.

| Category | Bioactivity | Bioassays AID/ Study | Control | |
|---|---|---|---|---|
| | | | Negative* (0% Efficacy) | Positive (100% Efficacy) |
| Anticancer | Leukemia (AML) cell viability reduction (Multiple cell lines) | AID: 1963876, 1963878 (NOMO1 cell line)<br>AID: 2202233 (MV-4-11 cell line) | Cell viability (DMSO treated) | Zero cell viability |
| | Ewing sarcoma cell viability reduction (Multiple cell lines) | AID: 1347089 (TC32)<br>AID: 1347092 (A673)<br>AID: 1347093 (SK-N-MC)<br>AID: 1347110 (A673)<br>AID: 1347111 (SK-N-MC)<br>AID: 1347118 (TC32) | Cell viability (DMSO treated) | Cell viability (treated with 2.3 uM Bortezomib) |
| | | AID: 1259253 (TC71)<br>AID: 1259255 (EW8) | Cell viability (DMSO treated) | Cell viability (treated with 9.2 uM Bortezomib) |
| | Carcinoma (MCC) cell viability reduction (Multiple cell lines) | AID: 1296009 (NIH3T3)<br>AID: 1296009 (WaGa)<br>AID: 1296009 (HACAT)<br>AID: 1296009 (MKL-1) | Cell viability (DMSO treated) | Zero cell viability |
| | Neuroblastoma cell viability reduction | AID: 1347095, 1347115 (NB-EBc1)<br>AID: 1347100 (LAN-5) | Cell viability (DMSO treated) | Cell viability (treated with 2.3 uM Bortezomib) |
| | Atypical Teratoid/Rhabdoid Tumor (AT/RT) cell viability reduction (BT37 cell line) | AID: 1347094, 1347117 | Cell viability (DMSO treated) | Cell viability (treated with 2.3 uM Bortezomib) |
| | Osteosarcoma cell viability reduction | AID: 1347103 (OHS-50) | Cell viability (DMSO treated) | Cell viability (treated with 2.3 uM Bortezomib) |
| | Myeloma cell viability reduction (MM1R_ATCC cell line) | AID: 1918981 | Cell viability (DMSO treated) | Zero cell viability |
| | Polymerase Eta inhibition | AID: 588591, 720502 | (Not Found) | Zero enzyme activity |
| | Human tyrosyl-DNA phosphodiesterase I (Tdp1) inhibition | AID: 485290 | 1 nM TDP1 activity without inhibition | Zero enzyme activity |
| | Polymerase Iota inhibition | AID: 588590, 720496 | (Not Found) | Zero enzyme activity |
| | Werner syndrome (WS) helicase (WRN) inhibition | AID: 651768 | 20 nM WRN activity without inhibition | Zero enzyme activity |
| | Jurkat Cell toxicity (caspase 3/7 activity) | AID: 655 | Toxicity of Jurkat Cell (DMSO treated) | Toxicity of Jurkat Cell (100 uM tamoxifen) |
| | Selenoprotein glutathione peroxidase 4 (GPX4) inhibition | AID: 1845192 | 333.33 nM GPX4 activity (DMSO treated) | Zero enzyme activity |
| | RecQ-Like DNA Helicase 1 (RECQ1) inhibition | AID: 2549 | 50 nM RECQ1 activity (DMSO treated) | Zero enzyme activity |
| | Histone Lysine Methyltransferase G9a inhibition | AID: 504332 | (Not Found) | 20 nM G9a activity inhibited by BIX-01294 (CID: 25150857) |
| Disease Therapy | GCN5L2 inhibition – multiple diseases therapy | AID: 504327 | 20 nM GCN5L2 (HAT) activity without inhibition | 20 nM GCN5L2 (HAT) activity inhibited by garcinol |
| | USP1 inhibition - multiple diseases therapy | AID: 504865 | (Not Found) | Zero enzyme activity |
| | Angiotensin-converting enzyme (ACE) inhibition - hypertension therapy | (46 Studies) | 0% ACE activity inhibition | 100% ACE activity inhibition* |
| | Dipeptidyl Peptidase IV (DPP IV) Inhibition - T2D therapy | (11 Studies) | 0% DPP-IV activity inhibition | 100% DPP-IV activity inhibition* |
| | Prolyl endopeptidase (PREP) inhibition - multiple diseases therapy | PMID: 1368255 | 0% Prolyl endopeptidase inhibition | 100% Prolyl endopeptidase inhibition |
| | Vitamin D receptor (VDR) inhibition - hyperparathyroidism therapy | AID: 504847 | (Not Found) | 800 nM VDR activity inhibited by CBT358 (3-dibutylamino-1-(4-hexylphenyl)propan-1-one) |
| | Tau aggregation inhibition - Alzheimer's therapy | AID: 1460 | Tau protein aggregation level (P301L cell, DMSO treated) | Tau protein aggregation level (K311D cell) |
| | Aldehyde dehydrogenase 1 (ALDH1A1) inhibition - metabolic disease therapy | AID: 1030 | 50 nM ALDH1A1 activity without inhibition | Zero enzyme activity |
| Anti-oxidation | ABTS reduction (Pubchem CID:16240279) | DOI: 10.1007/s00217-012-1894-5 | No $ABTS^{\cdot+}$ absorbance reduction | $ABTS^{\cdot+}$ absorbance reduction to 0%* |
| | XTT reduction | PMID: 10742656 | No superoxide anion reduction | Superoxide anion reduction to 0%* |
| | Radical scavenging – ORAC (Oxygen radical absorbance capacity) | DOI: 10.1016/j.idairyj.2013.05.008, 10.3390/foods9080991 | 0% oxygen radical absorbance | 100% oxygen radical absorbance |
| | 15-human lipoxygenase 1 (15-hLO-1) inhibition | AID: 887 | 50 nM 15hLO-1 activity without inhibition | Zero enzyme activity |
| | 15-human lipoxygenase 2 (15-hLO-2) inhibition | AID: 881 | 50 nM 15hLO-2 activity without inhibition | Zero enzyme activity |
| | Human lipoxygenase (12hLO) inhibition | AID: 1671190 | 150 nM 12hLO activity without inhibition | Zero enzyme activity |
| Antimicrobial | *S. mansoni* TGR (thioredoxin-glutathione reductase) inhibition for Schistosomiasis treatment | AID: 485364 | TGR activity (DMSO treated) | Zero enzyme activity |
| | *B. subtilis* Sfp-type PPTases inhibition | AID: 1490 | 15 nM Sfp-ppTase activity without inhibition | Zero enzyme activity |
| | Malaria parasite viability reduction | AID: 504749 | *P. falciparum* viability (DMSO treated) | *P. falciparum* viability inhibited by 0.29 uM artemisinin |
| Antivirus | Chikungunya virus (CHIKV) nsP2 protease inhibition | AID: 1964101, 1964105 | CHIKV nsP2 protease activity (DMSO treated) | CHIKV nsP2 protease activity (treated with 20 nM Zinc Acetate) |
| | Influenza A NS1 inhibition | AID: 2326 | (Not Found) | Cell growth of the positive control (Same yeast strain not expressing NS1) |

**Supplementary Table 3. Top six milk-derived peptides with the lowest $IC_{50}$ values against *A. baumannii* and *E. coli*.** The maximum tested concentration for all peptides was 1,024 µg/mL. All six selected peptides showed $IC_{50}$ values below 256 µg/mL against *A. baumannii*. To facilitate comparison with source-protein concentrations, $IC_{50}$ values were converted from µg/mL to µM.

| Peptide (Source Protein) | Average Concentration in Milk (µM) | | Inhibitory Concentration ($IC_{50}$) (µM) | | | | Selective Index $\left(\frac{IC_{50}(L. rhamnosus)}{IC_{50}(Pathogen)}\right)$ | |
|---|---|---|---|---|---|---|---|---|
| | Free Peptide | Source Protein | *A. baumannii* | *E. coli* | *L. plantarum* | *L. rhamnosus* | *A. baumannii* | *E. coli* |
| YLGYLEQLLRLKK (α-S1-casein) | 0.01 | 661.5 | 11.8 | 436.34 | > 625.9 | > 625.9 | > 53.0 | > 1.4 |
| LGPVRGPFPIIV (β-casein) | 0.01 | 269.7 | 10.0 | > 810.3 | > 810.3 | > 810.3 | > 81.0 | NA |
| NENLLRFFVAPFPEVFGKEKV (α-S1-casein) | $4.03*10^{-3}$ | 661.5 | 9.0 | > 413.0 | > 413.0 | > 413.0 | > 45.9 | NA |
| VRGPFPIIV (β-casein) | 0.02 | 269.7 | 110.2 | > 1027.5 | > 1027.5 | > 1027.5 | > 9.3 | NA |
| GYLEQLLRLKK (α-S1-casein) | 0.01 | 661.5 | 98.5 | > 753.0 | > 753.0 | > 753.0 | > 7.6 | NA |
| IAGAVIGAV (Platelet glycoprotein 4) | n.d. | 0.220 | 183.4 | > 1329.9 | > 1329.9 | > 1329.9 | > 7.3 | NA |
| Melittin | | | 0.5 | 1.3 | 0.3 | 4.4 | 8.8 | 3.4 |

n.d.: Not Detected

**Supplementary Table 4. The 35 peptides that significantly inhibit the growth of *A. baumannii*.**

| Sequence | Average $OD_{600}$ (*A. baumannii*) (N = 2) | | p-value | q-value |
|---|---|---|---|---|
| | No Peptide Added | Peptide Added (1024 µg/mL) | | |
| LYQEPVLGPVRGPFPIIV | 0.7470 | 0.5560 | 0.0041 | 0.0546 |
| RPKHPIKHQGLPQEVLN | 0.7475 | 0.6085 | 0.0058 | 0.0546 |
| FVAPFPEVFGKEKV | 0.7435 | 0.5385 | 0.0072 | 0.0546 |
| SAKPEKAKETEKDDTGTPITKIELVPSH | 0.7220 | 0.5570 | 0.0089 | 0.0546 |
| PVLGPVRGPFPIIV | 0.6915 | 0.4280 | 0.0092 | 0.0546 |
| FFVAPFPEVFGKEKVNELSK | 0.7380 | 0.4485 | 0.0095 | 0.0546 |
| GYLEQLLRLKK | 0.7200 | 0.4125 | 0.0111 | 0.0546 |
| FVAPFPEVFGKEKVNELSK | 0.7255 | 0.6075 | 0.0114 | 0.0546 |
| KLTEEEKNRLNFL | 0.7345 | 0.5295 | 0.0118 | 0.0546 |
| GPVRGPFPII | 0.7305 | 0.3990 | 0.0120 | 0.0546 |
| EPVLGPVRGPFPIIV | 0.7325 | 0.6020 | 0.0122 | 0.0546 |
| FFVAPFPEVFGKEKV | 0.7410 | 0.5465 | 0.0131 | 0.0546 |
| PGRPTGYSGSSKALVSTLVPLA | 0.7530 | 0.5125 | 0.0131 | 0.0546 |
| DTIAQAASTTTISDAVSKVKIQVNK | 0.6755 | 0.4075 | 0.0143 | 0.0553 |
| TKLTEEEKNRLNFLK | 0.7265 | 0.5290 | 0.0165 | 0.0577 |
| LTEEEKNRLNFLKKISQR | 0.7245 | 0.5320 | 0.0171 | 0.0577 |
| NPKLPLSILKEKHL | 0.7325 | 0.5595 | 0.0186 | 0.0590 |
| IIAEKTKIPAVFKIDALNENK | 0.7280 | 0.5075 | 0.0249 | 0.0682 |
| DTIAQAASTTTISDAVSK | 0.6950 | 0.4990 | 0.0263 | 0.0682 |
| ALNEINQFYQKFPQYLQYLYQGPIVLNPWDQVKR | 0.7360 | 0.5055 | 0.0267 | 0.0682 |
| KTKLTEEEKNRLN | 0.7365 | 0.5990 | 0.0275 | 0.0682 |
| EPVLGPVRGPFPII | 0.7210 | 0.5280 | 0.0278 | 0.0682 |
| YLEQLLRLKKY | 0.7020 | 0.4395 | 0.0301 | 0.0693 |
| VAPFPEVFGKEKVNELSK | 0.7165 | 0.5475 | 0.0308 | 0.0693 |
| YLGYLEQLLRLKK | 0.6975 | 0.5220 | 0.0373 | 0.0777 |
| LTEEEKNRLNFLKKISQ | 0.7555 | 0.6075 | 0.0381 | 0.0777 |
| RPKHPIKHQGLP | 0.7090 | 0.5770 | 0.0402 | 0.0777 |
| VRGPFPIIV | 0.7040 | 0.4775 | 0.0403 | 0.0777 |
| PGRPTGYSGSSKALVSTLVPL | 0.7145 | 0.5640 | 0.0419 | 0.0777 |
| RPKHPIKH | 0.7205 | 0.6080 | 0.0439 | 0.0777 |
| GAVGATIGGVA | 0.5374 | 0.2405 | 0.0446 | 0.0777 |
| AMKPWIQPKTKVIPYVRYL | 0.7385 | 0.5065 | 0.0479 | 0.0808 |
| SQNPKLPLSILKEKHL | 0.7110 | 0.5570 | 0.0500 | 0.0815 |
| TKLTEEEKNRLNFLKK | 0.7490 | 0.5395 | 0.0513 | 0.0815 |
| IAGAVIGAV | 0.5374 | 0.2967 | 0.0566 | 0.0873 |

**Supplementary Table 5. Compounds significantly correlated with purchase region temperature.** This table lists milk compounds whose concentrations (% w/w) showed significant Pearson and Spearman correlations with the purchase ambient temperature (°C) at the time of collection.

| Compound | Bioactivity | Content (% w/w) | q-val (ANOVA) | PCC | | Spearman | |
|---|---|---|---|---|---|---|---|
| | | | | Value | q-val | Value | q-val |
| Peptide:DEELNKLLG | N/A | $4.84 * 10^{-8}$ | 0.004 | -0.600 | 0.002 | -0.508 | 0.004 |
| Disintegrin and metalloproteinase domain-containing protein 10 (Uniprot:Q10741) | N/A | $1.66 * 10^{-6}$ | 0.068 | -0.501 | 0.022 | -0.421 | 0.020 |
| Peptide:LHQKHQRAKRAVSHEDQFLRLDF | N/A | $1.96 * 10^{-7}$ | 0.034 | 0.490 | 0.024 | 0.499 | 0.004 |
| Peptide:SATEQEKEKDPFHYD | N/A | $7.34 * 10^{-9}$ | 0.067 | -0.463 | 0.032 | -0.428 | 0.018 |
| C-C motif chemokine (Uniprot:A0A3Q1M5V9) | N/A | $2.01 * 10^{-7}$ | 0.003 | 0.462 | 0.032 | 0.375 | 0.049 |
| Peptide:IPLSPMGEDSASGDIETLHSK | N/A | $2.37 * 10^{-8}$ | 0.05 | -0.462 | 0.032 | -0.410 | 0.025 |
| Peptide: PWI | N/A | $3.31 * 10^{-9}$ | 0.055 | -0.428 | 0.056 | -0.371 | 0.051 |
| Peptide: AEDKENYKKF | N/A | $1.39 * 10^{-9}$ | 0.089 | -0.414 | 0.072 | -0.366 | 0.056 |

**Supplementary Table 6. Compounds significantly correlated with the estimated Jersey breed ratio in milk samples**. Shown are compounds whose concentrations (% w/w of milk) are significantly correlated with the estimated Jersey breed proportion (inferred from SNP data). Pearson (linear) and Spearman (rank) correlation coefficients are reported along with their q-values (FDR-adjusted p-values).

| Compound | Bioactivity | Content (% w/w) | PCC | | Spearman | |
|---|---|---|---|---|---|---|
| | | | Value | q-val | Value | q-val |
| κ-casein (Uniprot:P02668) | DPP-IV inhibition - T2D therapy, ACE inhibition - hypertension therapy, | 0.4255 | 0.574 | $2.00 * 10^{-4}$ | 0.435 | $4.88 * 10^{-2}$ |
| Urea | N/A | 0.0188 | 0.528 | $8.84 * 10^{-4}$ | 0.453 | $3.94 * 10^{-2}$ |
| Glutamic acid | B. subtilis PPTases inhibition | 0.0052 | -0.644 | $1.01 * 10^{-5}$ | -0.543 | $9.33 * 10^{-3}$ |
| Lactoperoxidase (Uniprot:P80025) | N/A | 0.0032 | 0.561 | $3.27 * 10^{-4}$ | 0.502 | $1.82 * 10^{-2}$ |
| 2-Hydroxyglutaric acid | N/A | 0.0008 | -0.587 | $1.13 * 10^{-4}$ | -0.477 | $3.02 * 10^{-2}$ |
| Pantothenic acid | N/A | 0.0005 | 0.541 | $6.03 * 10^{-4}$ | 0.446 | $4.00 * 10^{-2}$ |
| Serpin family G member 1 (Uniprot:E1BMJ0) | N/A | 0.0002 | 0.461 | $4.78 * 10^{-3}$ | 0.569 | $6.39 * 10^{-3}$ |
| Ig-like domain-containing protein (Uniprot:A0A3Q1LSF0) | N/A | 0.0001 | 0.534 | $7.64 * 10^{-4}$ | 0.508 | $1.61 * 10^{-2}$ |
| Serum amyloid A-3 protein (Uniprot:Q8SQ28) | N/A | 0.0001 | 0.667 | $3.40 * 10^{-6}$ | 0.535 | $1.52 * 10^{-2}$ |
| Ig-like domain-containing protein (Uniprot:F1MLW8) | N/A | 0.0001 | 0.776 | $8.25 * 10^{-10}$ | 0.566 | $6.39 * 10^{-3}$ |
| Isocitrate dehydrogenase [NADP] (Uniprot:A0A140T8A5) | N/A | $4.82 * 10^{-5}$ | 0.497 | $2.22 * 10^{-3}$ | 0.473 | $3.04 * 10^{-2}$ |
| Polymeric immunoglobulin receptor (Uniprot:F1MR22) | N/A | $3.45 * 10^{-5}$ | 0.861 | $5.57 * 10^{-15}$ | 0.619 | $1.45 * 10^{-3}$ |
| Polyubiquitin-B (Uniprot:P0CG53) | N/A | $2.67 * 10^{-5}$ | 0.431 | $9.71 * 10^{-3}$ | 0.460 | $3.34 * 10^{-2}$ |
| Peptide:LPIIQKLEPQIA | N/A | $1.67 * 10^{-5}$ | 0.611 | $4.53 * 10^{-5}$ | 0.438 | $4.78 * 10^{-2}$ |
| Peptide:LPQEVLNENLL | N/A | $1.21 * 10^{-5}$ | 0.541 | $5.98 * 10^{-4}$ | 0.449 | $3.94 * 10^{-2}$ |
| Chitinase 3 like 1 (Uniprot:G3X7D2) | N/A | $8.70 * 10^{-6}$ | 0.495 | $2.35 * 10^{-3}$ | 0.513 | $1.59 * 10^{-2}$ |
| [histone H3]-lysine(36) N-trimethyltransferase (Uniprot:A0A3Q1MJ95) | N/A | $6.49 * 10^{-6}$ | 0.675 | $2.31 * 10^{-6}$ | 0.457 | $3.28 * 10^{-2}$ |
| Peptide:RPKHPIKHQGLP | N/A | $4.90 * 10^{-6}$ | 0.401 | $1.92 * 10^{-2}$ | 0.430 | $4.88 * 10^{-2}$ |
| Peptide:VWEESLK | N/A | $3.41 * 10^{-6}$ | 0.538 | $6.62 * 10^{-4}$ | 0.472 | $3.04 * 10^{-2}$ |
| Peptide:SPPEINTVQVTSTAV | N/A | $3.01 * 10^{-6}$ | 0.404 | $1.82 * 10^{-2}$ | 0.460 | $3.16 * 10^{-2}$ |
| Membrane cofactor protein (Uniprot:A0A3Q1M3Z1) | N/A | $2.83 * 10^{-6}$ | 0.449 | $6.43 * 10^{-3}$ | 0.480 | $2.96 * 10^{-2}$ |
| Peptide:DPNHFRPAGLPDKY | N/A | $2.53 * 10^{-6}$ | 0.517 | $1.33 * 10^{-3}$ | 0.431 | $4.88 * 10^{-2}$ |
| Peptide:FALPQYLKTVYQHQ | N/A | $2.42 * 10^{-6}$ | 0.484 | $2.98 * 10^{-3}$ | 0.434 | $4.78 * 10^{-2}$ |
| Peptide:LYQEPVLGPVRGPFPII | N/A | $2.23 * 10^{-6}$ | 0.421 | $1.22 * 10^{-2}$ | 0.432 | $4.88 * 10^{-2}$ |
| Peptide:RLPQLVGVHPPLQGSSHGAAAIGQPSGELRL | N/A | $2.19 * 10^{-6}$ | 0.512 | $1.55 * 10^{-3}$ | 0.438 | $4.78 * 10^{-2}$ |
| Peptide:YLEQLLRL | N/A | $2.07 * 10^{-6}$ | 0.442 | $7.40 * 10^{-3}$ | 0.447 | $4.00 * 10^{-2}$ |
| Guanine nucleotide-binding protein G(o) subunit α (Uniprot:P08239) | N/A | $1.86 * 10^{-6}$ | 0.546 | $5.11 * 10^{-4}$ | 0.503 | $1.61 * 10^{-2}$ |
| Peptide:VPSERYLGYLEQL | N/A | $1.05 * 10^{-6}$ | 0.432 | $9.50 * 10^{-3}$ | 0.436 | $4.78 * 10^{-2}$ |
| Peptide:DVPSERY | N/A | $9.60 * 10^{-7}$ | 0.580 | $1.64 * 10^{-4}$ | 0.430 | $4.88 * 10^{-2}$ |
| Peptide:LKTTLSSEAPTTQQLSEYFK | N/A | $6.95 * 10^{-7}$ | 0.638 | $1.35 * 10^{-5}$ | 0.496 | $1.82 * 10^{-2}$ |
| Peptide:EVDDEALEKF | N/A | $6.12 * 10^{-7}$ | 0.496 | $2.35 * 10^{-3}$ | 0.449 | $3.94 * 10^{-2}$ |
| Peptide:LPIIQKLEPQI | N/A | $5.98 * 10^{-7}$ | 0.615 | $4.03 * 10^{-5}$ | 0.517 | $1.52 * 10^{-2}$ |
| Peptide:KEDVPSERYLGYLEQLL | N/A | $5.93 * 10^{-7}$ | 0.549 | $4.71 * 10^{-4}$ | 0.438 | $4.78 * 10^{-2}$ |
| Peptide:EPQLSVVTR | N/A | $5.74 * 10^{-7}$ | 0.558 | $3.52 * 10^{-4}$ | 0.506 | $1.61 * 10^{-2}$ |
| Peptide:LGYLEQLLRL | N/A | $5.59 * 10^{-7}$ | 0.486 | $2.84 * 10^{-3}$ | 0.434 | $4.78 * 10^{-2}$ |
| WAP four-disulfide core domain 2 (Uniprot:A0A3Q1LQ34) | N/A | $4.91 * 10^{-7}$ | 0.486 | $2.82 * 10^{-3}$ | 0.456 | $3.28 * 10^{-2}$ |
| Peptide:LLYQEPVLGPVRGPFPII | N/A | $4.79 * 10^{-7}$ | 0.509 | $1.68 * 10^{-3}$ | 0.471 | $3.04 * 10^{-2}$ |
| Peptide:LEIVPNS | N/A | $3.52 * 10^{-7}$ | -0.361 | $4.44 * 10^{-2}$ | -0.483 | $2.82 * 10^{-2}$ |
| Peptide:RPLPVLGGNVGINSF | N/A | $2.90 * 10^{-7}$ | 0.462 | $4.59 * 10^{-3}$ | 0.443 | $4.30 * 10^{-2}$ |
| Peptide:QEPVLGPVRGPFPII | N/A | $2.63 * 10^{-7}$ | 0.404 | $1.81 * 10^{-2}$ | 0.475 | $3.04 * 10^{-2}$ |
| Peptide:VLPVPQKAVPYPQRDMPIQAFLLYQEPVLGPVRGPFPII | N/A | $2.21 * 10^{-7}$ | 0.416 | $1.39 * 10^{-2}$ | 0.505 | $1.61 * 10^{-2}$ |
| Peptide:EEKLPILN | N/A | $1.07 * 10^{-7}$ | 0.674 | $2.31 * 10^{-6}$ | 0.522 | $1.52 * 10^{-2}$ |

**Supplementary Table 7. Bioactive compounds significantly correlated with specific SNPs**. Each entry represents a significant SNP–compound association meeting the criteria described (absolute correlation > 0.55 and q < 0.1). SNPs are reported by bovine genome coordinate and nucleotide change, with predicted impact and functional consequence annotated by SnpEff. Pearson and Spearman correlation coefficients are provided along with their corresponding q-values (FDR-adjusted p-values).

| Compound | | SNP | | | PCC | | Spearman | |
|---|---|---|---|---|---|---|---|---|
| Name | Bioactivity | ID (Chrom: Pos: Genotype) | Impact Level | Consequence Type | Value | q-val | Value | q-val |
| Alpha-S1-casein (Uniprot:P02662) | (**Peptide: EL**) XTT reduction - Antioxidant (Suetsuna et al. 2000)<br><br>(**Peptide: YFYPEL**) Radical scavenging - ORAC - Antioxidant (Contreras et al. 2013) | 23:48715238:C->T | Low | synonymous variant | -0.559 | 0.002 | -0.594 | 0.003 |
| Beta-casein (Uniprot:P02666) | (**Peptide: MAP**) ACE inhibition - hypertension therapy (Norris et al. 2014)<br><br>(**Peptide: VYPFPGPIP**) PREP inhibition - multiple diseases therapy (Asano et al. 1992)<br><br>(**Peptide: VYPFPGPIPN**): ABTS reduction - Antioxidant (Eisele et al. 2013) | 11:7125939:T->C | Moderate | missense variant | -0.558 | 0.002 | -0.575 | 0.006 |
| Kappa-casein (Uniprot:P02668) | (**Peptide: IPI**): DPP-IV inhibition - T2D therapy (Nongonierma et al. 2013) | 7:54153842:G->A | Low | synonymous variant | 0.574 | 0.001 | 0.599 | 0.003 |
| | | 7:54153881:C->T | Low | synonymous variant | 0.603 | $2.06*10^{-4}$ | 0.613 | 0.001 |
| | | 24:25617929:G->A | Low | synonymous variant | 0.568 | 0.002 | 0.568 | 0.014 |
| Citric acid | Jurkat Cell toxicity - Anticancer (AID:655) | 22:17613279:G->A | Low | synonymous variant | 0.565 | 0.001 | 0.578 | 0.004 |
| | ALDH1A1 inhibition - metabolic disease therapy (AID:1030) | | | | | | | |
| Cytidine | Leukemia (AML) cell viability reduction - anticancer (AID: 1963878)<br><br>Ewing sarcoma cell viability reduction - anticancer (AID: 1347089)<br><br>Carcinoma (MCC) cell viability reduction - anticancer (AID: 1296009)<br><br>Neuroblastoma cell viability reduction - anticancer (AID: 1347100)<br><br>Atypical Teratoid/Rhabdoid Tumor (AT/RT) cell viability reduction - anticancer (AID: 1347094)<br><br>Osteosarcoma cell viability reduction - anticancer (AID:1347103)<br><br>Myeloma cell viability reduction - anticancer (AID:1918981) | 24:50417613:G->A | Low | synonymous variant | 0.643 | $1.67*10^{-5}$ | 0.569 | 0.005 |
| Glutamic acid | *B. subtilis* PPTases inhibition (AID:1490) | 1:43714480:G->A | Moderate | missense variant | -0.559 | 0.001 | -0.564 | 0.006 |
| | | 2:17084690:A->C | Moderate | missense variant | -0.588 | 0.001 | -0.567 | 0.010 |
| | | 2:108198263:G->A | Low | synonymous variant | -0.652 | $9.91*10^{-6}$ | -0.608 | 0.001 |
| | | 3:26372199:C->T | Low | synonymous variant | -0.574 | 0.001 | -0.599 | 0.002 |
| | | 3:110133056:G->A | Low | synonymous variant | -0.556 | 0.002 | -0.562 | 0.009 |
| | | 4:10415042:G->A | Low | synonymous variant | -0.557 | 0.002 | -0.564 | 0.014 |
| | | 4:10415052:C->T | Moderate | missense variant | -0.561 | 0.002 | -0.563 | 0.016 |
| | | 6:26709811:T->C | Low | synonymous variant | 0.552 | 0.002 | 0.576 | 0.005 |
| | | 11:2600893:C->A | Moderate | missense variant | -0.569 | 0.001 | -0.572 | 0.007 |
| | | 12:82953875:C->T | Moderate | missense variant | -0.556 | 0.002 | -0.576 | 0.006 |
| | | 15:47127447:G->A | Low | synonymous variant | -0.611 | $2.16*10^{-4}$ | -0.573 | 0.008 |
| | | 15:55992192:A->G | Low | start gained | -0.552 | 0.001 | -0.554 | 0.009 |
| | | 19:51921602:C->T | Low | synonymous variant | -0.576 | 0.001 | -0.567 | 0.006 |
| | | 23:29581826:C->T | Moderate | missense variant | -0.568 | 0.001 | -0.558 | 0.010 |
| | | 29:50459997:C->T | Low | synonymous variant | 0.592 | $2.79*10^{-4}$ | 0.638 | $2.03*10^{-4}$ |
| Uridine | Influenza A NS1 inhibition (AID:2326) | 3:102508762:T->C | Moderate | missense variant | -0.585 | $4.56*10^{-4}$ | -0.616 | 0.001 |
| | | 3:102508769:T->C | Low | synonymous variant | 0.580 | 0.001 | 0.588 | 0.003 |

**Supplementary Table 8. Bioactive compounds significantly correlated with specific miRNAs.** Each entry represents a significant miRNA–compound association meeting the criteria described (absolute correlation > 0.35 and q < 0.1). miRNAs are reported by sequence together with the corresponding miRBase ID. Pearson and Spearman correlation coefficients are provided along with their q-values (FDR-adjusted p-values).

| Compound | | miRNA | | PCC | | Spearman | |
|---|---|---|---|---|---|---|---|
| Name | Bioactivity | Sequence | miRBase ID | Value | q-val | Value | q-val |
| Kappa-casein (Uniprot:P02668) | **(Peptide: IPI from Kappa-casein):** DPP-IV inhibition - T2D therapy (Nongonierma et al. 2013) | CCCAUUGGCAUAAACCCGUAGAUCCGAUCUUGUGGUGAAGUGGACCGCACAAGCUCGCUUCUAUGGGUCUGUGUCAGUGUG | MI0004751 | -0.488 | 0.008 | -0.473 | 0.031 |
| | | GAAAGCGCUUUGGAAUGACACGAUCACUCCCGUUGAGUGGGCACCCAAGAAGCCAUCGGGAAUGUCGUGUCCGCCCAGUGCUCUUUC | MI0005047 | -0.419 | 0.060 | -0.464 | 0.039 |
| | | GGAUUUAGUAUUGGGUUGGCUGAAAAGUUCGUUCGGGUUUUUCCAUAAGAACUCAAAUAAACUUUUUGGCCAACCCAGUAAUAACACUGAUGCACC | MI0014499 | -0.457 | 0.022 | -0.454 | 0.051 |
| Citric acid | Jurkat Cell toxicity - anticancer (AID:655)<br><br>ALDH1A1 inhibition - metabolic disease therapy (AID:1030) | AUUGGUCAAAAAUUUGUUUGGGUUUUUCCAGUUGAUGUUGAAAAACCAGAACGAACUUUUUGGCCAACC | MI0021133 | -0.429 | 0.048 | -0.456 | 0.043 |
| Glutamic acid | *B. subtilis* PPTases inhibition (AID:1490) | GCCGCCGACCUGGCAGGCCAGGAAGAGGAGGAAGCCCUGGAGGGGCUGGAGGUGAUGGAUGUUUUCCUCCGGUUCUCAGGGCUCCACCUUUUCCGGGCCGUGGAGCCAGGGCUGGUGC | MI0009887 | 0.396 | 0.098 | 0.499 | 0.012 |
| | | CUCCGGUGCCUACUGAGCUGAUAUCAGUUCUCAUUUUACACACUGGCUCAGUUCAGCAGGAACAGGAG | MI0009761 | 0.432 | 0.044 | 0.471 | 0.029 |
| Uridine | Influenza A NS1 inhibition (AID:2326) | AGGUGGGCCCAAAAGUUCGUUCGGGUUUUUCCAUACCAUCUUACAGAAAAACCCAAACAAAUUUUUUGGCCCACCUA | MI0022300 | 0.399 | 0.091 | 0.464 | 0.035 |

**Supplementary Table 9.** List of primers used for cow DNA amplification.

| Gene Name (fragment) | Forward Primer | Reverse Primer | Amplicon size | References |
|---|---|---|---|---|
| mtDNA (mt DNA encoding tRNALys and the subunit S and subunit 6 of the ATP synthase complex) | GCCATATACTCTCCT TGGTGACA | GTAGGCTTGGGA ATAGTACGA | 271 bp | Tartaglia et al. (1998)[26] |
| cytochrome b gene | GACCTCCCAGCTCC ATCAAACATCTCATC TTGATGAAA | CTAGAAAAGTGT AAGACCCGTAAT ATAAG | 274 bp | Mašková and Paulíčková (2006)[27] |

**Supplementary Table 10. AJS ESI ion source parameters for oxylipin analysis.**

| Parameter | Value (+) | Value (-) |
|---|---|---|
| Gas Temperature (ºC) | 300 | 300 |
| Gas Flow (L/min) | 10 | 10 |
| Nebulizer (psi) | 35 | 35 |
| Sheath Gas Heater | 350 | 350 |
| Sheath Gas Flow | 11 | 11 |
| Capillary (V) | 4000 | 4000 |
| Vcharging | 0 | 1000 |

## Supplementary Table 11. Dynamic MRM conditions for oxylipin analysis.

| Analyte | Internal standard (Free oxylipins) | Internal standard (Total oxylipins) | Precursor ion (*m/z*) | Product ion (*m/z*) | Fragmentor voltage (V) | Collision energy (V) | Collision Cell Accelerator Voltage (V) | Retention time (min) | Retention time window (min) | Polarity |
|---|---|---|---|---|---|---|---|---|---|---|
| **LA-derived oxylipins** | | | | | | | | | | |
| 9-HODE | d4-9-HODE | d4-9-HODE | 295.2 | 171.1 | 110 | 12 | 4 | 11.951 | 1.4 | Negative |
| 13-HODE | d4-9-HODE | d4-9-HODE | 295.2 | 195.2 | 100 | 7 | 4 | 11.89 | 1.5 | Negative |
| 9-oxo-ODE | d4-9-HODE | d4-9-HODE | 293.2 | 185.1 | 120 | 10 | 4 | 12.579 | 1.24 | Negative |
| 13-oxo-ODE | d4-9-HODE | d4-9-HODE | 293.2 | 195.1 | 120 | 10 | 4 | 12.336 | 1.24 | Negative |
| 12(13)-EpOME | d11-11(12)-EpETrE | d11-11(12)-EpETrE | 295.3 | 195.2 | 100 | 7 | 4 | 13.373 | 1.2 | Negative |
| 9(10)-EpOME | d11-11(12)-EpETrE | d11-11(12)-EpETrE | 295.3 | 171.2 | 100 | 10 | 4 | 13.49 | 1.2 | Negative |
| 12,13-DiHOME | d11-14,15-DiHETrE | d11-14,15-DiHETrE | 313.2 | 183.2 | 110 | 12 | 4 | 9.6 | 1.5 | Negative |
| 9,10-DiHOME | d11-14,15-DiHETrE | d11-14,15-DiHETrE | 313.2 | 201.2 | 120 | 12 | 4 | 9.871 | 1.4 | Negative |
| 9,12,13-TriHOME | d4-PGE2 | d4-TXB2 | 329.2 | 211.1 | 120 | 8 | 4 | 5.479 | 1.3 | Negative |
| 9,10,13-TriHOME | d4-PGE2 | d4-TXB2 | 329.2 | 171.1 | 120 | 8 | 4 | 5.581 | 1.4 | Negative |
| **DGLA-derived oxylipins** | | | | | | | | | | |
| 15(S)-HETrE | d8-5-HETE | d8-5-HETE | 321.2 | 221.2 | 80 | 10 | 4 | 12.8 | 1.47 | Negative |
| PGD1 | d4-PGE2 | d4-TXB2 | 353.3 | 317.2 | 80 | 5 | 4 | 5.8 | 1.5 | Negative |
| PGE1 | d4-PGE2 | d4-TXB2 | 353.3 | 317.2 | 80 | 5 | 4 | 5.852 | 1.2 | Negative |
| **AA-derived oxylipins** | | | | | | | | | | |
| 5-HETE | d8-5-HETE | d8-5-HETE | 319.2 | 115.1 | 80 | 5 | 4 | 12.96 | 1.2 | Negative |
| 8-HETE | d8-5-HETE | d8-5-HETE | 319.2 | 155.2 | 90 | 5 | 4 | 12.624 | 1.32 | Negative |
| 9-HETE | d8-5-HETE | d8-5-HETE | 319.2 | 167.2 | 100 | 10 | 4 | 12.414 | 1.26 | Negative |
| 11-HETE | d8-5-HETE | d8-5-HETE | 319.2 | 167.2 | 100 | 10 | 4 | 12.4 | 1.2 | Negative |
| 12-HETE | d8-5-HETE | d8-5-HETE | 319.2 | 179.2 | 100 | 10 | 4 | 12.62 | 1.4 | Negative |
| 15-HETE | d8-5-HETE | d8-5-HETE | 319.2 | 219.2 | 120 | 5 | 4 | 12.171 | 1.51 | Negative |
| 20-HETE | d6-20-HETE | d6-20-HETE | 319.2 | 275.1 | 110 | 7 | 4 | 11.3 | 2 | Negative |
| 5-oxo-ETE | d8-5-HETE | d8-5-HETE | 317.2 | 273.2 | 120 | 5 | 4 | 13.648 | 1.2 | Negative |
| 12-oxo-ETE | d8-5-HETE | d8-5-HETE | 317.2 | 153.1 | 70 | 7 | 4 | 12.885 | 1.2 | Negative |
| 15-oxo-ETE | d8-5-HETE | d8-5-HETE | 317.2 | 113.1 | 110 | 7 | 6 | 12.5 | 1.4 | Negative |
| 5(6)-EpETrE | d11-11(12)-EpETrE | d11-11(12)-EpETrE | 319.2 | 191.1 | 120 | 5 | 4 | 14.04 | 1.4 | Negative |
| 8(9)-EpETrE | d11-11(12)-EpETrE | d11-11(12)-EpETrE | 319.2 | 167.2 | 50 | 7 | 4 | 13.755 | 1.6 | Negative |
| 11(12)-EpETrE | d11-11(12)-EpETrE | d11-11(12)-EpETrE | 319.2 | 167.2 | 100 | 10 | 4 | 13.76 | 1.2 | Negative |
| 14(15)-EpETrE | d11-11(12)-EpETrE | d11-11(12)-EpETrE | 319.2 | 219.3 | 120 | 5 | 4 | 13.4 | 1.44 | Negative |
| 5,6-DiHETrE | d11-14,15-DiHETrE | d11-14,15-DiHETrE | 337.2 | 145.1 | 90 | 12 | 4 | 11.43 | 1.2 | Negative |
| 8,9-DiHETrE | d11-14,15-DiHETrE | d11-14,15-DiHETrE | 337.2 | 127.1 | 70 | 12 | 4 | 10.963 | 1.2 | Negative |
| 11,12-DiHErTE | d11-14,15-DiHETrE | d11-14,15-DiHETrE | 337.2 | 167.1 | 120 | 12 | 4 | 10.62 | 1.3 | Negative |
| 14,15-DiHETrE | d11-14,15-DiHETrE | d11-14,15-DiHETrE | 337.2 | 207.1 | 100 | 5 | 4 | 10.2 | 1.6 | Negative |
| LTB3 | d4-LTB4 | d4-LTB4 | 337.2 | 195.2 | 110 | 7 | 4 | 10.31 | 1.32 | Negative |
| LTB4 | d4-LTB4 | d4-LTB4 | 335.2 | 195.1 | 80 | 7 | 4 | 8.6 | 1.23 | Negative |
| LTC4 | d4-LTB4 | d4-LTB4 | 624.3 | 272.1 | 110 | 12 | 4 | 8 | 2 | Negative |
| LTD4 | d4-LTB4 | d4-LTB4 | 495.3 | 177.1 | 110 | 12 | 4 | 6.375 | 1.53 | Negative |
| LTE4 | d4-LTB4 | d4-LTB4 | 438.2 | 333.1 | 80 | 5 | 4 | 7.687 | 1.5 | Negative |
| LXA4 | d4-LTB4 | d4-LTB4 | 351.2 | 115.2 | 90 | 10 | 4 | 6.691 | 1.77 | Negative |
| 20-COOH-LTB4 | d4-6-keto-PGF1α | d4-6-keto-PGF1α | 365.2 | 347.2 | 110 | 10 | 4 | 3.78 | 2.2 | Negative |
| 20-OH-LTB4 | d4-6-keto-PGF1α | d4-6-keto-PGF1α | 351.2 | 195.2 | 130 | 12 | 4 | 4.08 | 1.23 | Negative |
| 6-trans-LTB4 | d4-LTB4 | d4-LTB4 | 335.2 | 195.1 | 80 | 7 | 4 | 8.865 | 1.76 | Negative |
| PGB2 | d4-PGE2 | d4-LTB4 | 333.3 | 175.1 | 80 | 10 | 4 | 8.042 | 1.5 | Negative |
| PGD2 | d4-PGE2 | d4-TXB2 | 351.2 | 271.3 | 80 | 10 | 4 | 6.1 | 2 | Negative |
| PGE2 | d4-PGE2 | d4-TXB2 | 351.2 | 271.3 | 80 | 10 | 4 | 5.837 | 2 | Negative |
| PGF2a | d4-PGE2 | d4-TXB2 | 353.2 | 309.2 | 80 | 12 | 4 | 5.601 | 1.8 | Negative |
| PGJ2 | d4-PGE2 | d4-TXB2 | 333.3 | 189.2 | 100 | 12 | 4 | 7.956 | 1.53 | Negative |
| 15-deoxy-PGJ2 | d4-PGE2 | d11-14,15-DiHETrE | 315.2 | 271.2 | 120 | 5 | 4 | 11.2 | 1.1 | Negative |
| 6-keto-PGF1α | d4-6-keto-PGF1α | d4-6-keto-PGF1α | 369.3 | 163.2 | 80 | 7 | 4 | 3.988 | 2.2 | Negative |
| TXB2 | d4-TXB2 | d4-TXB2 | 369.2 | 169.1 | 80 | 5 | 4 | 5.3 | 2.8 | Negative |
| **ALA-dervied oxylipins** | | | | | | | | | | |
| 9-HOTrE | d4-9-HODE | d4-9-HODE | 293.2 | 171.2 | 120 | 10 | 4 | 10.765 | 1.7 | Negative |
| 13-HOTrE | d4-9-HODE | d4-9-HODE | 293.2 | 195.1 | 100 | 10 | 4 | 10.95 | 1.5 | Negative |
| **EPA-derived oxylipins** | | | | | | | | | | |
| 5-HEPE | d6-20-HETE | d6-20-HETE | 317.2 | 115.1 | 80 | 5 | 4 | 11.78 | 1.63 | Negative |
| 8-HEPE | d6-20-HETE | d6-20-HETE | 317.2 | 155.2 | 100 | 7 | 4 | 11.448 | 1.53 | Negative |
| 12-HEPE | d6-20-HETE | d6-20-HETE | 317.2 | 179.2 | 110 | 7 | 4 | 11.53 | 1.23 | Negative |
| 15-HEPE | d6-20-HETE | d6-20-HETE | 317.2 | 219.2 | 110 | 5 | 4 | 11.28 | 1.5 | Negative |
| 17(18)-EpETE | d11-11(12)-EpETrE | d11-11(12)-EpETrE | 317.2 | 215.2 | 90 | 10 | 4 | 12.27 | 1.21 | Negative |
| 14(15)-EpETE | d11-11(12)-EpETrE | d11-11(12)-EpETrE | 317.2 | 207.2 | 90 | 5 | 4 | 12.575 | 1.46 | Negative |
| 11(12)-EpETE | d11-11(12)-EpETrE | d11-11(12)-EpETrE | 317.2 | 167.2 | 100 | 10 | 4 | 12.65 | 1.4 | Negative |
| 8(9)-EpETE | d11-11(12)-EpETrE | d11-11(12)-EpETrE | 317.2 | 127.2 | 100 | 10 | 4 | 12.752 | 1.7 | Negative |
| 17,18-DiHETE | d11-14,15-DiHETrE | d11-14,15-DiHETrE | 335.3 | 247.2 | 80 | 7 | 4 | 9 | 1.77 | Negative |
| 5,15-DiHETE | d11-14,15-DiHETrE | d11-14,15-DiHETrE | 335.2 | 173.2 | 90 | 7 | 4 | 8.875 | 2.5 | Negative |
| 5,6-DiHETE | d11-14,15-DiHETrE | d11-14,15-DiHETrE | 335.2 | 115.2 | 90 | 12 | 4 | 9.899 | 2 | Negative |
| 8,15-DiHETE | d11-14,15-DiHETrE | d11-14,15-DiHETrE | 335.2 | 235.2 | 50 | 7 | 4 | 8.606 | 1.4 | Negative |
| 8,9-DiHETE | d11-14,15-DiHETrE | d11-14,15-DiHETrE | 335.5 | 126.9 | 100 | 12 | 4 | 9.699 | 1.4 | Negative |
| 11,12-DiHETE | d11-14,15-DiHETrE | d11-14,15-DiHETrE | 335.5 | 167 | 100 | 10 | 4 | 9.46 | 1.4 | Negative |
| 14,15-DiHETE | d11-14,15-DiHETrE | d11-14,15-DiHETrE | 335.3 | 207.2 | 100 | 10 | 4 | 9.33 | 1.6 | Negative |
| Resolvin E1 | d4-6-keto-PGF1α | d4-6-keto-PGF1α | 349.3 | 195 | 90 | 10 | 4 | 3.902 | 1.8 | Negative |
| PGD3 | d4-PGE2 | d4-TXB2 | 349.3 | 269.2 | 80 | 5 | 4 | 5.25 | 1.8 | Negative |
| PGE3 | d4-PGE2 | d4-TXB2 | 349.3 | 269.2 | 80 | 5 | 4 | 4.986 | 1.8 | Negative |
| **DHA-derived oxylipins** | | | | | | | | | | |
| 19(20)-EpDPE | d11-11(12)-EpETrE | d11-11(12)-EpETrE | 343.2 | 241.2 | 120 | 5 | 4 | 13.23 | 1.1 | Negative |
| 16(17)-EpDPE | d11-11(12)-EpETrE | d11-11(12)-EpETrE | 343.2 | 233.2 | 70 | 5 | 4 | 13.5 | 1.33 | Negative |
| 13(14)-EpDPE | d11-11(12)-EpETrE | d11-11(12)-EpETrE | 343.2 | 193.2 | 70 | 5 | 4 | 13.53 | 1.2 | Negative |
| 10(11)-EpDPE | d11-11(12)-EpETrE | d11-11(12)-EpETrE | 343.2 | 153.2 | 90 | 5 | 4 | 13.62 | 1.2 | Negative |
| 7(8)-EpDPE | d11-11(12)-EpETrE | d11-11(12)-EpETrE | 343.2 | 113.1 | 50 | 5 | 4 | 13.761 | 1.2 | Negative |
| 17-HDoHE | d4-9-HODE | d4-9-HODE | 343.2 | 281.2 | 50 | 7 | 4 | 12.067 | 2 | Negative |
| 19,20-DiHDPA | d11-14,15-DiHETrE | d11-14,15-DiHETrE | 361.5 | 273.1 | 100 | 8 | 4 | 10.2 | 1.4 | Negative |
| 16,17-DiHDPA | d11-14,15-DiHETrE | d11-14,15-DiHETrE | 361.5 | 233.1 | 100 | 8 | 4 | 10.5 | 1.4 | Negative |
| Resolvin D1 | d4-6-keto-PGF1α | d4-6-keto-PGF1α | 375.2 | 141.1 | 90 | 10 | 4 | 6.5 | 1.8 | Negative |
| **Internal standards** | | | | | | | | | | |
| d11-11(12)EpETrE | N/A | N/A | 330.2 | 167.2 | 130 | 7 | 4 | 13.678 | 1.4 | Negative |
| d11-14,15-DiHETrE | N/A | N/A | 348.2 | 207.1 | 110 | 12 | 6 | 10.125 | 1.4 | Negative |
| d4-6-keto-PGF1a | N/A | N/A | 373.3 | 167.1 | 110 | 25 | 4 | 3.937 | 2.2 | Negative |
| d4-9-HODE | N/A | N/A | 299.2 | 172.3 | 80 | 10 | 4 | 11.9 | 1.4 | Negative |
| d4-LTB4 | N/A | N/A | 339.2 | 197.2 | 80 | 10 | 4 | 9.129 | 1.2 | Negative |
| d4-PGE2 | N/A | N/A | 355.2 | 275.3 | 100 | 10 | 4 | 5.817 | 1.6 | Negative |
| d4-TXB2 | N/A | N/A | 373.3 | 173.2 | 100 | 12 | 4 | 5.5 | 2.8 | Negative |
| d6-20-HETE | N/A | N/A | 325.2 | 281.2 | 70 | 12 | 4 | 11.279 | 1.2 | Negative |
| d8-5-HETE | N/A | N/A | 327.2 | 116.1 | 70 | 7 | 4 | 12.881 | 1.1 | Negative |
| d11-11,12-DiHErTE | N/A | N/A | 348.2 | 178.1 | 120 | 12 | 4 | 10.62 | 2 | Negative |

**Supplementary Table 12. Milk compounds with 28 bioactivities curated in PubChem Bioassay Database.**

| Category | Bioactivity | Pubchem Bioassays AID | Compound (Conc (μM)) | $EC_{50}$ (μM) | Max Efficacy (%) | Estimated Efficacy (%) | |
|---|---|---|---|---|---|---|---|
| | | | | | | In vitro | In vivo |
| Anticancer | Leukemia (AML) cell viability reduction (Multiple cell lines) | AID: 1963876 (NOMO1) | Cytidine (13.8 μM) | 0.07 | 87.8 | 87.7 | 0.7 |
| | | AID: 1963878 (NOMO1) | Cytidine (13.8 μM) | 0.02 | 82.4 | 80.9 | **18.0** |
| | | AID: 2202233 (MV-4-11) | Cytidine (13.8 μM) | 16.6 | 36.8 | 14.4 | 0.0 |
| | Ewing sarcoma cell viability reduction (Multiple cell lines) | AID: 1347089 (TC32) | Cytidine (13.8 μM) | 0.03 | 73.6 | 72.6 | **9.7** |
| | | AID: 1347092 (A673) | Cytidine (13.8 μM) | 0.02 | 35.1 | 34.4 | **6.5** |
| | | AID: 1347093 (SK-N-MC) | Betaine (59.7 μM) | 16.6 | 76.3 | 70.8 | 0.0 |
| | | AID: 1347110 (A673) | Cytidine (13.8 μM) | 0.7 | 79.7 | 79.6 | 0.0 |
| | | AID: 1347111 (SK-N-MC) | Cytidine (13.8 μM) | 2.6 | 111.4 | 108.9 | 0.0 |
| | | AID: 1347118 (TC32) | Cytidine (13.8 μM) | 0.3 | 107.7 | 107.0 | 0.1 |
| | | AID: 1259253 (TC71) | Cytidine (13.8 μM) | 0.03 | 86.0 | 85.3 | **8.2** |
| | | AID: 1259255 (EW8) | Cytidine (13.8 μM) | 0.1 | 80.5 | 79.0 | 3.6 |
| | Carcinoma (MCC) cell viability reduction (Multiple cell lines) | AID: 1296009 (NIH3T3) | Cytidine (13.8 μM) | 0.5 | 81.8 | 74.0 | 1.6 |
| | | AID: 1296009 (WaGa) | Cytidine (13.8 μM) | 0.1 | 76.5 | 69.5 | **8.3** |
| | | AID: 1296009 (HACAT) | Cytidine (13.8 μM) | 29.6 | 100.0 | 5.1 | 0.0 |
| | | AID: 1296009 (MKL-1) | Cytidine (13.8 μM) | 9.3 | 81.1 | 44.1 | 1.2 |
| | Neuroblastoma cell viability reduction | AID: 1347095 (NB-EBc1) | Cytidine (13.8 μM) | 33.2 | 88.3 | 23.9 | 0.0 |
| | | AID: 1347115 (NB-EBc1) | Cytidine (13.8 μM) | 3.3 | 86.1 | 82.1 | 0.0 |
| | | AID: 1347100 (LAN-5) | Cytidine (13.8 μM) | 0.1 | 74.6 | 65.6 | **13.1** |
| | | AID: 1347094 | Cytidine (13.8 μM) | 0.0 | 53.5 | 53.4 | **9.9** |

| | | | | | | |
|---|---|---|---|---|---|---|
| Atypical Teratoid/Rhabdoid Tumor (AT/RT) cell viability reduction (BT37 cell line) | AID: 1347117 | Cytidine (13.8 µM) | 0.3 | 77.6 | 77.5 | 0.0 |
| Osteosarcoma cell viability reduction | AID: 1347103 (OHS-50) | Cytidine (13.8 µM) | 13.7 | 73.3 | 36.4 | 0.9 |
| Myeloma cell viability reduction (MM1R_ATCC cell line) | AID: 1918981 | Cytidine (13.8 µM) | 0.1 | 28.9 | 28.9 | 0.0 |
| Polymerase Eta inhibition | AID: 588591 | Octadecatrienoic acid-(9Z,12Z,15Z) (288.4 µM) | 7.9 | 96.1 | 96.1 | 0.0 |
| | AID: 720502 | Octadecatrienoic acid-(9Z,12Z,15Z) (288.4 µM) | 23.9 | 100.0 | 100.0 | 0.0 |
| Human tyrosyl-DNA phosphodiesterase I (Tdp1) inhibition | AID: 485290 | Octadecatrienoic acid-(9Z,12Z,15Z) (288.4 µM) | 5.0 | 91.5 | 89.8 | 0.8 |
| Polymerase Iota inhibition | AID: 588590 | Octadecatrienoic acid-(9Z,12Z,15Z) (288.4 µM) | 12.6 | 81.0 | 81.0 | 0.0 |
| | AID: 720496 | Octadecatrienoic acid-(9Z,12Z,15Z) (288.4 µM) | 29.9 | 84.5 | 83.8 | 0.0 |
| Werner syndrome (WS) helicase (WRN) inhibition | AID: 651768 | Octadecatrienoic acid-(9Z,12Z,15Z) (288.4 µM) | 17.8 | 79.9 | 78.5 | 0.0 |
| Jurkat Cell toxicity (caspase 3/7 activity) | AID: 655 | Citric acid (8585.4 µM) | 39.8 | 99.4 | 99.4 | 0.0 |
| Selenoprotein glutathione peroxidase 4 (GPX4) inhibition | AID: 1845192 | 9-octadecenoic acid (18879.3 µM) | 11.2 | 82.5 | 82.5 | 0.1 |
| | | 9Z-hexadecenoic acid (1336.0 µM) | 14.1 | 100.0 | 99.3 | 0.8 |
| RecQ-Like DNA Helicase 1 (RECQ1) inhibition | AID: 2549 | Octadecatrienoic acid-(9Z,12Z,15Z) (288.4 µM) | 1.8 | 65.3 | 65.3 | 0.0 |
| | | Icosatetraenoic acid-(5Z,8Z,11Z,14Z) (76.9 µM) | 2.2 | 64.6 | 63.5 | 0.1 |
| Histone Lysine Methyltransferase G9a inhibition | AID: 504332 | Octadecatrienoic acid-(9Z,12Z,15Z) (288.4 µM) | 17.8 | 119.4 | 119.2 | 0.0 |
| | | Icosatetraenoic acid-(5Z,8Z,11Z,14Z) (76.9 µM) | 15.8 | 125.7 | 111.4 | 0.0 |
| | | Docosahexaenoic acid-(4Z,7Z,10Z,13Z,16Z,19Z) (7.4 µM) | 7.1 | 111.1 | 12.7 | 0.1 |
| | | Docosahexaenoic acid- | 25.1 | 146.3 | 13.6 | 0.0 |

| | | | | | | | |
|---|---|---|---|---|---|---|---|
| | | | (4Z,7Z,10Z,13Z,16Z, 19Z) (7.4 μM) | | | | |
| | | | Glutamic acid (366.2 μM) | 10.6 | 124.3 | 123.8 | 0.0 |
| Anti-oxidation | 15-human lipoxygenase 1 (15-hLO-1) inhibition | AID: 887 | Octadecatrienoic acid-(9Z,12Z,15Z) (288.4 μM) | 0.1 | 82.3 | 82.3 | **5.6** |
| | | | Icosatetraenoic acid-(5Z,8Z,11Z,14Z) (76.9 μM) | 0.3 | 81.6 | 81.6 | 0.0 |
| | | | Icosatetraenoic acid-(5Z,8Z,11Z,14Z) (76.9 μM) | 0.1 | 86.7 | 86.7 | 1.3 |
| | | | Octadecadienoic acid-(9Z,12Z) (2777.0 μM) | 0.2 | 85.8 | 85.8 | **73.4** |
| | 15-human lipoxygenase 2 (15-hLO-2) inhibition | AID: 881 | Octadecatrienoic acid-(9Z,12Z,15Z) (288.4 μM) | 14.1 | 73.9 | 73.6 | 0.0 |
| | Human lipoxygenase (12hLO) inhibition | AID: 1671190 | Tyrosine (4.6 μM) | 1.8 | 100.0 | 70.8 | 0.0 |
| | | | Acetylcarnitine (41.3 μM) | 35.5 | 100.0 | 53.6 | 0.0 |
| Disease Therapy | GCN5L2 inhibition - multiple diseases therapy | AID: 504327 | Octadecatrienoic acid-(9Z,12Z,15Z) (288.4 μM) | 10.0 | 70.0 | 70.0 | 0.0 |
| | | | Icosatetraenoic acid-(5Z,8Z,11Z,14Z) (76.9 μM) | 22.4 | 94.9 | 80.6 | 0.0 |
| | | | 9-octadecenoic acid (18879.3 μM) | 31.6 | 107.5 | 107.5 | 0.5 |
| | USP1 inhibition - multiple diseases therapy | AID: 504865 | Icosatetraenoic acid-(5Z,8Z,11Z,14Z) (76.9 μM) | 50.1 | 74.4 | 45.1 | 0.0 |
| | | | Docosahexaenoic acid-(4Z,7Z,10Z,13Z,16Z, 19Z) (7.4 μM) | 50.1 | 69.3 | 4.2 | 0.0 |
| | | | 9-octadecenoic acid (18879.3 μM) | 35.5 | 94.0 | 94.0 | 0.0 |
| | VDR inhibition - hyperparathyroidism therapy | AID: 504847 | Octadecatrienoic acid-(9Z,12Z,15Z) (288.4 μM) | 25.1 | 110.1 | 107.8 | 0.0 |
| | | | Icosatetraenoic acid-(5Z,8Z,11Z,14Z) (76.9 μM) | 5.6 | 94.6 | 93.5 | 0.0 |
| | | | Docosahexaenoic acid-(4Z,7Z,10Z,13Z,16Z, 19Z) (7.4 μM) | 28.2 | 96.6 | 9.3 | 0.0 |
| | | | Icosapentaenoic acid-(5Z,8Z,11Z,14Z,17Z) (61.1 μM) | 39.8 | 92.4 | 54.0 | 0.0 |
| | Tau aggregation inhibition - Alzheimer's therapy | AID: 1460 | Octadecatrienoic acid-(9Z,12Z,15Z) (288.4 μM) | 28.2 | 96.9 | 95.0 | 0.0 |
| | | | Icosatetraenoic acid-(5Z,8Z,11Z,14Z) (76.9 μM) | 28.2 | 96.3 | 73.5 | 0.0 |
| | | | Docosahexaenoic acid-(4Z,7Z,10Z,13Z,16Z,19Z) (7.4 μM) | 15.8 | 82.4 | 9.1 | 0.0 |
| | ALDH1A1 inhibition - metabolic disease therapy | AID: 1030 | Octadecatrienoic acid-(9Z,12Z,15Z) (288.4 μM) | 3.5 | 100.0 | 100.0 | 0.0 |
| | | | Icosatetraenoic acid-(5Z,8Z,11Z,14Z) (76.9 μM) | 2.8 | 100.0 | 98.2 | 0.0 |
| | | | Docosahexaenoic acid-(4Z,7Z,10Z,13Z,16Z,19Z) (7.4 μM) | 22.4 | 100.0 | 9.6 | 0.0 |
| | | | Citric acid (8585.4 μM) | 12.6 | 100.0 | 100.0 | 0.5 |
| | | | Octadecadienoic acid-(9Z,12Z) (2777.0 μM) | 17.8 | 100.0 | 100.0 | 0.0 |
| Antimicrobial | *S. mansoni* TGR inhibition - Schistosomiasis treatment | AID: 485364 | Octadecatrienoic acid-(9Z,12Z,15Z) (288.4 μM) | 35.5 | 95.2 | 90.0 | 0.0 |
| | | | Icosatetraenoic acid-(5Z,8Z,11Z,14Z) (76.9 μM) | 17.8 | 90.0 | 77.2 | 0.0 |
| | *B. subtilis* PPTases inhibition | AID: 1490 | Octadecatrienoic acid-(9Z,12Z,15Z) (288.4 μM) | 79.4 | 100.0 | 94.1 | 0.0 |
| | | | Icosatetraenoic acid-(5Z,8Z,11Z,14Z) (76.9 μM) | 22.4 | 100.0 | 82.8 | 0.0 |
| | | | Docosahexaenoic acid-(4Z,7Z,10Z,13Z,16Z,19Z) (7.4 μM) | 79.4 | 100.0 | 3.8 | 0.0 |
| | | | Glutamic acid (366.2 μM) | 56.2 | 81.0 | 80.1 | 0.0 |
| | Malaria parasite viability reduction | AID: 504749 | Docosahexaenoic acid-(4Z,7Z,10Z,13Z,16Z,19Z) (7.4 μM) | 12.6 | 80.9 | 9.3 | 0.0 |
| Antivirus | Chikungunya virus (CHIKV) nsP2 protease inhibition | AID: 1964101, 1964105 | Icosapentaenoic acid-(5Z,8Z,11Z,14Z,17Z) (61.1 μM) | 6.7 | 102.7 | 85.2 | 0.1 |
| | | | Octadecanoic acid (9998.3 μM) | 3.7 | 96.7 | 96.6 | **25.8** |
| | Influenza A NS1 inhibition | AID: 2326 | Uridine (34.1 μM) | 12.6 | 62.3 | 61.4 | 0.0 |

* The highest efficacy will be recorded if more than one efficacy values are reported for the same compound – AID pairs.

**Supplementary Table 13. Milk compounds with 6 bioactivities curated in Milk Bioactive Peptide Database (MBPDB).**

| Category | Bioactivity | Study | Compounds | | | | Estimated Efficacy (%) | |
|---|---|---|---|---|---|---|---|---|
| | | | Peptide Sequence | Source Protein (Conc. (μM)) | Available Conc. (μM)* | $EC_{50}$ (μM) | In vitro | In vivo |
| Anti-oxidation | ABTS reduction (Pubchem CID:16240279) | DOI: 10.1007/s00217-012-1894-5 | VYPFPGPIPN | Beta-casein (269.7 μM) | 269.7 | 6.2 | 99.8 | 0.0 |
| | XTT reduction | PMID: 10742656 | EL | Alpha-S1-casein (661.5 μM) | 1984.4 | 63.1 | 99.6 | 0.0 |
| | | | PEL | | 661.5 | 306.0 | 77.6 | 0.0 |
| | Radical scavenging – ORAC (Oxygen radical absorbance capacity) | DOI: 10.1016/j.idairyj.2013.05.008 | FYPEL | Alpha-S1-casein (661.5 μM) | 661.5 | 127.5 | 93.5 | 0.0 |
| | | DOI: 10.3390/foods9080991 | YFYPEL | | 661.5 | 79.2 | 96.9 | 0.0 |
| Disease Therapy | ACE inhibition - hypertension therapy | DOI: 10.1016/j.foodchem.2014.01.056 | MAP | Beta-casein (269.7 μM) | 269.7 | 0.4 | 100.0 | 0.0 |
| | | | LPP | | 269.7 | 9.6 | 99.6 | 0.0 |
| | | | IQA | | 269.7 | 32.9 | 96.8 | 0.0 |
| | | | VEP | | 269.7 | 63.7 | 91.1 | 0.0 |
| | | | HQP | | 539.4 | 134.7 | 90.4 | 0.0 |
| | | | FLQP | | 269.7 | 68.7 | 90.1 | 0.0 |
| | | | GP | | 809.1 | 252.6 | 86.7 | 0.0 |
| | | | VP | | 1078.8 | 420.0 | 82.0 | 0.0 |
| | | | VLGP | | 269.7 | 153.5 | 71.1 | 0.0 |
| | | | KYP | | 269.7 | 363.9 | 37.9 | 0.0 |
| | | | FA | | 269.7 | 487.1 | 27.6 | 0.0 |
| | | | NSLP | | 269.7 | 755.1 | 15.7 | 0.0 |
| | | | IP | Serum albumin (5.3 μM) | 5.3 | 130.0 | 0.5 | 0.0 |
| | | | FP | | 5.3 | 315.0 | 0.1 | 0.0 |
| | | | YP | Lactotransferrin (3.4 μM) | 3.4 | 720.0 | 0.0 | 0.0 |
| | | DOI: 10.1271/bbb1961.54.1101 | IHPFAQTQSLVYP | Beta-casein (269.7 μM) | 269.7 | 19.0 | 98.7 | 0.0 |
| | | | FAQTQSLVYP | | 269.7 | 25.0 | 97.9 | 0.0 |
| | | | HPFAQTQSLVYP | | 269.7 | 26.0 | 97.8 | 0.0 |
| | | | KIHPFAQTQSLVYP | | 269.7 | 39.0 | 95.8 | 0.0 |

| | | | | | | |
|---|---|---|---|---|---|---|
| | SLVYP | | 269.7 | 40.0 | 95.6 | 0.0 |
| | QSLVYP | | 269.7 | 41.0 | 95.5 | 0.0 |
| | TQSLVYP | | 269.7 | 64.0 | 91.1 | 0.0 |
| | QTQSLVYP | | 269.7 | 73.0 | 89.2 | 0.0 |
| | AQTQSLVYP | | 269.7 | 76.0 | 88.5 | 0.0 |
| | LVYP | | 269.7 | 170.0 | 67.6 | 0.0 |
| DOI: 10.3168/jds.S0022-0302(94)77026-0 | AYFYPE | Alpha-S1-casein (661.5 µM) | 661.5 | 106.0 | 95.1 | 0.0 |
| | GTQYTDAPSFSDIPNPIGSENSEKTTMPLW | | 661.5 | 346.0 | 73.9 | 0.0 |
| | DELQDKIHPFAQTQSLVYPFPGPIPNS | Beta-casein (269.7 µM) | 269.7 | 4.0 | 99.9 | 0.0 |
| | LLYQEPVLGPVRGPFPIIV | | 270.0 | 21.0 | 98.4 | 0.0 |
| | PPQSVLSLSQSKVLPVPQ | | 269.7 | 25.0 | 97.9 | 0.0 |
| | SKVLPVPQ | | 269.7 | 39.0 | 95.8 | 0.0 |
| | KYPVEPFTESQSLTL | | 269.7 | 93.0 | 84.8 | 0.0 |
| | YQEPVLGPVRGPFPIIV | | 269.9 | 101.0 | 83.0 | 0.0 |
| | LPQNIPPLTQTPVVVPPFLQPEVMGVSK | | 269.7 | 144.0 | 73.2 | 0.0 |
| | RDMPIQAF | | 269.7 | 209.0 | 59.9 | 0.0 |
| DOI: 10.1016/s0014-5793(02)03576-7 | TTMPLW | Alpha-S1-casein (661.5 µM) | 661.6 | 12.0 | 99.9 | 0.0 |
| | FFVAPFPEVFGK | | 661.6 | 18.0 | 99.7 | 0.0 |
| | FALPQYLK | Alpha-S2-casein (288.0 µM) | 288.2 | 4.3 | 99.9 | 0.0 |
| | FALPQY | | 288.0 | 4.3 | 99.9 | 0.0 |
| | FPQYLQY | | 288.0 | 14.0 | 99.3 | 0.0 |
| | TVY | | 288.0 | 15.0 | 99.2 | 0.0 |
| | NMAINPSK | | 288.0 | 60.0 | 92.6 | 0.0 |
| | ALNEINQFY | | 288.0 | 219.0 | 60.6 | 0.0 |
| | ALNEINQFYQK | | 288.0 | 264.0 | 53.2 | 0.0 |
| DOI: 10.1017/s0022029999003982 | WLAHK | Alpha-lactalbumin (145.7 µM) | 145.7 | 0.6 | 100.0 | 0.0 |
| | VGINYWLAHK | | 145.7 | 327.0 | 21.1 | 0.0 |
| | YGL | | 145.7 | 409.0 | 15.7 | 0.0 |
| | ALPMH | Beta-lactoglobulin (445.5 µM) | 445.5 | 521.0 | 43.5 | 0.0 |
| | LAMA | | 445.5 | 556.0 | 40.9 | 0.0 |
| | LDAQSAPLR | | 445.5 | 635.0 | 35.8 | 0.0 |
| | VAGTWY | | 445.5 | 767.5 | 29.1 | 0.0 |
| | CMENSA | | 445.5 | 788.0 | 28.3 | 0.0 |
| | VLDTDYK | | 445.5 | 946.0 | 22.7 | 0.0 |
| DOI: 10.1016/j.idairyj.2013.05.008 | LGY | Alpha-S1-casein (661.5 µM) | 661.5 | 21.5 | 99.6 | 0.0 |
| | YLGY | | 661.5 | 41.9 | 98.9 | 0.0 |
| | RY | | 661.5 | 54.4 | 98.3 | 0.0 |
| | FYPEL | | 661.5 | 80.6 | 96.8 | 0.0 |
| | RYL | | 661.5 | 106.6 | 95.1 | 0.0 |
| | RYLG | | 661.5 | 224.7 | 85.1 | 0.0 |
| | YQK | Alpha-S2-casein (288.0 µM) | 864.1 | 312.2 | 83.7 | 0.0 |
| | FPQY | | 288.0 | 300.7 | 48.0 | 0.0 |
| DOI: 10.1016/j.idairyj.2007.04.008 | YGGVSLPEW | Alpha-lactalbumin (145.7 µM) | 145.7 | 16.0 | 97.2 | 0.0 |
| | GVSLPEW | | 145.7 | 30.0 | 92.6 | 0.0 |
| | VSLPEW | | 145.7 | 57.0 | 81.5 | 0.0 |
| | LKGYGGVSLPEW | | 145.7 | 83.0 | 70.7 | 0.0 |
| | LVYPFPGPIPNSLPQNIPP | Beta-casein (269.7 µM) | 269.7 | 73.0 | 89.2 | 0.0 |
| | VYPFPGPIPNSLPQNIPP | | 269.7 | 87.0 | 86.1 | 0.0 |
| | LVYPFPGP | | 269.7 | 147.0 | 72.6 | 0.0 |
| DOI: 10.1016/j.jff.2015.11.038 | GYGGVSLPEW | Alpha-lactalbumin (145.7 µM) | 145.7 | 2.0 | 99.9 | 0.0 |
| | WLAHKAL | | 145.7 | 29.0 | 92.9 | 0.0 |
| | LKALPMH | Beta-lactoglobulin (445.5 µM) | 445.5 | 11.0 | 99.8 | 0.0 |
| | IQKVAGTW | | 445.5 | 51.0 | 97.1 | 0.0 |
| | LSFNPTQ | | 445.5 | 106.0 | 91.0 | 0.0 |
| | LDTDY | | 445.5 | 121.0 | 89.1 | 0.0 |
| | VLDTDY | | 445.5 | 128.0 | 88.2 | 0.0 |
| DOI: 10.1016/S0958-6946(98)00048-X | LAHKAL | Alpha-lactalbumin (145.7 µM) | 145.7 | 621.0 | 8.7 | 0.0 |
| | LAYFYP | Alpha-S1-casein (661.5 µM) | 661.5 | 65.0 | 97.7 | 0.0 |
| | DAYPSGAW | | 661.5 | 98.0 | 95.7 | 0.0 |
| | AVPYPQR | Beta-casein (269.7 µM) | 269.7 | 274.0 | 49.1 | 0.0 |
| | YQEPVL | | 269.7 | 280.0 | 48.2 | 0.0 |
| | EMPFPK | | 269.7 | 565.6 | 23.0 | 0.0 |
| | GLDIQK | Beta-lactoglobulin (445.5 µM) | 445.5 | 580.0 | 39.3 | 0.0 |
| DOI: 10.1021/jf049510t | NENLLRFFVAPFPEVFG | Alpha-S1-casein (661.5 µM) | 661.5 | 55.0 | 98.3 | 0.0 |
| | ENLLRFFVAPFPEVFG | | 661.5 | 250.0 | 82.8 | 0.0 |
| | LNENLLRFFVAPFPEVFG | | 661.5 | 280.0 | 80.0 | 0.0 |
| | FVAPFPEVFG | | 661.5 | 650.0 | 50.5 | 0.0 |
| | NLHLPLPLL | Beta-casein (269.7 µM) | 269.7 | 15.0 | 99.1 | 0.0 |
| | VENLHLPLPLL | | 269.7 | 175.0 | 66.6 | 0.0 |
| | ENLHLPLPLL | | 269.7 | 250.0 | 52.8 | 0.0 |
| DOI: 10.3168/jds.S0022-0302(96)76487-1 | YKVPQL | Alpha-S1-casein (661.5 µM) | 661.5 | 22.0 | 99.6 | 0.0 |
| | MKPWIQPK | Alpha-S2-casein (288.0 µM) | 288.0 | 300.0 | 48.1 | 0.0 |
| | TKVIP | | 288.0 | 400.0 | 36.8 | 0.0 |
| | AMKPW | | 288.0 | 580.0 | 24.2 | 0.0 |
| | AMKPWIQPK | | 288.1 | 600.0 | 23.2 | 0.0 |
| | LQSW | Beta-casein (269.7 µM) | 269.7 | 500.0 | 26.7 | 0.0 |
| | KVLPVPQ | | 269.7 | 1000 | 10.6 | 0.0 |
| DOI: 10.3390/foods9080991 | RYLGY | Alpha-S1-casein (661.5 µM) | 661.5 | 3.1 | 100.0 | 0.0 |
| | YFYPEL | | 661.5 | 8.8 | 99.9 | 0.0 |
| | YLGYLE | | 661.5 | 85.8 | 96.5 | 0.0 |
| | YPFPGPIP | Beta-casein (269.7 µM) | 269.7 | 224.1 | 57.2 | 0.0 |
| | YPFPGPIPN | | 269.7 | 378.7 | 36.4 | 0.0 |
| | YPFPGPI | | 269.7 | 685.9 | 17.9 | 0.0 |
| | YLL | Beta-lactoglobulin (445.5 µM) | 445.5 | 518.5 | 43.7 | 0.0 |
| DOI: 10.1016/j.idairyj.2007.02.009 | TYKQE | Alpha-S2-casein (288.0 µM) | 288.0 | 18.6 | 98.8 | 0.0 |
| | VPP | Beta-casein (269.7 µM) | 269.7 | 8.4 | 99.6 | 0.0 |
| | SLPQN | | 269.7 | 9.5 | 99.6 | 0.0 |
| | RINKK | | 269.7 | 18.3 | 98.7 | 0.0 |
| | IPP | Kappa-casein (207.0 µM) | 207.0 | 11.6 | 98.9 | 0.0 |
| | ARHPH | | 207.0 | 15.2 | 98.3 | 0.0 |
| DOI: 10.1021/jf100899u | CRRWQWR | Lactotransferrin (3.4 µM) | 3.4 | 2.3 | 64.7 | 0.0 |
| | KCRRWQWR | | 3.4 | 5.8 | 29.7 | 0.0 |
| | FKCRRWQW | | 3.4 | 10.5 | 14.0 | 0.0 |
| | FKCRRWQWRMKKLGAP | | 3.4 | 11.0 | 13.2 | 0.0 |
| | FKCRRW | | 3.4 | 26.7 | 3.5 | 0.0 |
| | RRWQWR | | 3.4 | 32.0 | 2.6 | 0.0 |
| DOI: 10.1021/jf4053868 | LRP | Lactotransferrin (3.4 µM) | 10.3 | 0.4 | 99.6 | 0.0 |
| | PYKLRP | | 3.4 | 10.2 | 14.6 | 0.0 |
| | YKLRP | | 3.4 | 16.5 | 7.3 | 0.0 |
| | DPYKLRP | | 3.4 | 30.5 | 2.8 | 0.0 |
| | GILRP | | 3.4 | 90.7 | 0.5 | 0.0 |
| | KLRP | | 3.4 | 91.6 | 0.5 | 0.0 |
| DOI: 10.1128/AEM.00096-07 | TTMP | Alpha-S1-casein (661.5 µM) | 661.5 | 51.0 | 98.5 | 0.0 |
| | IGSENSEKTTMP | | 661.5 | 773.1 | 43.5 | 0.0 |
| | MPFPKYPVEP | Beta-casein (269.7 µM) | 269.7 | 83.0 | 87.0 | 0.0 |
| | SQSKVLPVPQ | | 269.7 | 92.0 | 85.0 | 0.0 |
| | NIPPLTQTPVVVPPFLQ | | 269.7 | 450.0 | 30.2 | 0.0 |
| | EPVLGPVRGPFP | | 270.2 | 790.0 | 14.8 | 0.0 |
| DOI: 10.1016/j.foodchem.2011.09.052 | KGYGGVSLPEW | Alpha-lactalbumin (145.7 µM) | 145.7 | 0.7 | 100.0 | 0.0 |
| | DKVGINYW | | 145.7 | 25.4 | 94.2 | 0.0 |
| | DKVGINY | | 145.7 | 104.2 | 62.7 | 0.0 |
| | KGYGGVSL | | 145.7 | 325.2 | 21.3 | 0.0 |
| | DAQSAPLRVY | Beta-lactoglobulin (445.5 µM) | 445.5 | 11.6 | 99.7 | 0.0 |
| DOI: 10.1021/jf063427j | WY | Beta-lactoglobulin (445.5 µM) | 445.5 | 38.3 | 98.2 | 0.0 |
| | WYSLAM | | 445.5 | 56.3 | 96.6 | 0.0 |
| | WYSLAMA | | 445.5 | 59.9 | 96.3 | 0.0 |
| | WYSLA | | 445.5 | 86.9 | 93.4 | 0.0 |
| | WYSL | | 445.5 | 90.4 | 92.9 | 0.0 |
| DOI: 10.1080/1475636031000138723 | YY | Alpha-S1-casein (661.5 µM) | 661.5 | 9.0 | 99.9 | 0.0 |
| | LW | | 661.5 | 15.0 | 99.8 | 0.0 |
| | SY | Kappa-casein (207.0 µM) | 207.0 | 41.0 | 92.0 | 0.0 |
| | AY | Lactotransferrin (3.4 µM) | 3.4 | 40.0 | 1.8 | 0.0 |
| | LY | | 3.4 | 44.0 | 1.6 | 0.0 |
| DOI: 10.1016/j.idairyj.2013.DOI: 10.009 | IVP | Alpha-S1-casein (661.5 µM) | 1322.9 | 49.7 | 99.5 | 0.0 |
| | VIP | Alpha-S2-casein (288.0 µM) | 288.0 | 26.1 | 98.0 | 0.0 |
| | INP | | 288.0 | 71.2 | 90.5 | 0.0 |
| | IQP | | 288.0 | 106.7 | 83.1 | 0.0 |
| DOI: 10.1515/bchm3.1996.377.4.259 | YGLF | Alpha-lactalbumin (145.7 µM) | 145.7 | 733.0 | 6.8 | 0.0 |
| | YL | Alpha-S1-casein (661.5 µM) | 1322.9 | 122.1 | 97.9 | 0.0 |
| | YLLF | Beta-lactoglobulin (445.5 µM) | 445.5 | 172.0 | 82.2 | 0.0 |
| | LF | | 445.5 | 349.1 | 59.5 | 0.0 |
| DOI: 10.3168/jds.S0022-0302(98)75878-3 | VYPFPG | Beta-casein (269.7 µM) | 269.7 | 221.0 | 57.8 | 0.0 |
| | VYP | | 269.7 | 288.0 | 47.1 | 0.0 |
| | TPVVVPPFLQP | | 269.7 | 749.0 | 15.9 | 0.0 |
| | IPA | Beta-lactoglobulin (445.5 µM) | 445.5 | 141.0 | 86.5 | 0.0 |
| DOI: 10.1002/elps.200700324 | IAK | Kappa-casein (207.0 µM) | 207.0 | 15.7 | 98.2 | 0.0 |
| | LPYPY | | 207.0 | 28.9 | 95.3 | 0.0 |
| | VR | Serum albumin (5.3 µM) | 5.3 | 52.8 | 2.3 | 0.0 |
| DOI: 10.1016/j.chroma.2004.05.022 | PKHP | Alpha-S1-casein (661.5 µM) | 661.5 | 709.1 | 47.0 | 0.0 |
| | PFP | Beta-casein (269.7 µM) | 809.1 | 144.4 | 94.2 | 0.0 |
| | QP | Kappa-casein (207.0 µM) | 414.0 | 598.1 | 34.6 | 0.0 |
| DOI: 10.1016/j.foodchem.2011.08.076 | LIWKL | Lactotransferrin (3.4 µM) | 3.4 | 0.5 | 95.9 | 0.0 |
| | RPYL | | 3.4 | 56.5 | 1.1 | 0.0 |
| | LNNSRAP | | 3.4 | 105.3 | 0.4 | 0.0 |
| DOI: 10.1016/j.idairyj.2005.12.011 | VLGPVRGPFP | Beta-casein (269.7 µM) | 269.7 | 137.0 | 74.8 | 0.0 |
| | LHLPLPL | | 269.7 | 425.0 | 32.2 | 0.0 |
| | VRGPFPIIV | | 269.7 | 600.0 | 21.3 | 0.0 |
| DOI: 10.1021/acs.jafc.9b06237 | IPAVF | Beta-lactoglobulin (445.5 µM) | 445.5 | 9.1 | 99.8 | 0.0 |
| | MHI | | 445.5 | 11.6 | 99.7 | 0.0 |
| | IAEK | | 445.5 | 25.1 | 99.1 | 0.0 |
| DOI: 10.1128/AEM.66.9.3898-3904.2000 | NIPPLTQTPV | Beta-casein (269.7 µM) | 269.7 | 173.3 | 67.0 | 0.0 |
| | DKIHPF | | 269.7 | 256.8 | 51.7 | 0.0 |
| | LNVPGEIVE | | 269.7 | 300.1 | 45.5 | 0.0 |
| DOI: 10.3168/jds.S0022-0302(04)70013-2 | IR | Beta-lactoglobulin (445.5 µM) | 445.5 | 695.0 | 32.6 | 0.0 |
| | ALPM | | 445.5 | 928.0 | 23.2 | 0.0 |
| | HIR | | 445.5 | 953.0 | 22.4 | 0.0 |

| | | | | | | | |
|---|---|---|---|---|---|---|---|
| | DOI: 10.1016/j.chroma.2006.03.006 | LDIQK | Beta-lactoglobulin (445.5 μM) | 445.5 | 27.6 | 98.9 | 0.0 |
| | | MKG | | 445.5 | 71.8 | 95.0 | 0.0 |
| | DOI: 10.1016/j.foodchem.2014.09.098 | MKP | Alpha-S2-casein (288.0 μM) | 288.0 | 0.4 | 100.0 | 0.0 |
| | | ALPMHIR | Beta-lactoglobulin (445.5 μM) | 445.5 | 14.0 | 99.6 | 0.0 |
| | DOI: 10.1016/j.foodchem.2015.04.130 | DERF | Kappa-casein (207.0 μM) | 207.0 | 37.1 | 93.1 | 0.0 |
| | | RYPSYG | | 207.0 | 65.7 | 84.5 | 0.0 |
| | DOI: 10.1016/j.idairyj.2006.07.004 | GAW | Alpha-S1-casein (661.5 μM) | 661.5 | 240.0 | 83.7 | 0.0 |
| | | GTW | Beta-lactoglobulin (445.5 μM) | 445.5 | 464.4 | 48.1 | 0.0 |
| | DOI: 10.1016/j.idairyj.2009.07.003 | LVYPFPGPIPNSLPQN | Beta-casein (269.7 μM) | 269.7 | 71.0 | 89.6 | 0.0 |
| | | LVYPFPGPIP | | 269.7 | 89.0 | 85.7 | 0.0 |
| | DOI: 10.1016/j.idairyj.2015.09.004 | IW | Alpha-lactalbumin (145.7 μM) | 145.7 | 0.7 | 100.0 | 0.0 |
| | | WL | | 291.4 | 10.0 | 99.6 | 0.0 |
| | DOI: 10.1016/j.jff.2014.04.002 | IIAEK | Beta-lactoglobulin (445.5 μM) | 445.5 | 63.7 | 95.9 | 0.0 |
| | | IPAVFK | | 445.5 | 144.8 | 86.0 | 0.0 |
| | DOI: 10.1016/j.jff.2015.04.043 | LRFF | Alpha-S1-casein (661.5 μM) | 661.5 | 116.9 | 94.3 | 0.0 |
| | | VLSRYP | Kappa-casein (207.0 μM) | 207.0 | 36.7 | 93.2 | 0.0 |
| | DOI: 10.1016/S1671-2927(11)60026-6 | KVLILA | Beta-casein (269.7 μM) | 269.7 | 20.9 | 98.4 | 0.0 |
| | | VVPP | | 269.7 | 258.2 | 51.5 | 0.0 |
| | DOI: 10.1051/lait:2007019 | FPEVFGK | Alpha-S1-casein (661.5 μM) | 661.5 | 140.0 | 92.5 | 0.0 |
| | | AVPYP | Beta-casein (269.7 μM) | 269.7 | 80.0 | 87.7 | 0.0 |
| | DOI: 10.3168/jds.2015-9569 | YQEPVLGPVR | Beta-casein (269.7 μM) | 269.7 | 300.0 | 45.5 | 0.0 |
| | | QEPVLGPVRGPFPIIV | | 269.9 | 600.0 | 21.4 | 0.0 |
| | DOI: 10.1002/jsfa.3394 | AIPPKKNQD | Kappa-casein (207.0 μM) | 207.0 | 19.9 | 97.4 | 0.0 |
| | DOI: 10.1007/s00217-011-1430-z | KLLSHSL | Hemoglobin subunit alpha (0.0 μM) | 0.0 | 42.5 | 0.0 | 0.0 |
| | DOI: 10.1007/s00217-012-1894-5 | VYPFPGPIPN | Beta-casein (269.7 μM) | 269.7 | 325.0 | 42.3 | 0.0 |
| | DOI: 10.1016/j.bcab.2019.101258 | TTFHTSGY | Alpha-lactalbumin (145.7 µM) | 145.7 | 142.0 | 50.6 | 0.0 |
| | DOI: 10.1016/j.foodchem.2015.05.121 | VLPVPQK | Beta-casein (269.7 µM) | 269.8 | 15.0 | 99.1 | 0.0 |
| | DOI: 10.1016/j.foodchem.2018.02.107 | NMAINPSKENLCSTFCK | Alpha-S2-casein (288.0 µM) | 288.0 | 129.1 | 78.3 | 0.0 |
| | DOI: 10.1016/j.idairyj.2009.05.004 | AYFYPEL | Alpha-S1-casein (661.5 µM) | 661.5 | 6.6 | 99.9 | 0.0 |
| | DOI: 10.1016/j.ijfoodmicro.2013.09.002 | LVYPFP | Beta-casein (269.7 µM) | 269.7 | 132.0 | 75.9 | 0.0 |
| | DOI: 10.1016/j.jchromb.2020.122042 | LLYQEPVLGPVR | Beta-casein (269.7 µM) | 269.7 | 274.0 | 49.1 | 0.0 |
| | DOI: 10.1016/S0141-0229(97)00261-5 | LVYPFPGPIPNSLPQ | Beta-casein (269.7 µM) | 269.7 | 18.0 | 98.8 | 0.0 |
| | DOI: 10.1017/S0022029908003452 | LQP | Beta-casein (269.7 µM) | 269.7 | 3.4 | 99.9 | 0.0 |
| | DOI: 10.1080/00021369.1985.10866901 | FFVAP | Alpha-S1-casein (661.5 µM) | 661.5 | 6.0 | 100.0 | 0.0 |
| | DOI: 10.3168/jds.2015-10437 | PFPEVFGK | Alpha-S1-casein (661.5 µM) | 661.5 | 108.0 | 95.0 | 0.0 |
| | DOI: 10.3168/jds.S0022-0302(00)75013-2 | RPKHPIKHQ | Alpha-S1-casein (661.5 µM) | 661.5 | 13.4 | 99.8 | 0.0 |
| | DOI: 10.3168/jds.S0022-0302(06)72370-0 | RPK | Alpha-S1-casein (661.5 µM) | 661.5 | 91.8 | 96.1 | 0.0 |
| | DOI: 10.3390/foods13213392 | PYVRYL | Alpha-S2-casein (288.0 µM) | 288.2 | 7.0 | 99.8 | 0.0 |
| | PMID: 16830788 | LRPVAA | Lactotransferrin (3.4 µM) | 3.4 | 4.1 | 41.9 | 0.0 |
| Dipeptidyl Peptidase IV (DPP IV) inhibition - T2D therapy | DOI: 10.1016/j.peptides.2014.01.002 | WLAHKALCSEKLDQ | Alpha-lactalbumin (145.7 µM) | 145.7 | 141.0 | 50.8 | 0.0 |
| | | LAHKALCSEKL | | 145.7 | 165.0 | 44.6 | 0.0 |
| | | TKCEVFRE | | 145.7 | 166.0 | 44.3 | 0.0 |
| | | LCSEKLDQ | | 145.7 | 186.0 | 39.9 | 0.0 |
| | | ILDKVGINY | | 145.7 | 263.0 | 27.5 | 0.0 |
| | | IVQNNDSTEYGLF | | 145.7 | 337.0 | 20.3 | 0.0 |
| | | LKPTPEGDL | Beta-lactoglobulin (445.5 µM) | 445.5 | 45.0 | 97.6 | 0.0 |
| | | LKPTPEGDLE | | 445.5 | 45.0 | 97.6 | 0.0 |
| | | LKPTPEGDLEIL | | 445.5 | 57.0 | 96.5 | 0.0 |
| | | IPAVFKIDA | | 445.5 | 191.0 | 79.6 | 0.0 |

| | | | | | | |
|---|---|---|---|---|---|---|
| | VPITPTL | Alpha-S2-casein (288.0 μM) | 288.0 | 110.0 | 82.4 | 0.0 |
| | VPITPT | | 288.0 | 130.0 | 78.1 | 0.0 |
| DOI: 10.1016/j.idairyj.2011.08.002 | LPQ | Beta-casein (269.7 μM) | 269.7 | 82.0 | 87.2 | 0.0 |
| | LPQNIPP | | 269.7 | 160.0 | 69.8 | 0.0 |
| | FPGPIPN | | 269.7 | 260.0 | 51.2 | 0.0 |
| | YPFPGPIPN | | 269.7 | 670.0 | 18.5 | 0.0 |
| | PGPIPNS | | 269.7 | 1000 | 10.6 | 0.0 |
| | PQNIPPL | | 269.7 | 1500 | 5.8 | 0.0 |
| DOI: 10.1016/j.peptides.2016.03.005 | LPVPQ | Beta-casein (269.7 μM) | 269.7 | 43.8 | 95.0 | 0.0 |
| | YPVEPF | | 269.7 | 124.7 | 77.5 | 0.0 |
| | LPQNIPPLT | | 269.7 | 205.2 | 60.7 | 0.0 |
| | VPGEIVE | | 269.7 | 224.5 | 57.1 | 0.0 |
| | LPLPL | | 269.7 | 325.0 | 42.3 | 0.0 |
| | LPLPLL | | 269.7 | 371.5 | 37.1 | 0.0 |
| | IPPLTQT | | 269.7 | 465.1 | 29.1 | 0.0 |
| | YPFPGP | | 269.7 | 749.2 | 15.9 | 0.0 |
| DOI: 10.1016/j.jff.2013.09.012 | WIQP | Alpha-S2-casein (288.0 μM) | 288.0 | 237.3 | 57.5 | 0.0 |
| | FLQP | Beta-casein (269.7 μM) | 269.7 | 65.3 | 90.8 | 0.0 |
| | VLGP | | 269.7 | 580.4 | 22.3 | 0.0 |
| | IPI | Kappa-casein (207.0 μM) | 207.0 | 3.5 | 99.8 | 0.0 |
| | IP | Serum albumin (5.3 μM) | 5.3 | 149.6 | 0.4 | 0.0 |
| | LP | | 21.1 | 712.5 | 0.3 | 0.0 |
| DOI: 10.1016/j.foodchem.2013.08.097 | LPL | Beta-casein (269.7 μM) | 269.7 | 241.4 | 54.2 | 0.0 |
| | IPIQY | Kappa-casein (207.0 μM) | 207.0 | 35.2 | 93.6 | 0.0 |
| | LPYPY | | 207.0 | 108.3 | 71.6 | 0.0 |
| | YPY | | 207.0 | 243.7 | 42.0 | 0.0 |
| | YP | Lactotransferrin (3.4 μM) | 3.4 | 658.1 | 0.0 | 0.0 |
| DOI: 10.1016/j.foodchem.2020.126333 | KILDKVGINYWLAHK | Alpha-lactalbumin (145.7 μM) | 145.7 | 11.4 | 98.4 | 0.0 |
| | EQLTKCEVFR | | 145.7 | 12.0 | 98.2 | 0.0 |
| | VGINYWLAHK | | 145.7 | 13.8 | 97.8 | 0.0 |
| | ILDKVGINYWLAHK | | 145.7 | 23.1 | 95.0 | 0.0 |
| | LDQWLCEKL | | 145.7 | 59.6 | 80.4 | 0.0 |

| | | | | | | | |
|---|---|---|---|---|---|---|---|
| | DOI: 10.1016/j.jff.2015.11.038 | WLAHKAL | Alpha-lactalbumin (145.7 µM) | 145.7 | 286.0 | 24.9 | 0.0 |
| | | LKGYGGVSLPE | | 145.7 | 486.0 | 12.4 | 0.0 |
| | | LKALPMH | Beta-lactoglobulin (445.5 µM) | 445.5 | 193.0 | 79.4 | 0.0 |
| | | IQKVAGTW | | 445.5 | 329.0 | 61.8 | 0.0 |
| | | VLDTDY | | 445.5 | 471.0 | 47.5 | 0.0 |
| | DOI: 10.1016/j.jff.2014.04.002 | VAGTWY | Beta-lactoglobulin (445.5 µM) | 445.5 | 74.9 | 94.7 | 0.0 |
| | | IPAVFK | | 445.5 | 149.1 | 85.4 | 0.0 |
| | | TPEVDDEALEK | | 445.5 | 578.7 | 39.4 | 0.0 |
| | DOI: 10.1016/j.foodchem.2013.03.056 | IPAVF | Beta-lactoglobulin (445.5 µM) | 445.5 | 44.7 | 97.6 | 0.0 |
| | | VLVLDTDYK | | 445.5 | 424.4 | 51.8 | 0.0 |
| | DOI: 10.1016/j.foodres.2015.09.012 | WL | Alpha-lactalbumin (145.7 µM) | 291.4 | 43.6 | 95.4 | 0.0 |
| | | WI | Lactotransferrin (3.4 µM) | 3.4 | 138.7 | 0.2 | 0.0 |
| | DOI: 10.1016/j.peptides.2012.11.016 | WV | Alpha-lactalbumin (145.7 µM) | 145.7 | 65.7 | 77.9 | 0.0 |
| | | VR | Serum albumin (5.3 µM) | 5.3 | 826.1 | 0.0 | 0.0 |
| | DOI: 10.1016/j.foodchem.2017.DOI: 10.033 | LPVP | Beta-casein (269.7 µM) | 269.7 | 87.0 | 86.1 | 0.0 |
| | DOI: 10.1016/j.peptides.2011.01.002 | IPA | Beta-lactoglobulin (445.5 µM) | 445.5 | 49.0 | 97.3 | 0.0 |
| | DOI: 10.1111/1471-0307.12868 | LKDLKGYGG | Alpha-lactalbumin (145.7 µM) | 145.7 | 493 | 0.3 | 0.0 |
| Prolyl endopeptidase (PREP) inhibition - multiple diseases therapy | PMID: 1368255 | IHPFAQTQ | Beta-casein (269.7 µM) | 269.7 | 150.0 | 71.9 | 0.0 |
| | | VYPFPGPI | Beta-casein (269.7 µM) | 269.7 | 110.0 | 80.9 | 0.0 |
| | | VYPFPGPIP | Beta-casein (269.7 µM) | 269.7 | 50.0 | 93.8 | 0.0 |

[*] The bioactivity efficacies of peptides were evaluated based on the following formula which takes the concentration of the protein, the number of intervals that contain the peptide sequences (which is equal to 1 for the most cases except few peptides), and the concentration of the free peptide (which is negligible compared to the concentration of the source proteins):

$$Conc_{Final} = Conc_{Protein} \times n_{Interval} + Conc_{Free\ Peptide}$$

**Supplementary Table 14.** List of 54 milk peptides selected for antibacterial activity measurement.

| Peptide | Source Protein | Average Content (% w/w) | | Average Concentration (μM) | |
|---|---|---|---|---|---|
| | | Free Peptide | Source Protein | Free Peptide | Source Protein |
| FFVAPFPEVFGKEKV | α-S1-casein | $7.55 * 10^{-6}$ | 1.58 | 0.05 | 661.46 |
| FFVAPFPEVFGKEKVNELSK | | $7.53 * 10^{-6}$ | | 0.03 | |
| FVAPFPEVFGKEKV | | $6.88 * 10^{-6}$ | | 0.04 | |
| FVAPFPEVFGKEKVNELSK | | $2.64 * 10^{-6}$ | | 0.01 | |
| GYLEQLLRLKK | | $9.06 * 10^{-7}$ | | 0.01 | |
| LGYLEQLLRLK | | $2.74 * 10^{-6}$ | | 0.02 | |
| LNENLLRFFVAPFPEVFGKEKV | | $8.97 * 10^{-7}$ | | $3.86 * 10^{-3}$ | |
| NENLLRFFVAPFPEVFGKEKV | | $9.35 * 10^{-7}$ | | $4.03 * 10^{-3}$ | |
| RPKHPIKH | | $1.34 * 10^{-6}$ | | 0.01 | |
| RPKHPIKHQ | | $3.85 * 10^{-6}$ | | 0.04 | |
| RPKHPIKHQGLP | | $4.88 * 10^{-6}$ | | 0.04 | |
| RPKHPIKHQGLPQEVLN | | $1.04 * 10^{-5}$ | | 0.06 | |
| VAPFPEVFGKEKVNELSK | | $1.64 * 10^{-6}$ | | 0.01 | |
| YLEQLLRLKKY | | $9.86 * 10^{-7}$ | | 0.01 | |
| YLGYLEQLLRLKK | | $2.14 * 10^{-6}$ | | 0.01 | |
| ALNEINQFYQKFPQYLQYLYQGPIVLN PWDQVKR | α-S2-casein | $7.04 * 10^{-6}$ | 0.73 | 0.02 | 288.03 |
| AMKPWIQPKTKVIPYVRYL | | $1.93 * 10^{-5}$ | | 0.09 | |
| KLTEEEKNRLNFL | | $1.57 * 10^{-5}$ | | 0.10 | |
| KTKLTEEEKNRLN | | $1.90 * 10^{-6}$ | | 0.01 | |
| KTKLTEEEKNRLNFLK | | $3.58 * 10^{-5}$ | | 0.19 | |
| LTEEEKNRLNFLK | | $1.63 * 10^{-5}$ | | 0.10 | |
| LTEEEKNRLNFLKKISQ | | $4.14 * 10^{-6}$ | | 0.02 | |
| LTEEEKNRLNFLKKISQR | | $2.75 * 10^{-6}$ | | 0.01 | |
| LTEEEKNRLNFLKKISQRYQ | | $5.58 * 10^{-6}$ | | 0.02 | |
| TKLTEEEKNRLNFL | | $6.39 * 10^{-5}$ | | 0.38 | |
| TKLTEEEKNRLNFLK | | $1.30 * 10^{-4}$ | | 0.72 | |
| TKLTEEEKNRLNFLKK | | $2.83 * 10^{-5}$ | | 0.15 | |
| TKLTEEEKNRLNFLKKISQ | | $2.39 * 10^{-5}$ | | 0.11 | |
| TKLTEEEKNRLNFLKKISQR | | $5.76 * 10^{-6}$ | | 0.02 | |
| EPVLGPVRGPFPII | | $1.76 * 10^{-5}$ | | 0.12 | |
| EPVLGPVRGPFPIIV | | $8.59 * 10^{-5}$ | | 0.55 | |
| GPVRGPFPII | | $4.54 * 10^{-6}$ | | 0.05 | |
| GPVRGPFPIIV | | $4.00 * 10^{-5}$ | | 0.36 | |
| LGPVRGPFPIIV | | $1.24 * 10^{-6}$ | | 0.01 | |
| LYQEPVLGPVRGPFPIIV | β-casein | $1.77 * 10^{-4}$ | 0.66 | 0.91 | 269.70 |
| PVLGPVRGPFPIIV | | $4.76 * 10^{-6}$ | | 0.03 | |
| RELEELNVPGEIVESLSSSEESITRINKKIEK | | $3.40 * 10^{-6}$ | | 0.01 | |
| SKVKEAMAPKHK | | $8.48 * 10^{-7}$ | | 0.01 | |
| VRGPFPIIV | | $1.78 * 10^{-6}$ | | 0.02 | |
| IIAEKTKIPAVFKIDALNENK | β-lactoglobulin | $6.72 * 10^{-7}$ | 0.86 | $4.25 * 10^{-3}$ | 445.53 |
| SRSKKF | Osteopontin | $5.41 * 10^{-6}$ | $2.31 * 10^{-3}$ | 0.08 | 0.77 |
| DTIAQAASTTTISDAVSK | | $1.05 * 10^{-6}$ | | 0.01 | |
| DTIAQAASTTTISDAVSKVK | Lactoperoxidase | $5.21 * 10^{-6}$ | $3.23 * 10^{-3}$ | 0.02 | 0.41 |
| DTIAQAASTTTISDAVSKVKIQVNK | | $1.87 * 10^{-6}$ | | 0.01 | |
| NPKLPLSILKEKHL | Glycosylation-dependent cell adhesion molecule 1 | $4.83 * 10^{-6}$ | 0.05 | 0.03 | 27.15 |
| SQNPKLPLSILKEKHL | | $3.57 * 10^{-6}$ | | 0.02 | |
| AGEIQNKALLDPSFFAKESVKDAAGGPGAPADPGRPTGYSGSSKA | | $2.64 * 10^{-5}$ | | 0.06 | |
| PGRPTGYSGSSKALVSTLVPL | Polymeric immunoglobulin receptor | $6.69 * 10^{-6}$ | 0.01 | 0.03 | 0.93 |
| PGRPTGYSGSSKALVSTLVPLA | | $3.10 * 10^{-5}$ | | 0.15 | |
| AKPEKAKETEKDDTGTPITKIELVPSH | Sodium-dependent phosphate transport protein 2B | $2.45 * 10^{-5}$ | $1.15 * 10^{-3}$ | 0.08 | 0.16 |
| SAKPEKAKETEKDDTGTPITKIELVPSH | | $3.57 * 10^{-6}$ | | 0.01 | |
| IAGAVIGAV | Platelet glycoprotein 4 | n.d. | $1.13 * 10^{-3}$ | n.d. | 0.22 |
| IGIGGIIK | Uncharacterized protein (UniProt ID: F1N1Z8) | n.d. | $3.72 * 10^{-5}$ | n.d. | 0.023 |
| GAVGATIGGVA | Transmembrane protein 263 | n.d. | $9.11 * 10^{-8}$ | n.d. | $8.0 * 10^{-5}$ |

## Supplementary Figures

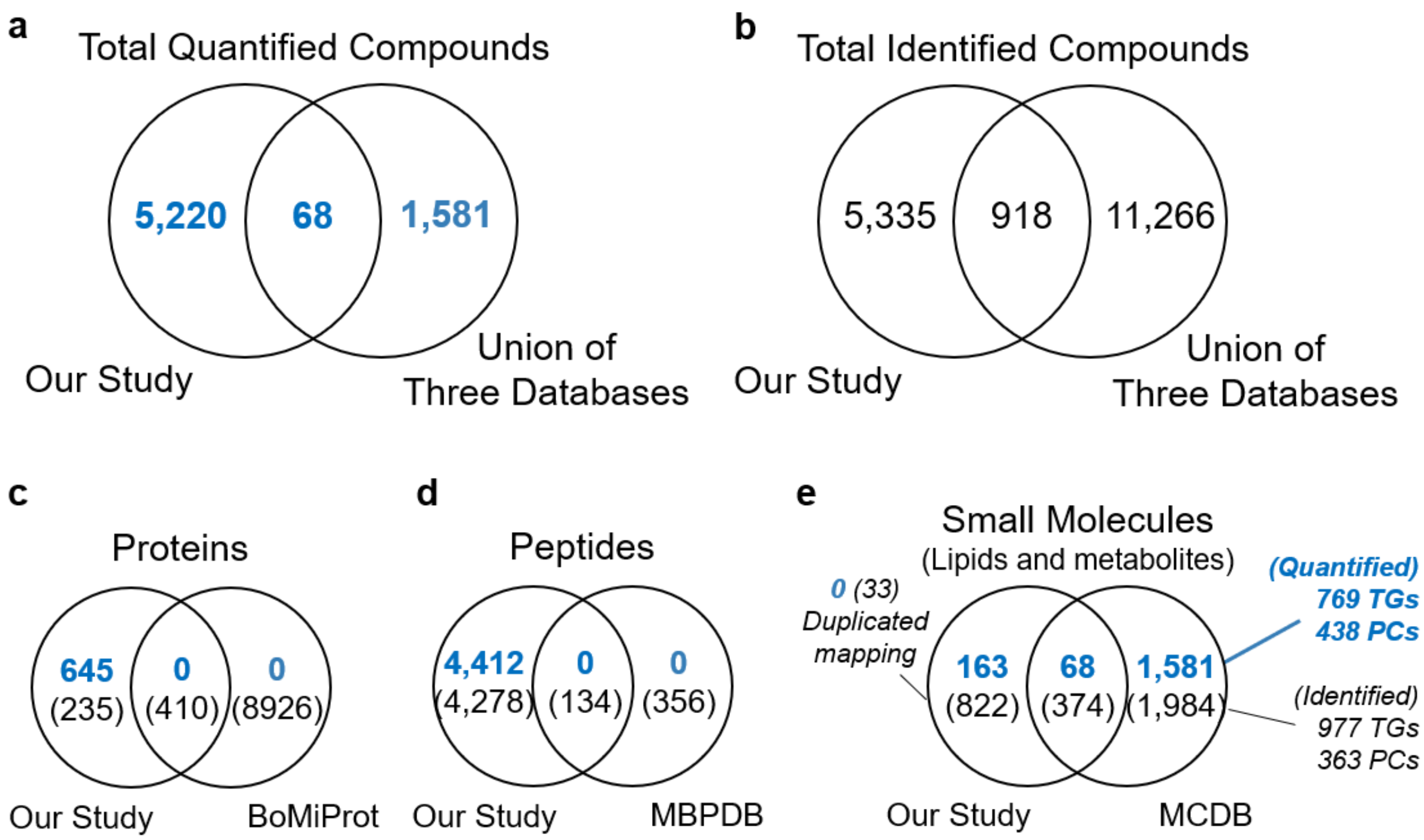

**Supplementary Figure 1. Comparison of identified and quantified compounds between the present study and curated milk molecular databases.**

The numbers of proteins, peptides, and small molecules identified or quantified in the present study were compared with those reported in BoMiProt, MBPDB, and MCDB, respectively. For each molecular class, values shown in bold blue indicate quantified compounds, whereas values shown in black indicate identified compounds.

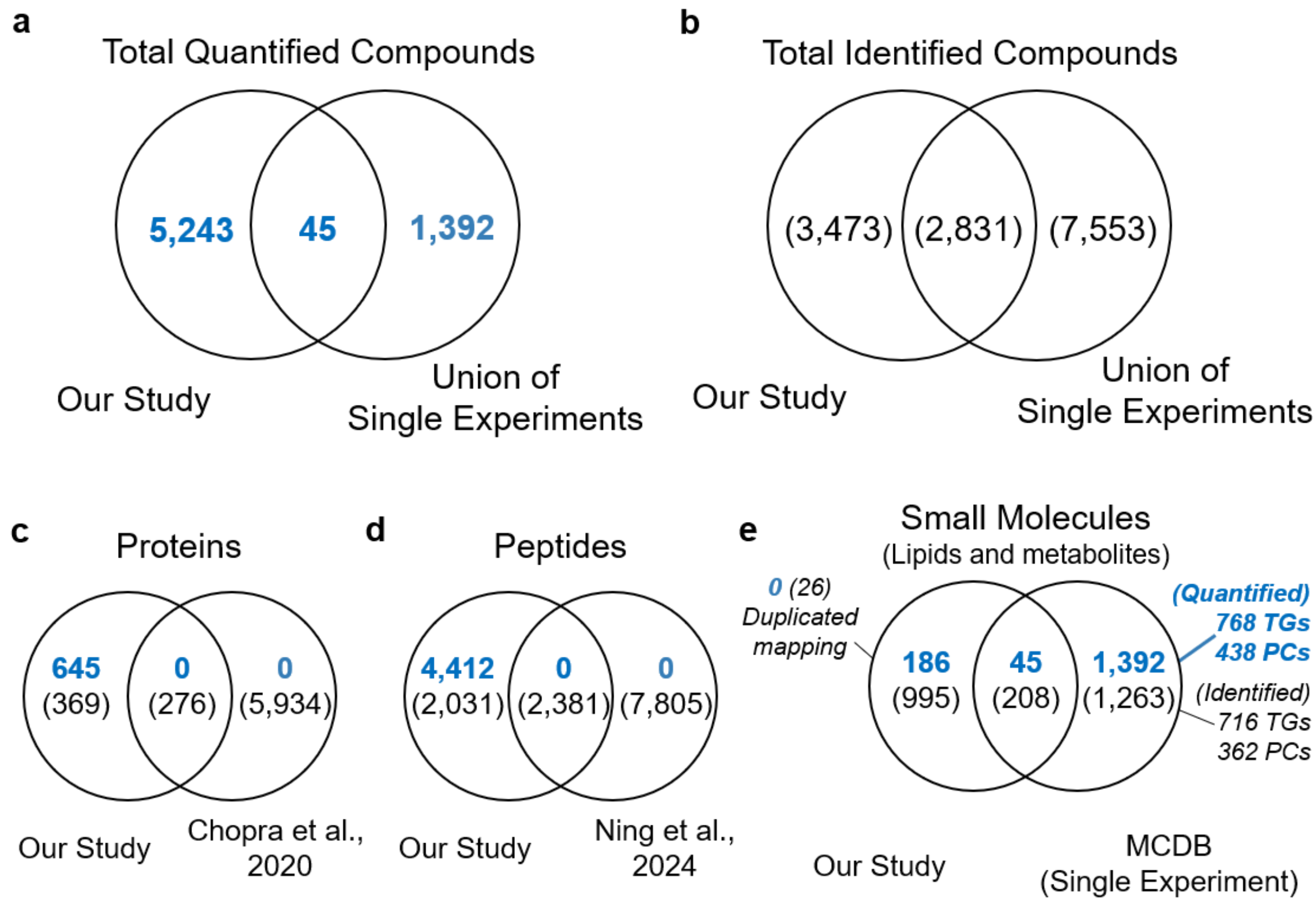

**Supplementary Figure 2. Comparison of identified and quantified compounds between the present study and previous single-omics studies.**

The numbers of proteins, peptides, and small molecules identified or quantified in the present study were compared with those reported in the largest available single-omics studies for proteomics, peptidomics, and small-molecule profiling, respectively. For each molecular class, bold blue values indicate quantified compounds, whereas black values indicate identified compounds.

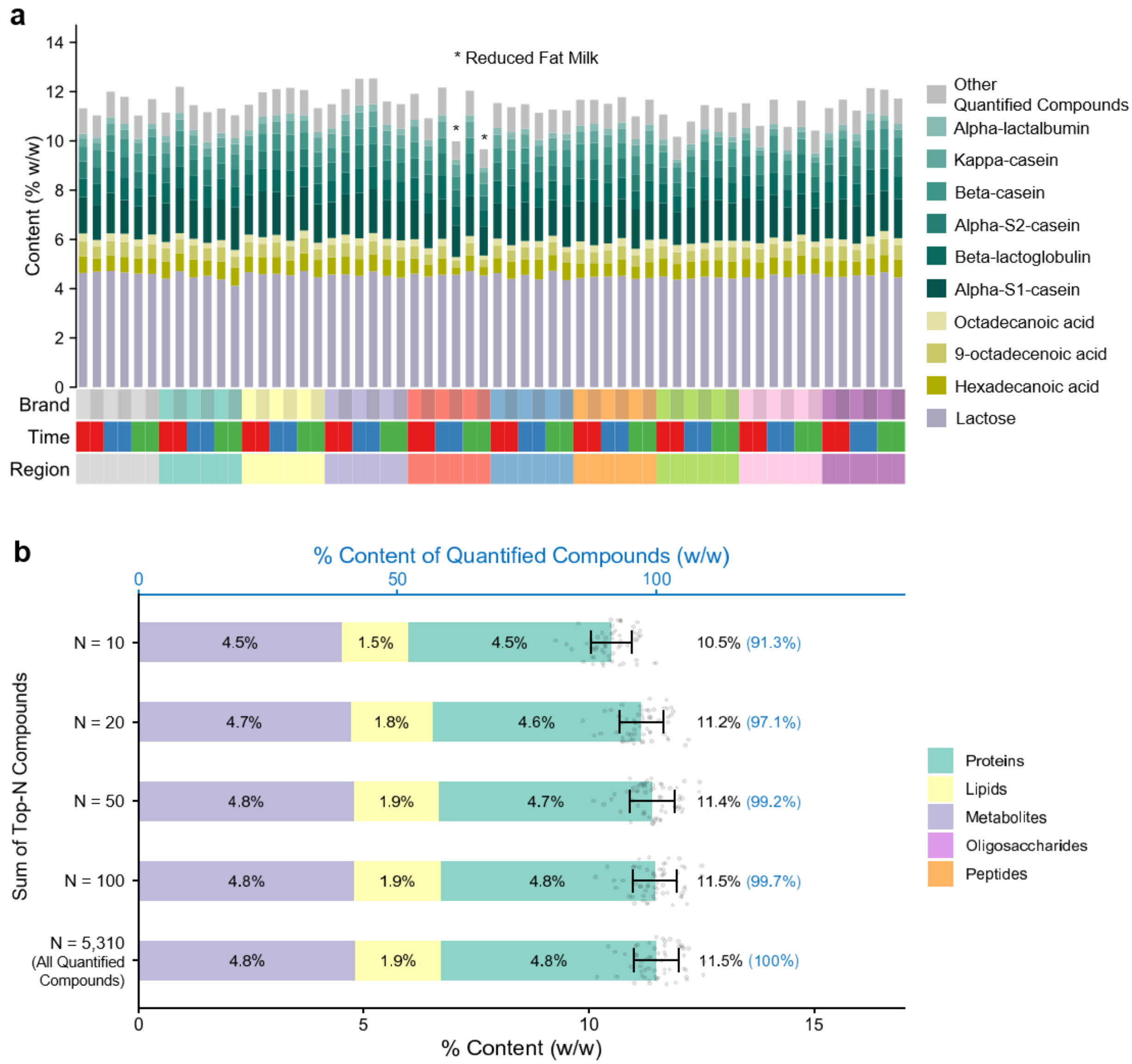

**Supplementary Figure 3. Composition of milk compounds across samples. a.** Overview of the concentration of the top 10 compounds in 60 milk samples. **b.** Contribution of high-abundance compounds, showing the summed contributions of the top 10, 20, 50, and 100 highest-concentration compounds and all compounds to whole milk and to dry milk content.

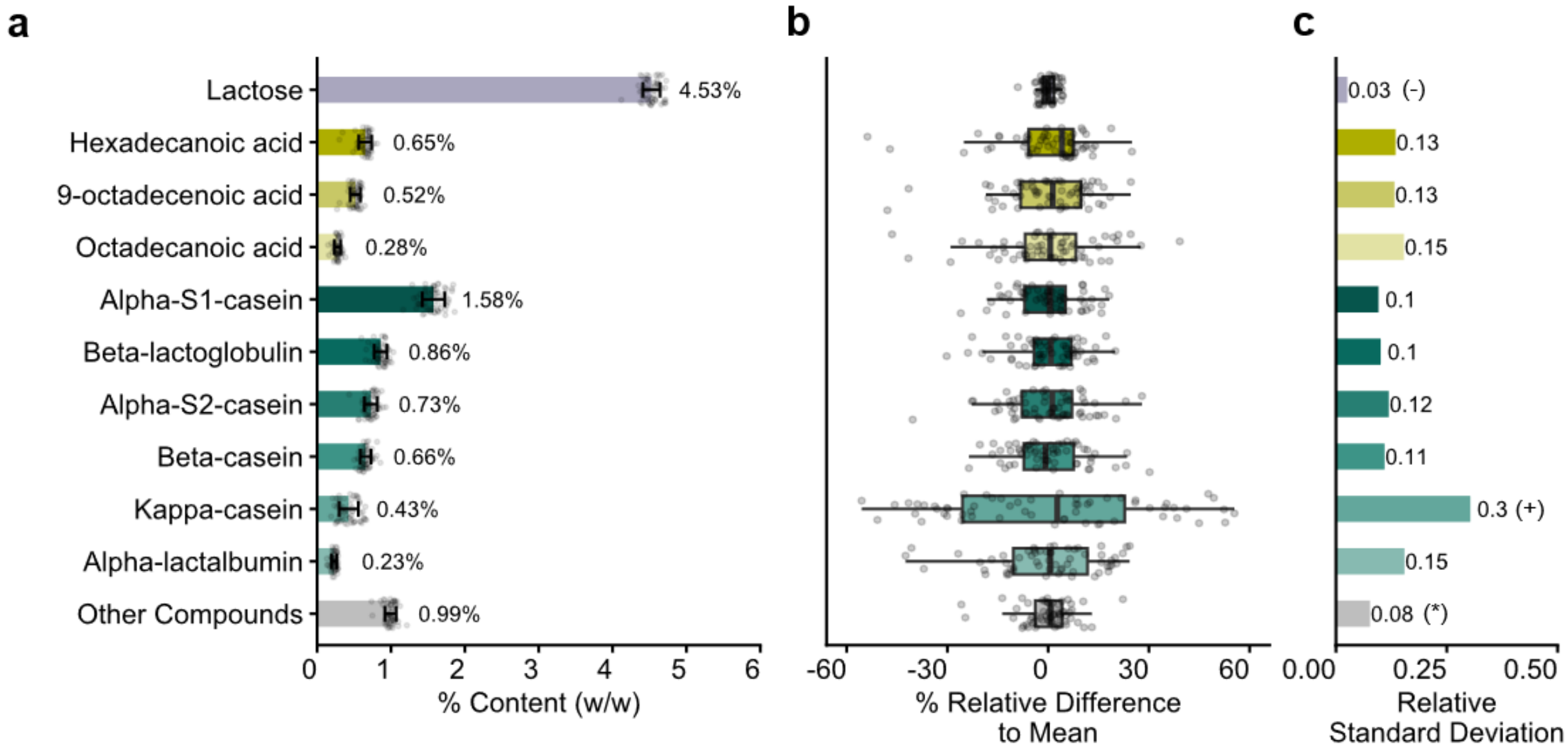

**Supplementary Figure 4. Concentrations and relative variability of major quantified milk compounds. a,** Mean concentrations of representative high-abundance quantified compounds in milk, including lactose, major fatty acids, major milk proteins, and the summed abundance of other quantified compounds. **b,** Sample-level relative differences from the mean concentration for each compound or compound group. **c,** Relative standard deviation of each compound or compound group. Symbols indicate compounds with significantly different variance based on Levene's test after multiple-testing correction, with q-value < 0.05. The RSD of the sum of other compounds is significantly higher than the RSD of lactose but significantly lower than all other compounds.

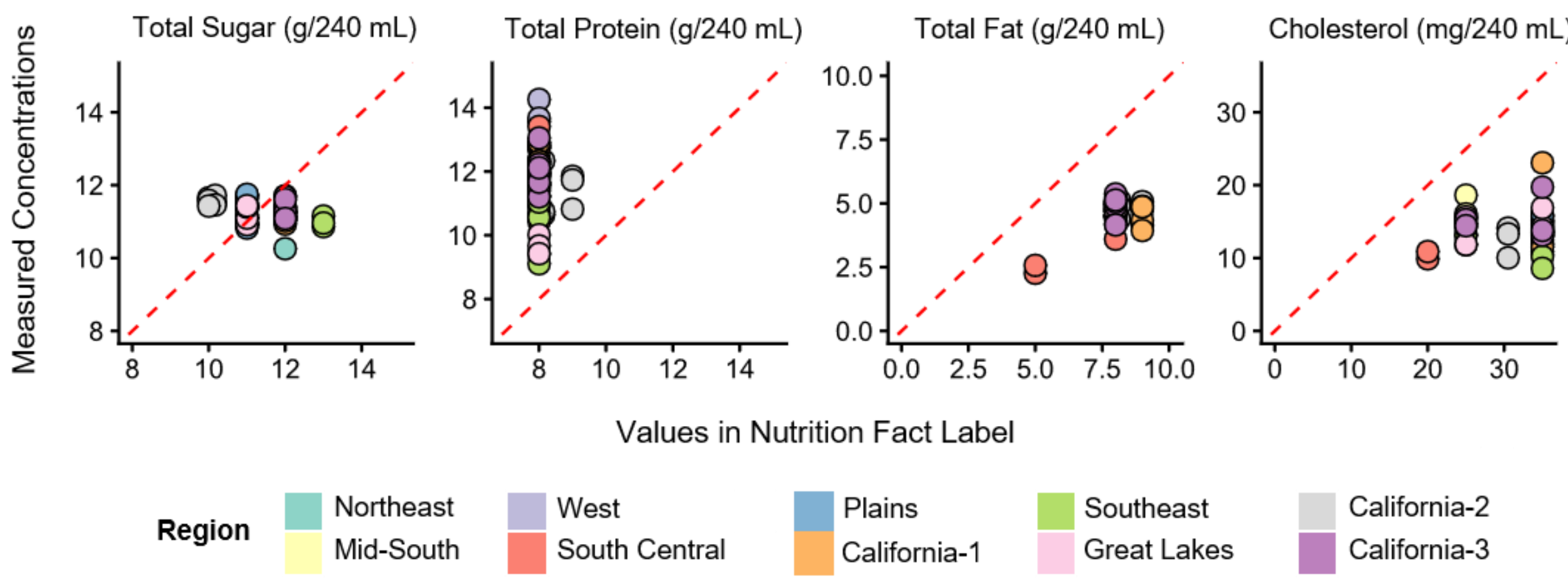

**Supplementary Figure 5. Quantified concentrations of total sugar, total protein, total fat, and cholesterol compared with Nutrition Facts label values.**

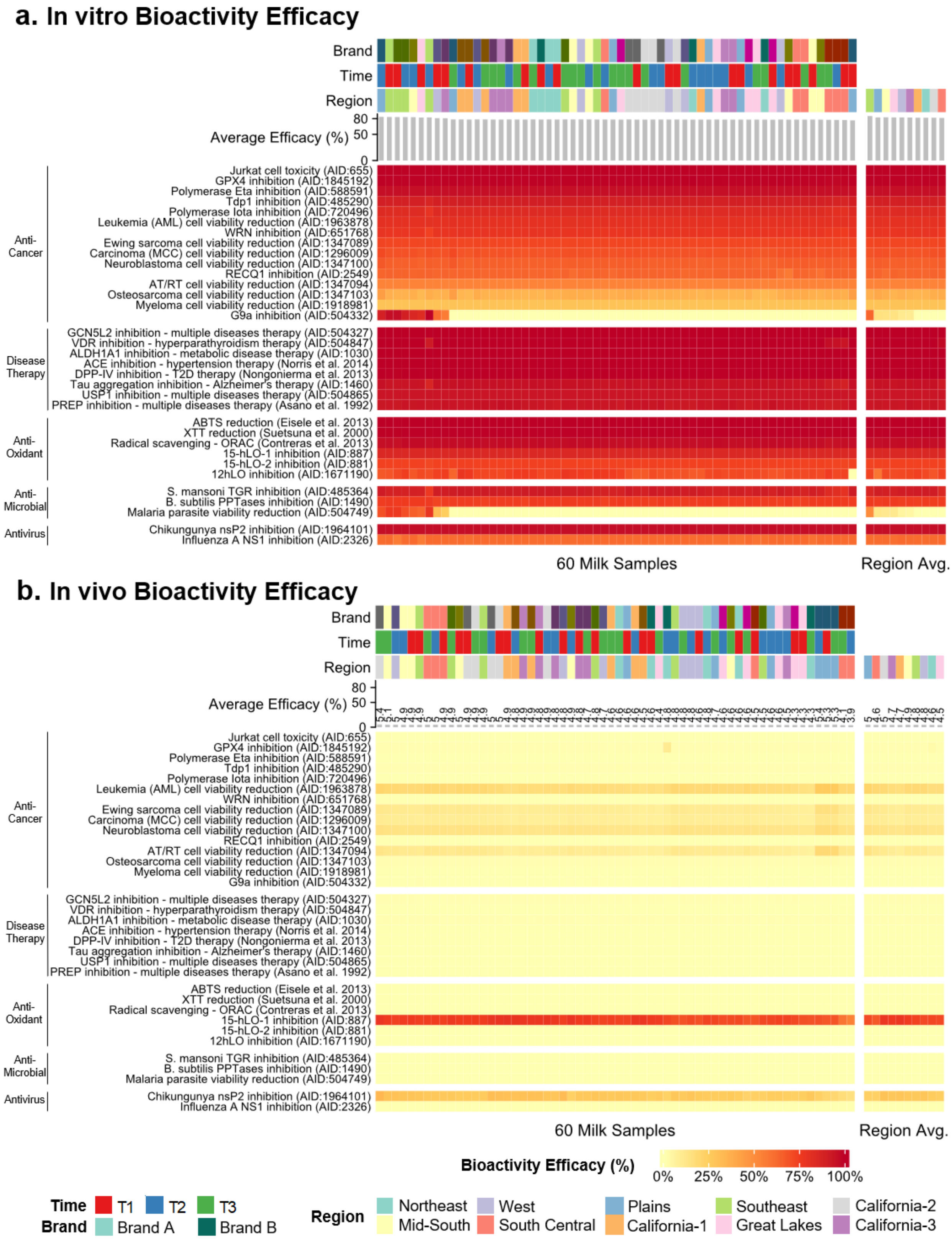
a. In vitro Bioactivity Efficacy
Brand
Time
Region
Average Efficacy (%)
80
50
0
Anti-Cancer
Jurkat cell toxicity (AID:655)
GPX4 inhibition (AID:1845192)
Polymerase Eta inhibition (AID:588591)
Tdp1 inhibition (AID:485290)
Polymerase Iota inhibition (AID:720496)
Leukemia (AML) cell viability reduction (AID:1963878)
WRN inhibition (AID:651768)
Ewing sarcoma cell viability reduction (AID:1347089)
Carcinoma (MCC) cell viability reduction (AID:1296009)
Neuroblastoma cell viability reduction (AID:1347100)
RECQ1 inhibition (AID:2549)
AT/RT cell viability reduction (AID:1347094)
Osteosarcoma cell viability reduction (AID:1347103)
Myeloma cell viability reduction (AID:1918981)
G9a inhibition (AID:504332)
Disease Therapy
GCN5L2 inhibition - multiple diseases therapy (AID:504327)
VDR inhibition - hyperparathyroidism therapy (AID:504847)
ALDH1A1 inhibition - metabolic disease therapy (AID:1030)
ACE inhibition - hypertension therapy (Norris et al. 2014)
DPP-IV inhibition - T2D therapy (Nongonierma et al. 2013)
Tau aggregation inhibition - Alzheimer's therapy (AID:1460)
USP1 inhibition - multiple diseases therapy (AID:504865)
PREP inhibition - multiple diseases therapy (Asano et al. 1992)
Anti-Oxidant
ABTS reduction (Eisele et al. 2013)
XTT reduction (Suetsuna et al. 2000)
Radical scavenging - ORAC (Contreras et al. 2013)
15-hLO-1 inhibition (AID:887)
15-hLO-2 inhibition (AID:881)
12hLO inhibition (AID:1671190)
Anti-Microbial
S. mansoni TGR inhibition (AID:485364)
B. subtilis PPTases inhibition (AID:1490)
Malaria parasite viability reduction (AID:504749)
Antivirus
Chikungunya nsP2 inhibition (AID:1964101)
Influenza A NS1 inhibition (AID:2326)
60 Milk Samples
Region Avg.
b. In vivo Bioactivity Efficacy
Brand
Time
Region
Average Efficacy (%)
80
50
0
Anti-Cancer
Jurkat cell toxicity (AID:655)
GPX4 inhibition (AID:1845192)
Polymerase Eta inhibition (AID:588591)
Tdp1 inhibition (AID:485290)
Polymerase Iota inhibition (AID:720496)
Leukemia (AML) cell viability reduction (AID:1963878)
WRN inhibition (AID:651768)
Ewing sarcoma cell viability reduction (AID:1347089)
Carcinoma (MCC) cell viability reduction (AID:1296009)
Neuroblastoma cell viability reduction (AID:1347100)
RECQ1 inhibition (AID:2549)
AT/RT cell viability reduction (AID:1347094)
Osteosarcoma cell viability reduction (AID:1347103)
Myeloma cell viability reduction (AID:1918981)
G9a inhibition (AID:504332)
Disease Therapy
GCN5L2 inhibition - multiple diseases therapy (AID:504327)
VDR inhibition - hyperparathyroidism therapy (AID:504847)
ALDH1A1 inhibition - metabolic disease therapy (AID:1030)
ACE inhibition - hypertension therapy (Norris et al. 2014)
DPP-IV inhibition - T2D therapy (Nongonierma et al. 2013)
Tau aggregation inhibition - Alzheimer's therapy (AID:1460)
USP1 inhibition - multiple diseases therapy (AID:504865)
PREP inhibition - multiple diseases therapy (Asano et al. 1992)
Anti-Oxidant
ABTS reduction (Eisele et al. 2013)
XTT reduction (Suetsuna et al. 2000)
Radical scavenging - ORAC (Contreras et al. 2013)
15-hLO-1 inhibition (AID:887)
15-hLO-2 inhibition (AID:881)
12hLO inhibition (AID:1671190)
Anti-Microbial
S. mansoni TGR inhibition (AID:485364)
B. subtilis PPTases inhibition (AID:1490)
Malaria parasite viability reduction (AID:504749)
Antivirus
Chikungunya nsP2 inhibition (AID:1964101)
Influenza A NS1 inhibition (AID:2326)
60 Milk Samples
Region Avg.
Bioactivity Efficacy (%)
0%
25%
50%
75%
100%
Time
T1
T2
T3
Brand
Brand A
Brand B
Region
Northeast
West
Plains
Southeast
California-2
Mid-South
South Central
California-1
Great Lakes
California-3

**Supplementary Figure 6. Estimated bioactivity efficacy profiles of milk samples.**

**a,** Estimated in vitro bioactivity efficacy of milk samples, and **b,** Estimated in vivo bioactivity efficacy of milk samples.

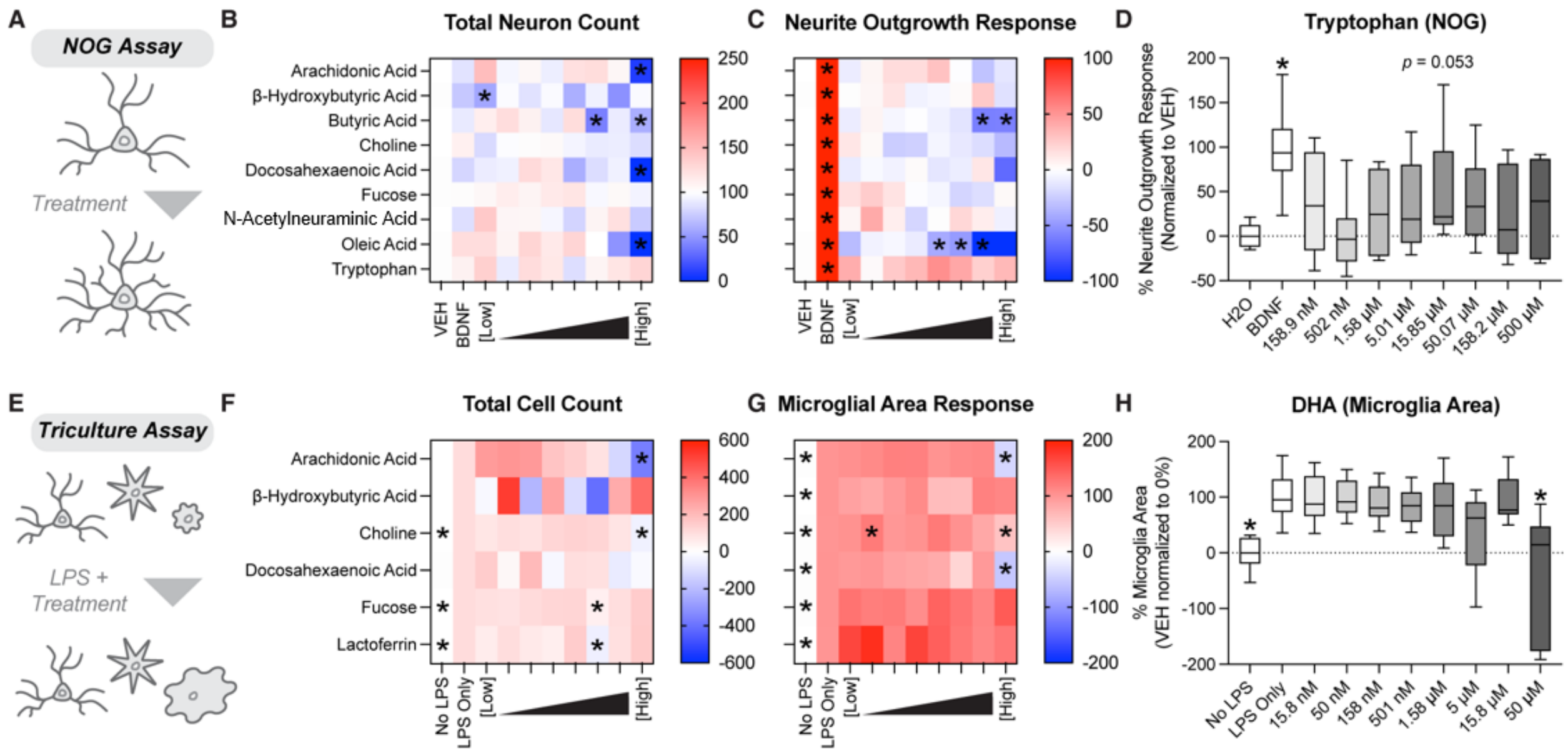

**Supplementary Figure 7. Bioactivity screening of selected milk compounds in neuronal outgrowth and neuroinflammation assays. a,** Schematic overview of the neurite outgrowth assay (NOG), in which cells were treated with candidate milk-derived compounds and assessed for neuronal responses. **b,** Heatmap showing the effects of selected compounds on total neuron count across tested concentrations, with vehicle and BDNF controls included for comparison. **c,** Heatmap showing neurite outgrowth responses after compound treatment, normalized to the vehicle control. **d,** Dose–response analysis of tryptophan in the NOG assay, showing a trend toward increased neurite outgrowth relative to vehicle control. **e,** Schematic overview of the triculture neuroinflammation assay, in which cells were co-treated with LPS and candidate compounds. **f,** Heatmap showing total cell count responses across compounds and concentrations. **g,** Heatmap showing microglial area responses, normalized to the LPS-treated control. **h,** Dose–response analysis of docosahexaenoic acid (DHA), showing reduced microglial area at the highest tested concentration. Data represent mean ±

SEM following a 10% ROUT outlier test (N = 3–12 wells total from 3 independent cultures). Asterisks indicate statistically significant differences relative to the corresponding control condition VEH group for NOG; LPS only group for triculture) following a Brown-Forsythe and Welch ANOVA with Dunnett post hoc test.

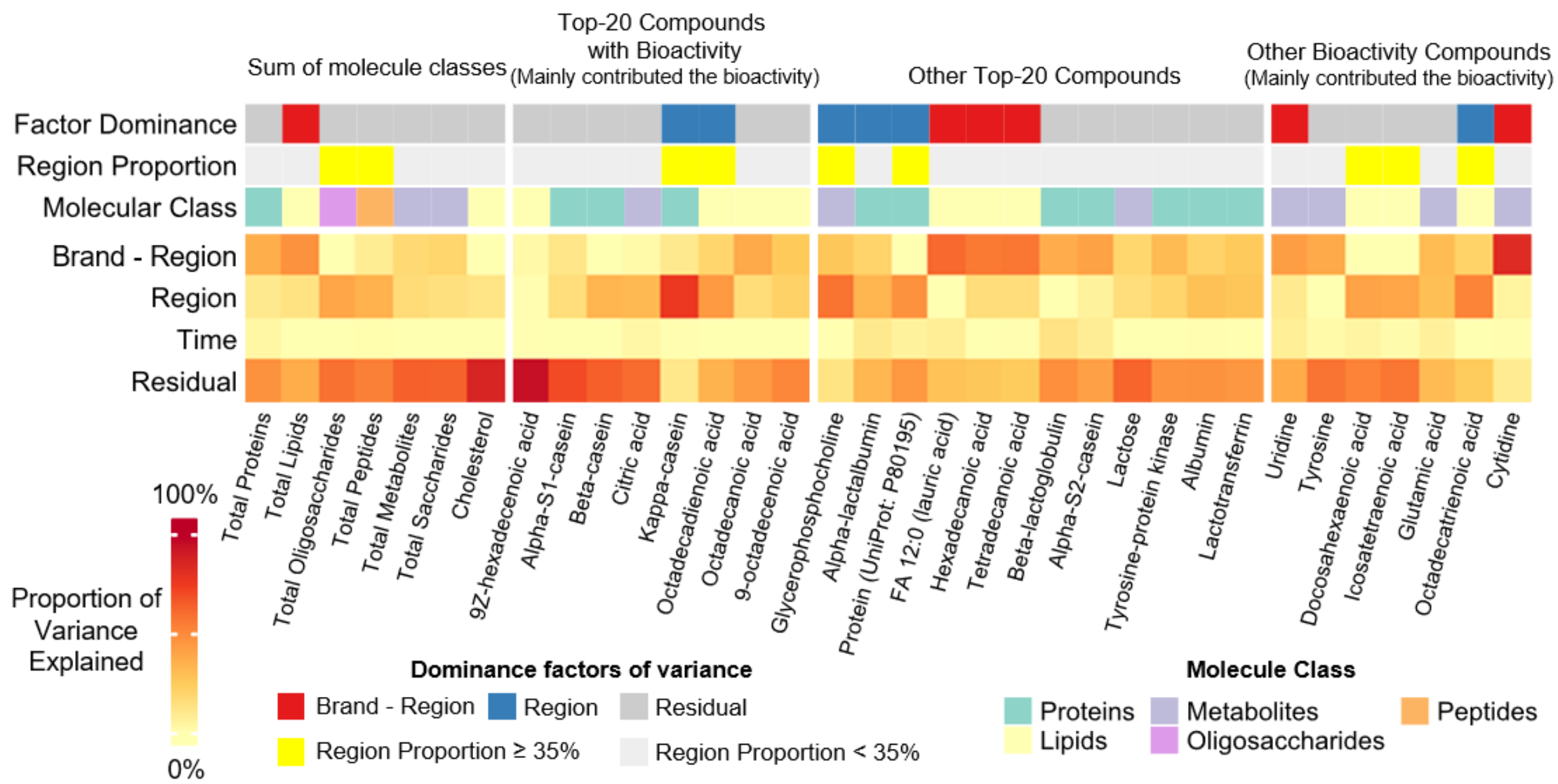

**Supplementary Figure 8. Variance decomposition analysis results of selected compound features.** The heatmap showed the proportions of variance explained by region, time, brand nested within region, and residual variation. The selected compound features include total molecular classes, the top 20 highest-concentration compounds, and bioactive compounds contributing the highest efficacy in at least one bioactivity.

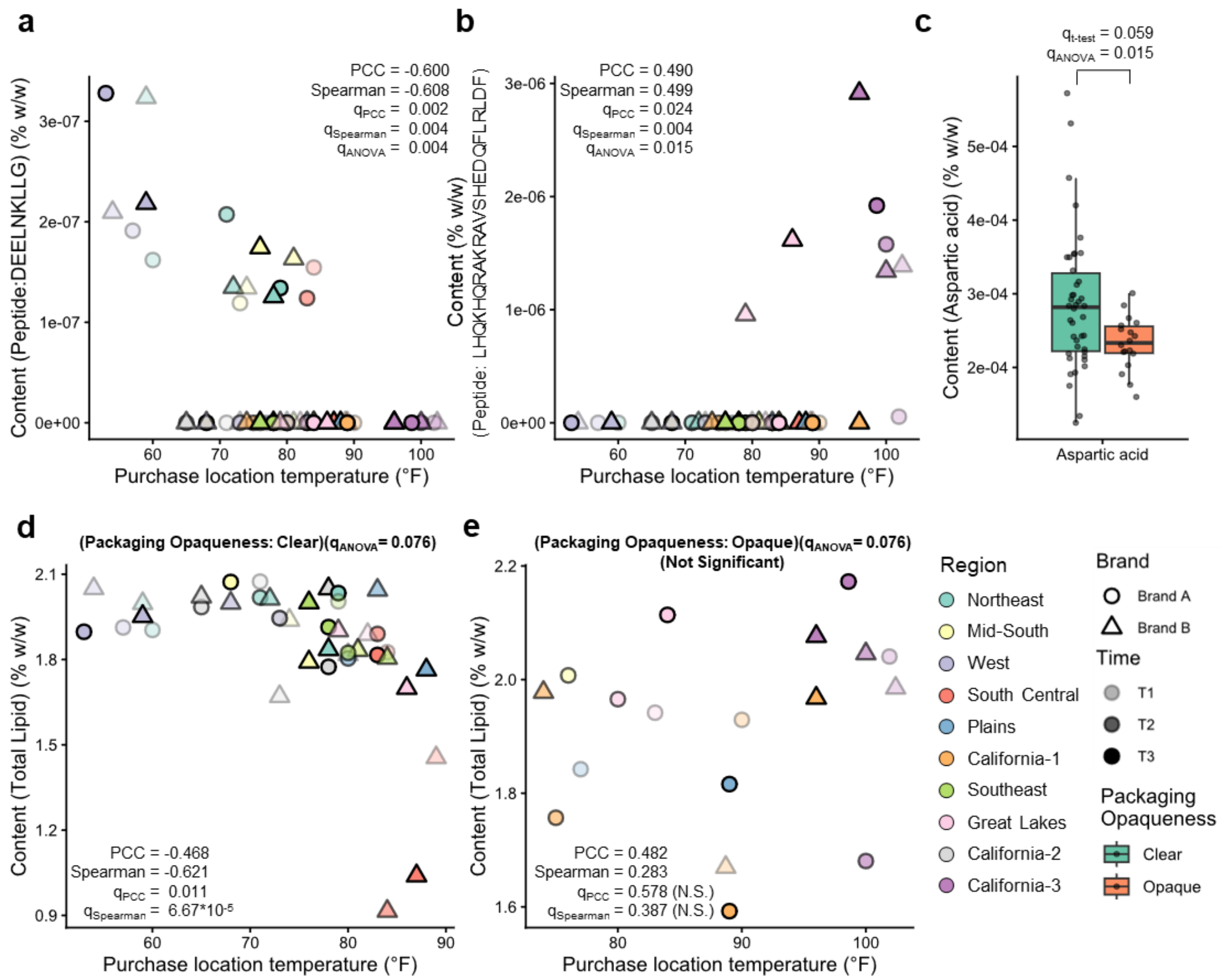

**Supplementary Figure 9. Additional factors affecting compound concentrations.** **a.** Negative correlation between the concentration of the peptide DEELNKLLG and the temperature of the purchase region, **b.** positive correlation between the peptide LHQKHQRAKRAVSHEDQFLRLDF and the temperature of the purchase region, **c.** Significantly higher aspartic acid concentration in the clear packaging compared to opaque packaging, **d.** Negative correlation between the concentration of the total lipid and the temperature of the purchase region for only the samples with the clear packaging. **e.** No significant correlation between the the concentration of the total lipid and the temperature of the purchase region for the samples with the opaque packaging.

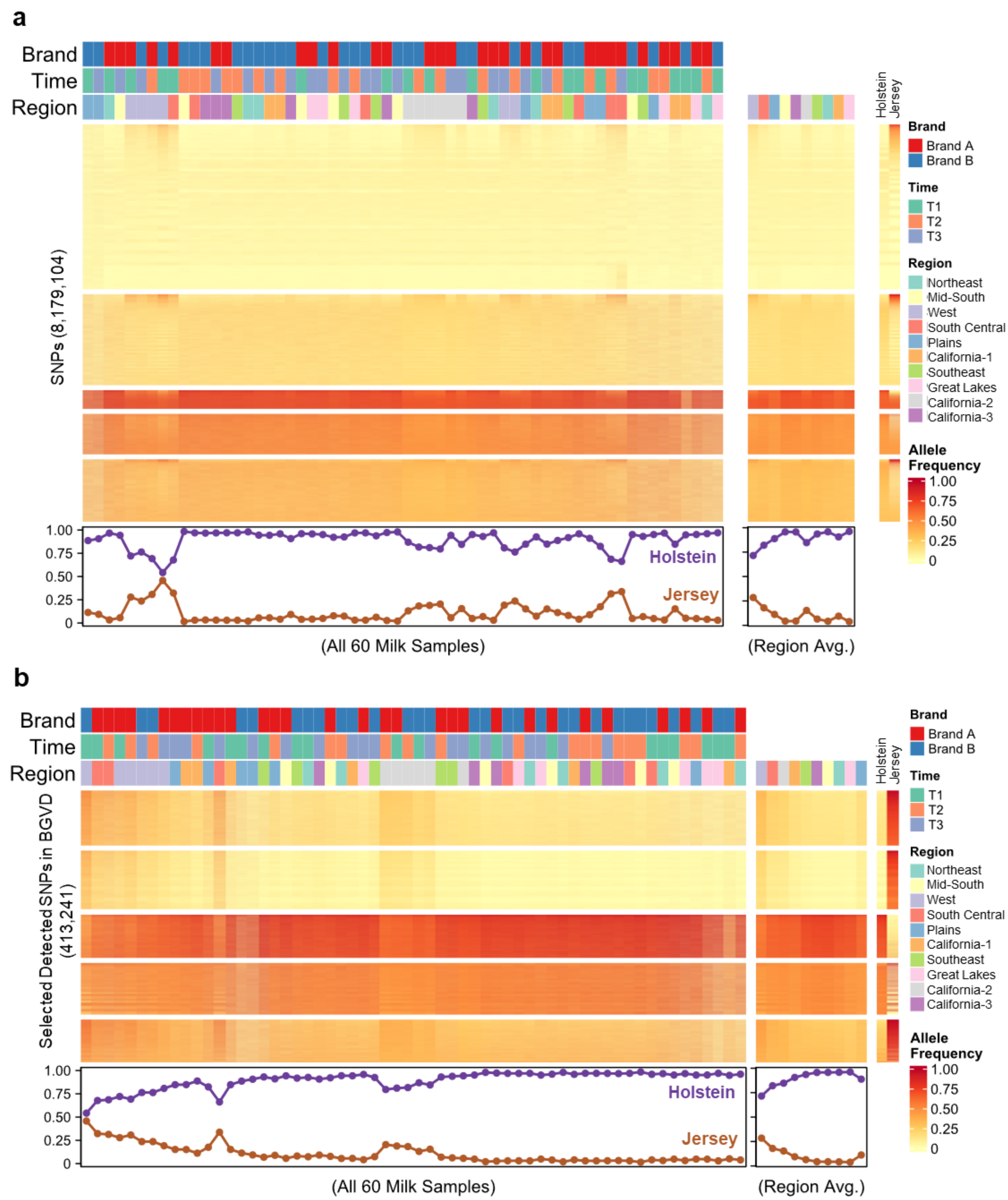

**Supplementary Figure 10. SNP data summary across milk samples.** Heatmap of **a.** All SNPs and **b.** the SNPs with the top 5% of the absolute reference allele frequency

difference in Holstein and Jersey breeds detected in the milk samples that are recorded in the BGVD database, shown alongside reference allele frequencies for Holstein and Jersey breeds.

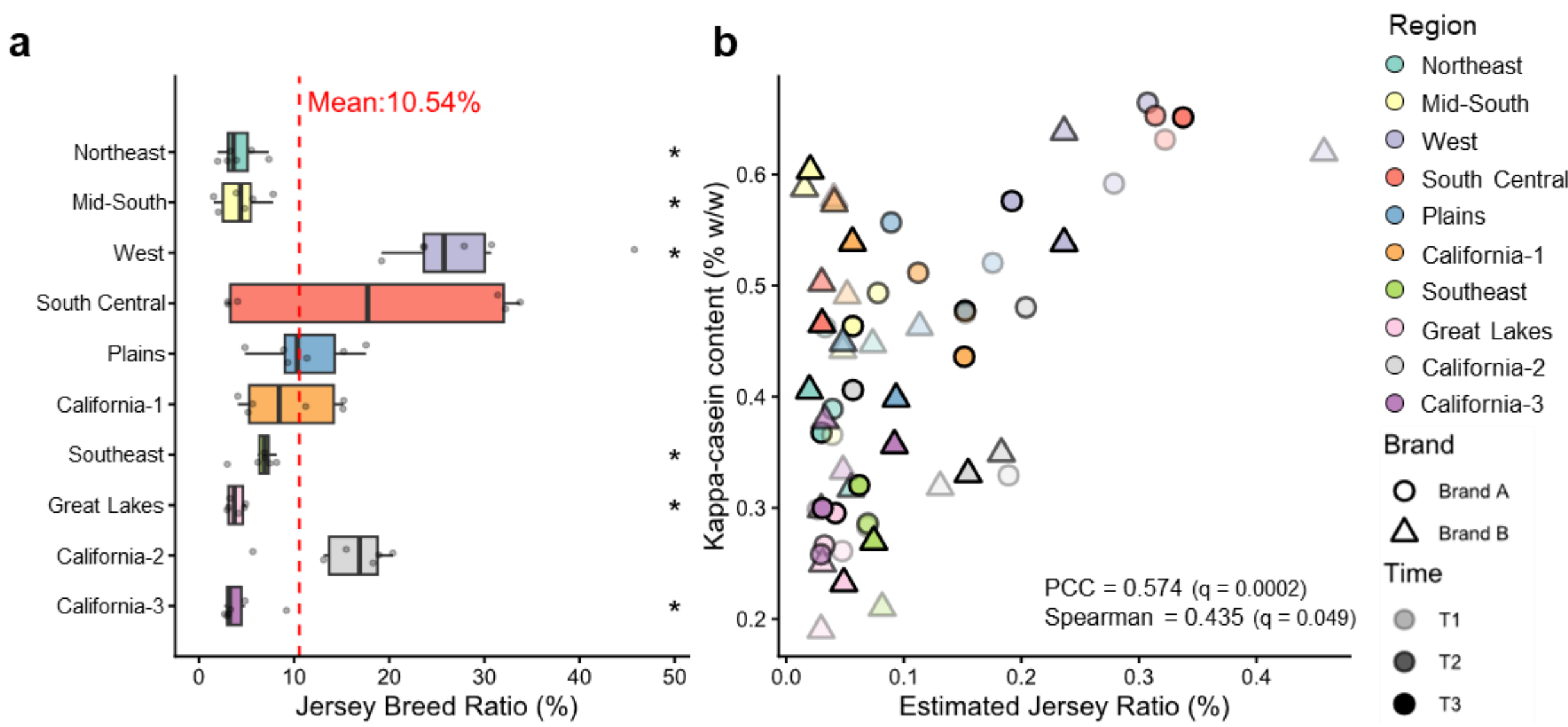

**Supplementary Figure 11. a.** Estimated Jersey breed proportion for milk samples from the ten regions and **b.** correlation between the estimated Jersey breed proportion and κ-casein concentration across milk samples.

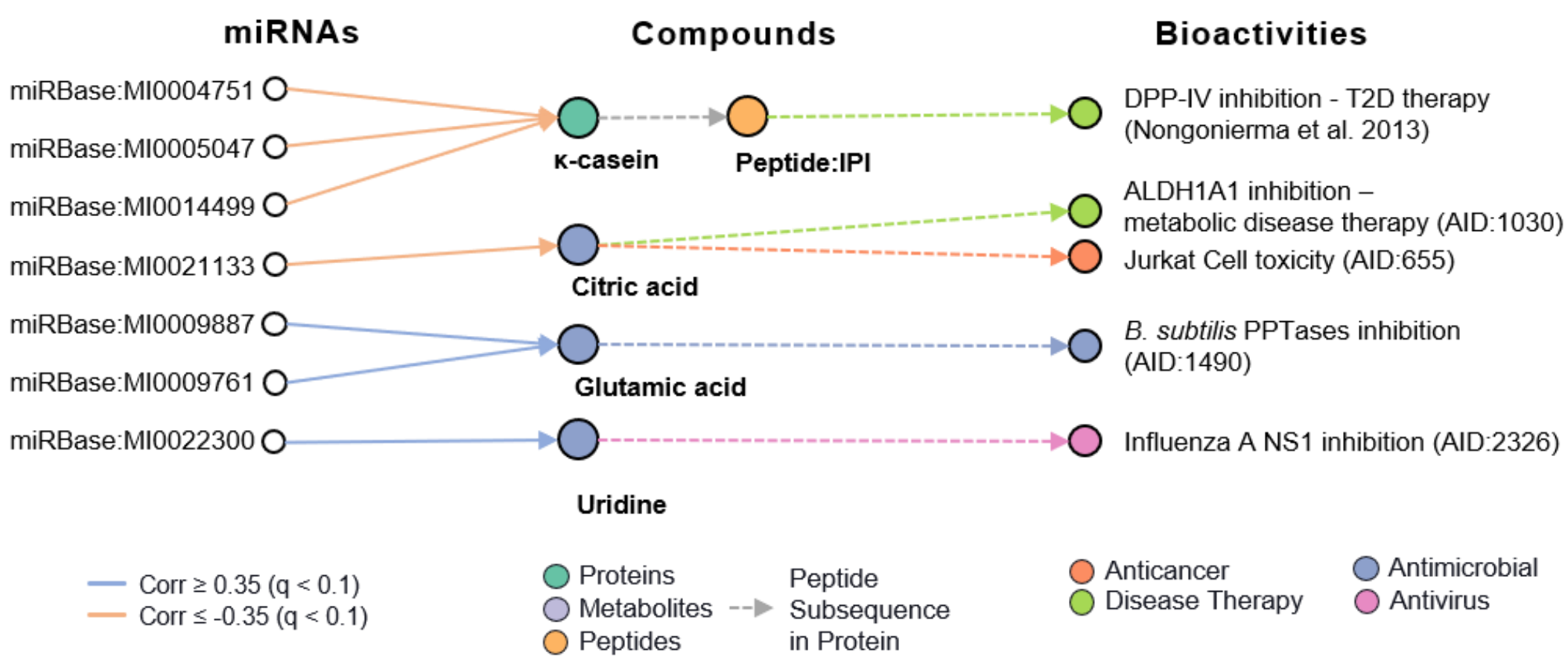

**Supplementary Figure 12. miRNAs associate with milk compound profiles and bioactivity efficacies.**

**a. Bioactivity Data Collection**

| Bioassay/Study | Compound | IC$_{50}$ (μM) | Hill Curve Parameters | | |
|---|---|---|---|---|---|
| | | | Slope | Min Efficacy | Max Efficacy |
| Anticancer (p53 Activation) AID: 651631 | Linoleic Acid SID: 144213397 | 68.60 | 4.954 | 0.04% | 90.06% |
| Antimicrobial (Malaria 7G8) AID: 504749 | Docosahexaenoic acid SID: 50110834 | 12.59 | 1.558 | -3.88% | -79.03% |
| Anticancer (L1210) AID: 406142 | Docosahexaenoic acid SID: 103445456 | 0.013 | NA | | |
| | … | | | | |
| ACE-inhibitory (Abubakar et al. 1998) | Peptide: IPA | 141 | NA | | |
| | … | | | | |

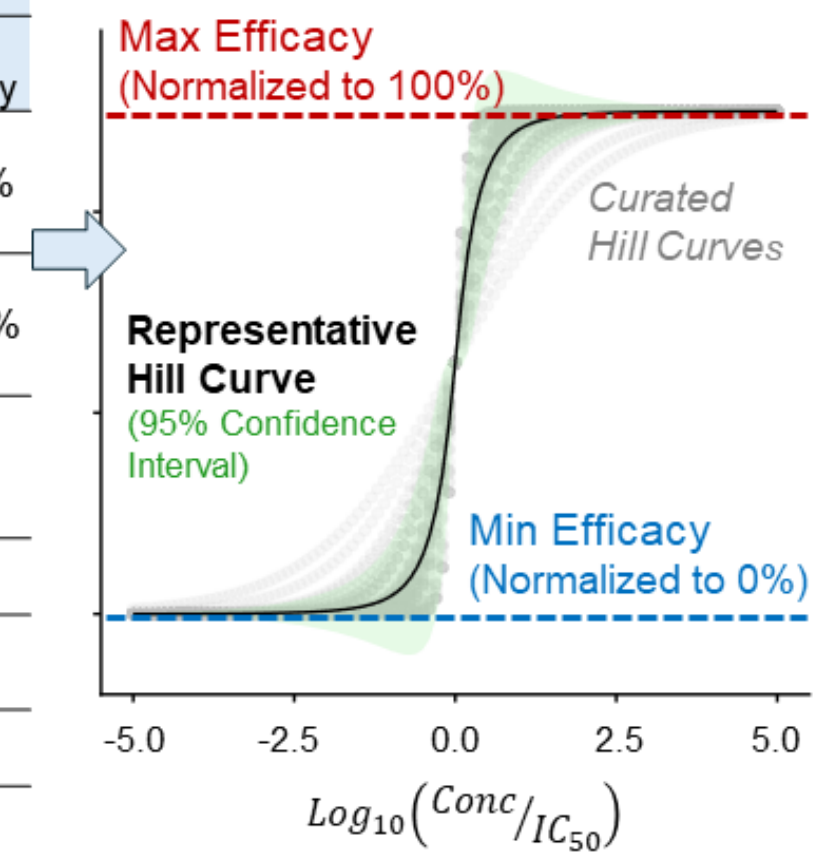

**Curated Bioactivity Entries** | **Representative Hill Curve**

**b. *In vitro* Bioactivity Efficacy Estimation**

| Bioassay AID/ Study | Compound | IC$_{50}$ (μM) | Milk Concentration (μM) |
|---|---|---|---|
| 651631 | Linoleic Acid | 68.60 | 2,776.9 |
| 504749 | Docosahexaenoic acid | 12.59 | 7.37 |
| 406142 | Docosahexaenoic acid | 0.013 | 7.37 |
| | … | | … |
| Abubakar et al. 1998 | Peptide: IPA | 141 | 445.5 |
| | … | | … |

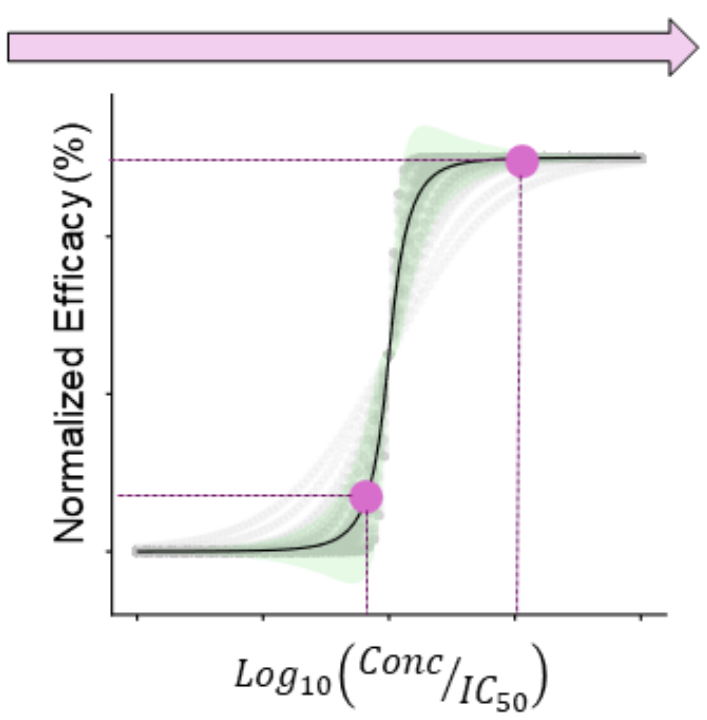

| Estimated *in vitro* Normalized Efficacy (%) |
|---|
| 100.0 |
| 11.45 |
| 100.0 |
| … |
| 88.31 |
| … |

**c. *In vivo* Bioactivity Efficacy Estimation**

| Milk Concentration (μM) |
|---|
| 2,776.9 |
| 7.37 |
| 7.37 |
| … |
| 445.5 |
| … |

*Apply Estimated Conversion Ratio*

**Peptides:** $6*10^{-6}$

**Small Molecules:** $1.47*10^{-4}$

| Estimated Postprandial Serum Concentration (μM) |
|---|
| 0.41 |
| 7.37 |
| 7.37 |
| … |
| 445.5 |
| … |

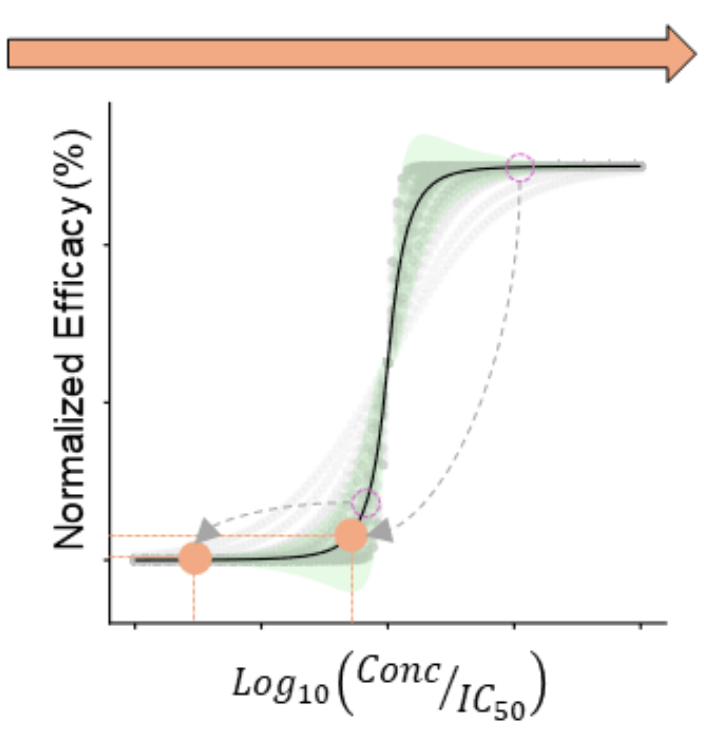

| Estimated *in vivo* Normalized Efficacy (%) |
|---|
| 100.0 |
| 11.45 |
| 100.0 |
| … |
| 88.31 |
| … |

**Supplementary Figure 13. Flowchart of bioactivity efficacy estimation in milk. a.**

Collection of bioactivity entries. Data points from assays with reported IC$_{50}$ values and Hill slopes were compiled (points), and a representative Hill curve was fitted (solid line). **b.** In vitro bioactivity efficacy was estimated using the representative Hill curve together with the measured compound concentration in milk. **c.** In vivo bioactivity efficacy was estimated using the predicted postprandial serum concentration derived from the milk concentration and the estimated milk-to-serum conversion ratio.

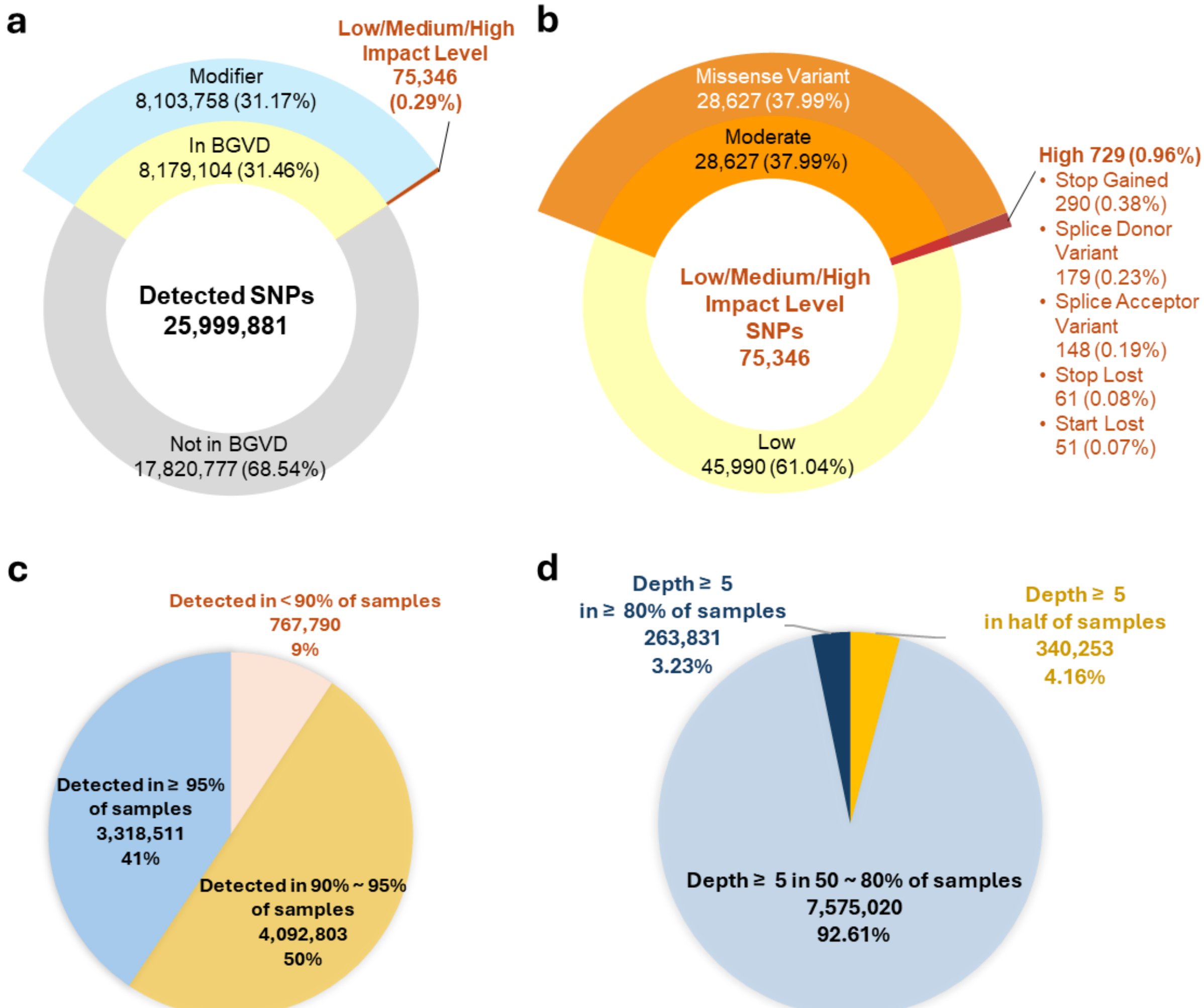

**Supplementary Figure 14. Summary of detected SNPs. a,** Total number of SNPs detected per sample and the subset annotated in the BGVD database. **b,** Numbers of detected SNPs and BGVD-annotated SNPs stratified by predicted impact (high, moderate, low) based on SnpEff consequence annotations. **c,** Number of SNPs detected across different proportions of samples. **d,** Number of SNPs with sequencing depth ≥ 5 across different proportions of samples.

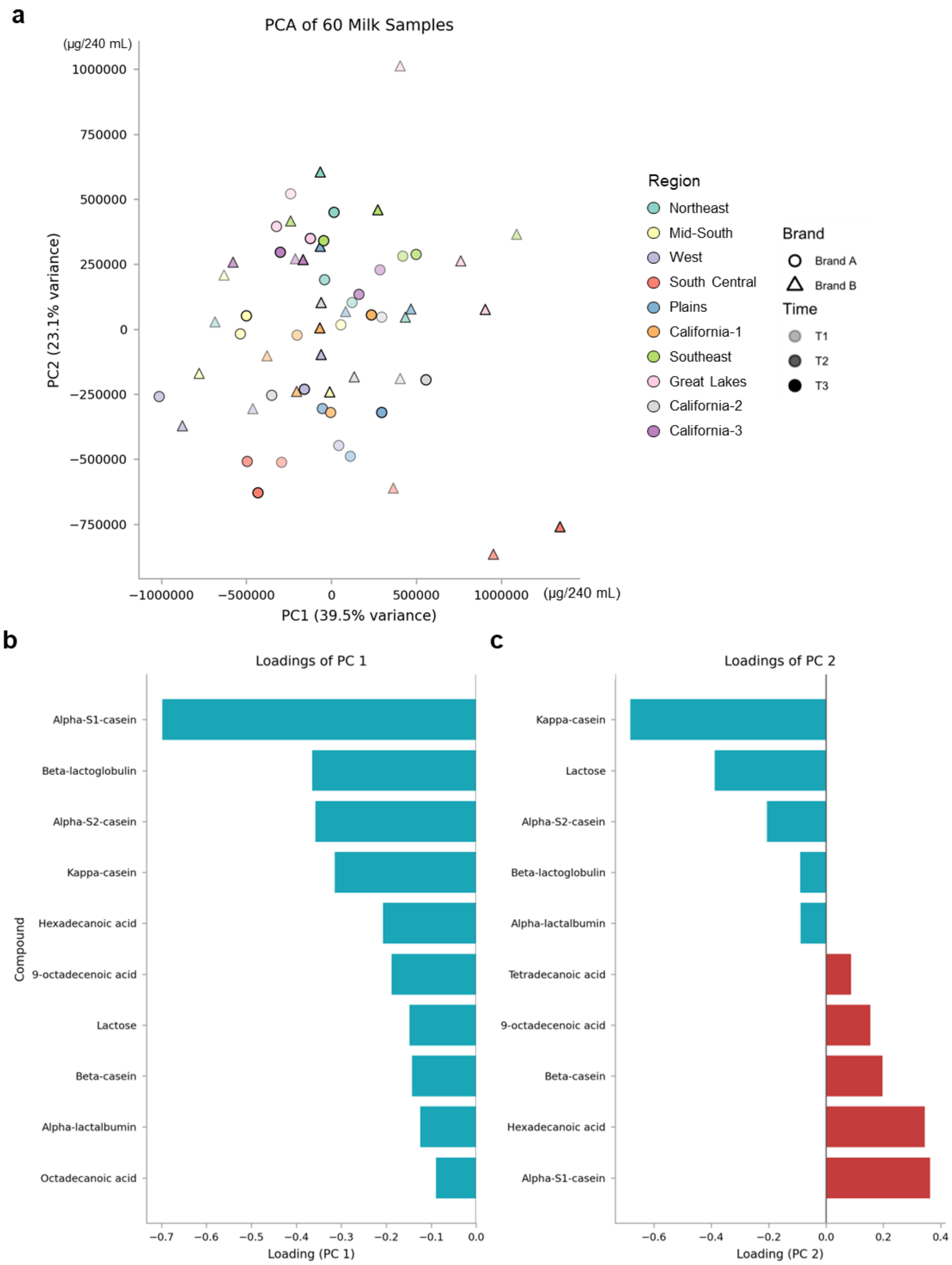

**Supplementary Figure 15. Principal component analysis (PCA) of milk samples based on compound profiles and corresponding loadings. a,** PCA score plot for the first two principal components (PC1 and PC2), showing all 60 milk samples and colored by region (ten regions). **b,** Loading plot for PC1. **c,** Loading plot for PC2.

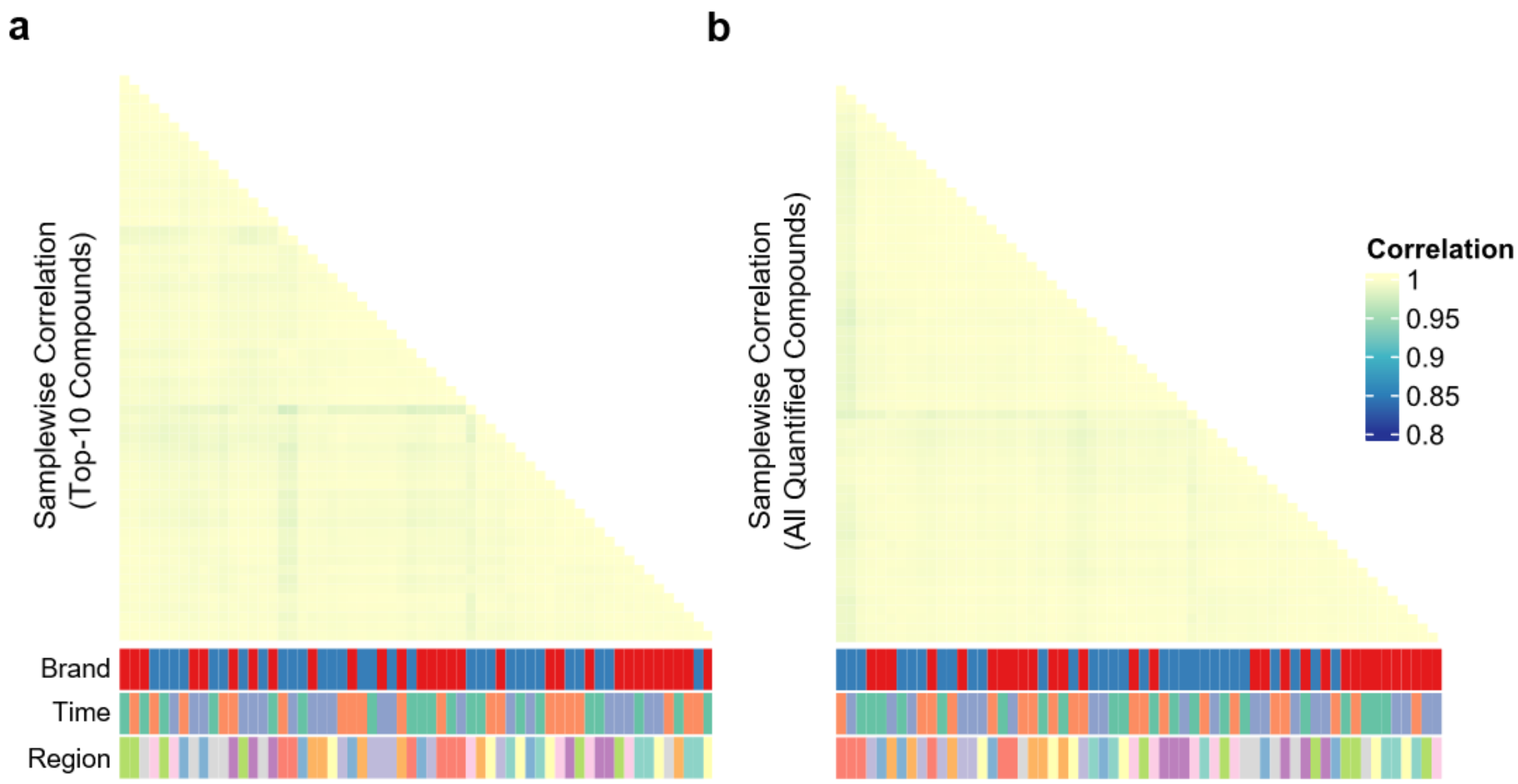

**Supplementary Figure 16. Sample-wise correlations based on a.** the top 10 highest-concentration compounds and **b.** all compounds.

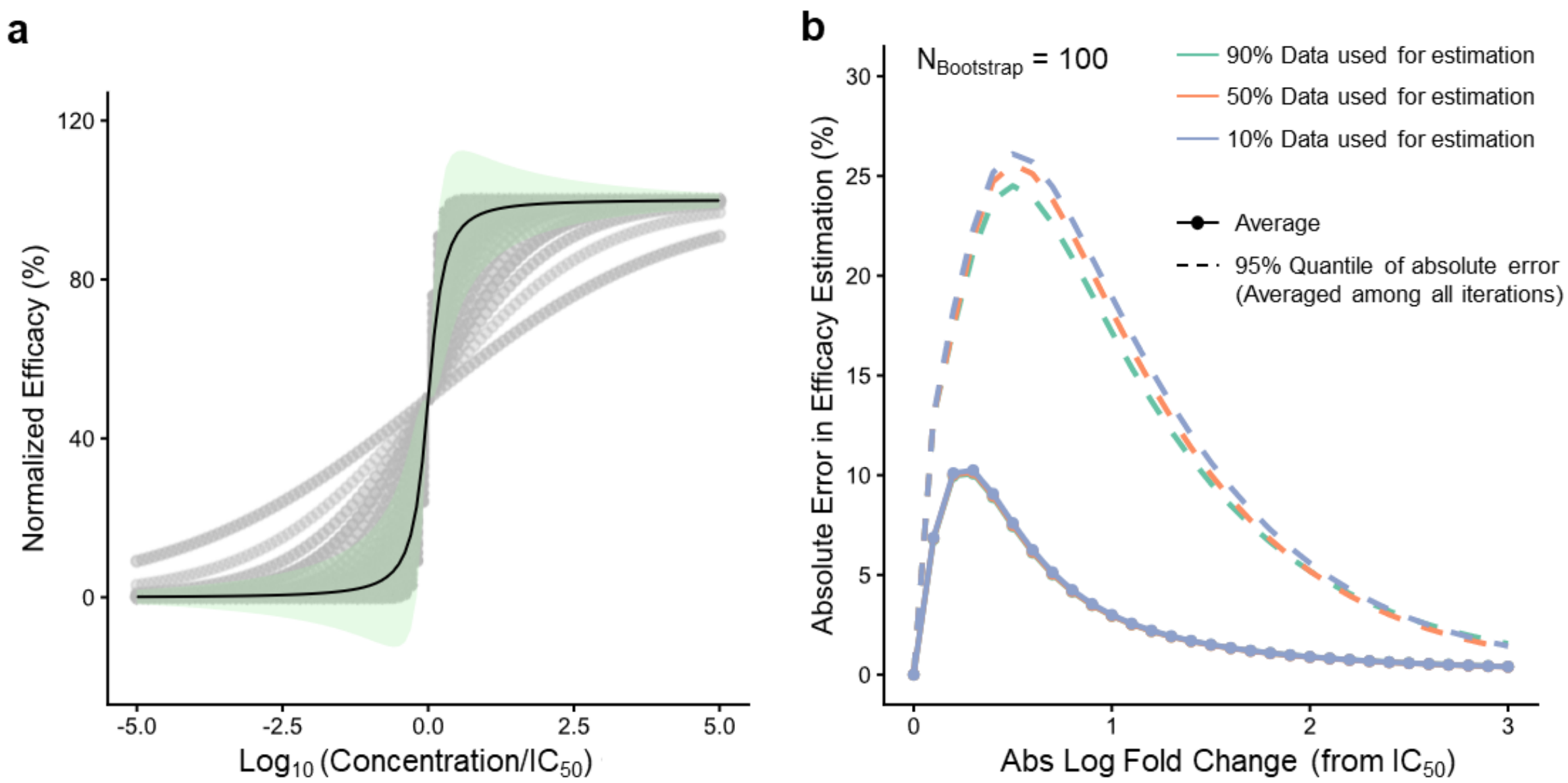

**Supplementary Figure 17. Curated Hill curves and absolute error of efficacy estimation. a,** The 114 curated Hill curves with 95% confidence intervals. The x-axis shows $\log_{10}$ of the ratio of compound concentration to the $IC_{50}$ for the assay, and the y-axis shows normalized efficacy (0–100%). Curves were normalized such that 0% indicates no effect and 100% indicates the maximal effect for that assay. **b,** Absolute error of efficacy estimation assessed by bootstrapping (100 iterations). In each iteration, 10%, 50%, or 90% of the curated Hill curves were randomly sampled to fit the representative curve, and the remaining curves were used for evaluation. The mean absolute error and the 95th percentile of absolute error (averaged across iterations) are shown.

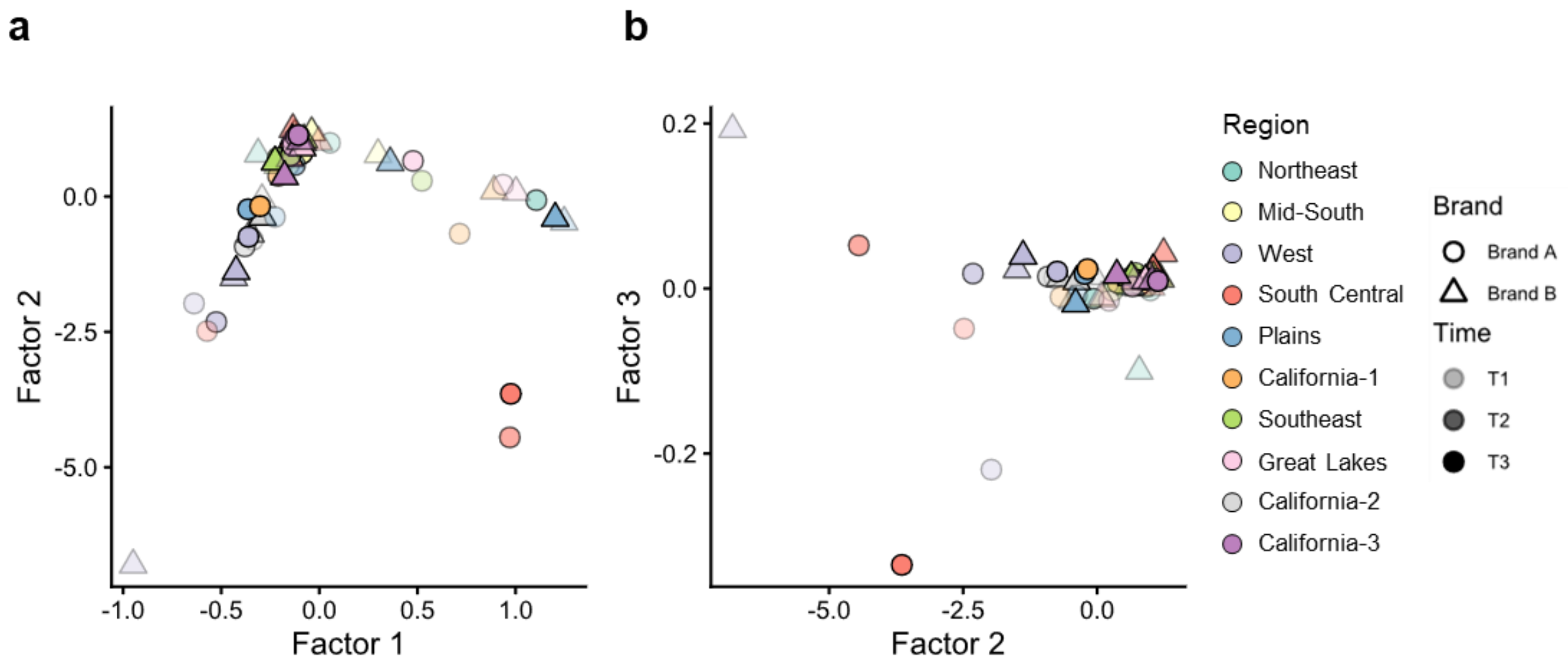

**Supplementary Figure 18. The factors of the integrated omics layers extracted by MOFA analysis. a.** The scatter plot of the factor 1 and the factor 2, and **b.** The scatter plot of the factor 2 and the factor 3.

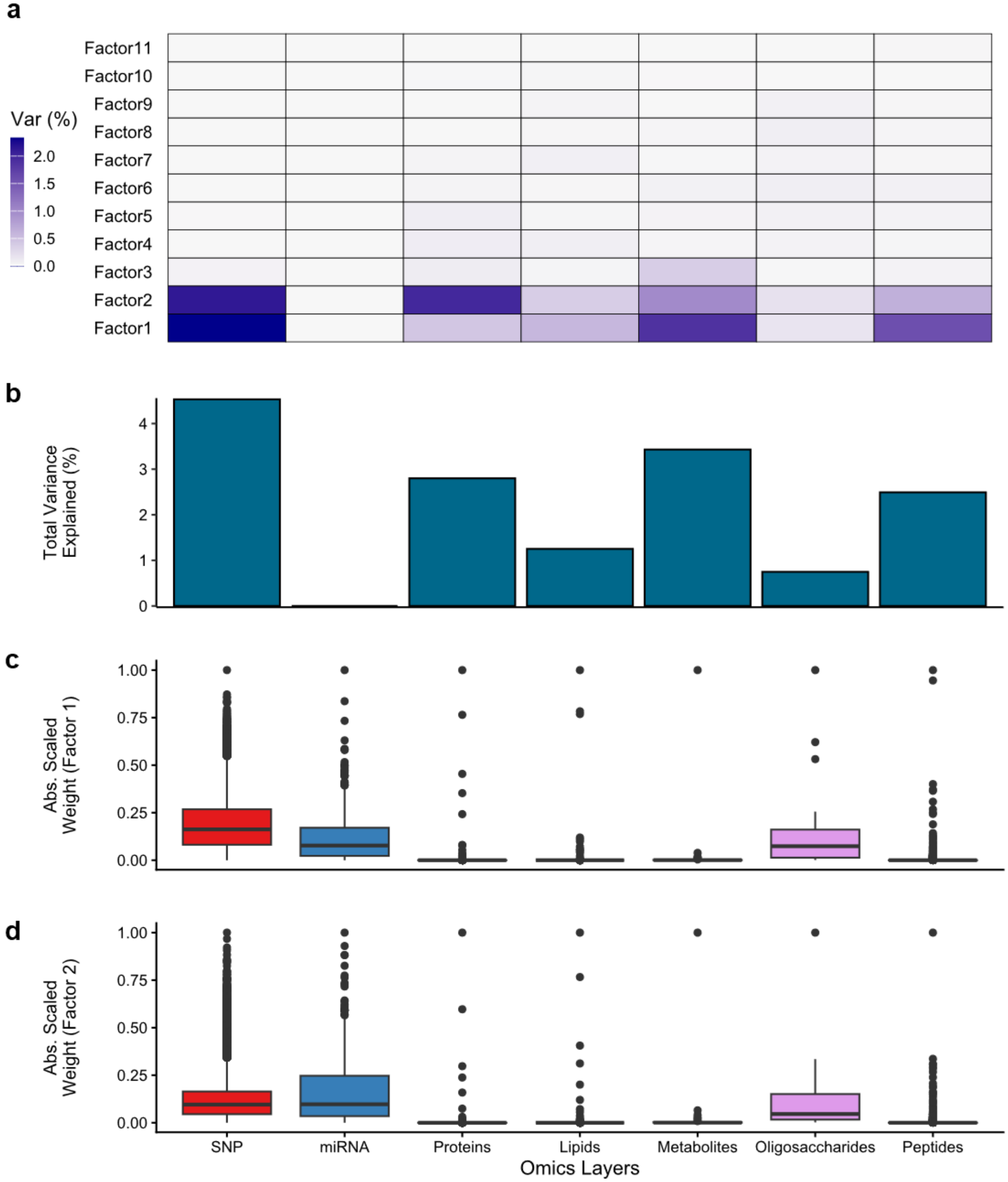

**Supplementary Figure 19. The summary of the MOFA analysis. a.** The explained variance for each factor in different omics layers, **b.** The sum explained variance for all factors in different omics layers, **c.** The absolute scaled weights of Factor 1 (linearly

scaled by the maximum absolute weight), and **d.** The absolute scaled weights of Factor 2 (linearly scaled by the maximum absolute weight).

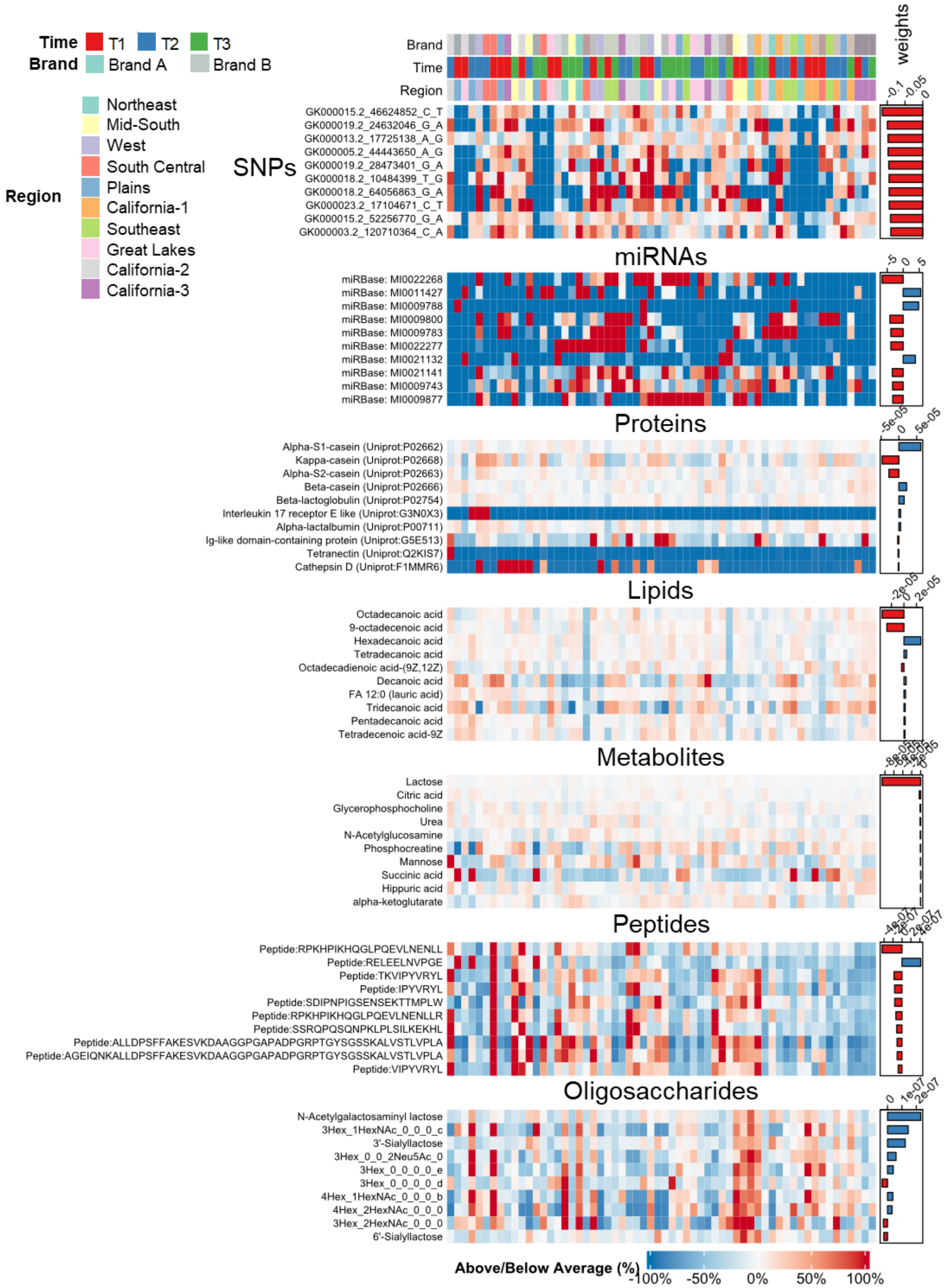

**Supplementary Figure 20. The omics profile subset with the top 10 absolute weights of Factor 1 in each omics layer.**

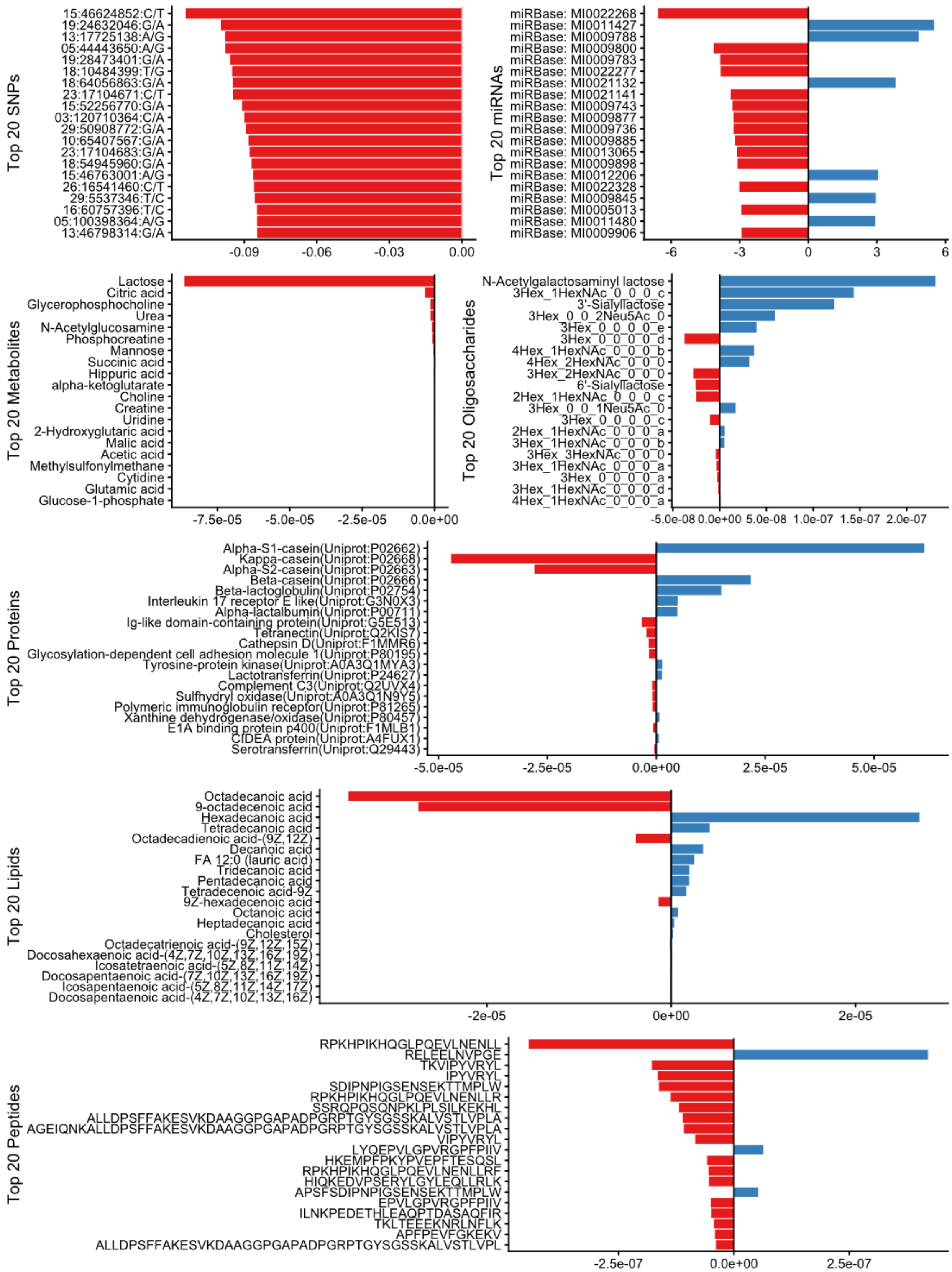
Top 20 SNPs
15:46624852:C/T
19:24632046:G/A
13:17725138:A/G
05:44443650:A/G
19:28473401:G/A
18:10484399:T/G
18:64056863:G/A
23:17104671:C/T
15:52256770:G/A
03:120710364:C/A
29:50908772:G/A
10:65407567:G/A
23:17104683:G/A
18:54945960:G/A
15:46763001:A/G
26:16541460:C/T
29:5537346:T/C
16:60757396:T/C
05:100398364:A/G
13:46798314:G/A
-0.09
-0.06
-0.03
0.00
Top 20 miRNAs
miRBase: MI0022268
miRBase: MI0011427
miRBase: MI0009788
miRBase: MI0009800
miRBase: MI0009783
miRBase: MI0022277
miRBase: MI0021132
miRBase: MI0021141
miRBase: MI0009743
miRBase: MI0009877
miRBase: MI0009736
miRBase: MI0009885
miRBase: MI0013065
miRBase: MI0009898
miRBase: MI0012206
miRBase: MI0022328
miRBase: MI0009845
miRBase: MI0005013
miRBase: MI0011480
miRBase: MI0009906
-6
-3
0
3
6
Top 20 Metabolites
Lactose
Citric acid
Glycerophosphocholine
Urea
N-Acetylglucosamine
Phosphocreatine
Mannose
Succinic acid
Hippuric acid
alpha-ketoglutarate
Choline
Creatine
Uridine
2-Hydroxyglutaric acid
Malic acid
Acetic acid
Methylsulfonylmethane
Cytidine
Glutamic acid
Glucose-1-phosphate
-7.5e-05
-5.0e-05
-2.5e-05
0.0e+00
Top 20 Oligosaccharides
N-Acetylgalactosaminyl lactose
3Hex_1HexNAc_0_0_c
3'-Sialyllactose
3Hex_0_0_2Neu5Ac_0
3Hex_0_0_0_e
3Hex_0_0_0_d
4Hex_1HexNAc_0_0_b
4Hex_2HexNAc_0_0_0
3Hex_2HexNAc_0_0_0
6'-Sialyllactose
2Hex_1HexNAc_0_0_c
3Hex_0_0_1Neu5Ac_0
3Hex_0_0_0_c
2Hex_1HexNAc_0_0_a
3Hex_1HexNAc_0_0_b
3Hex_3HexNAc_0_0_0
3Hex_1HexNAc_0_0_a
3Hex_0_0_0_a
3Hex_1HexNAc_0_0_d
4Hex_1HexNAc_0_0_a
-5.0e-08
0.0e+00
5.0e-08
1.0e-07
1.5e-07
2.0e-07
Top 20 Proteins
Alpha-S1-casein(Uniprot:P02662)
Kappa-casein(Uniprot:P02668)
Alpha-S2-casein(Uniprot:P02663)
Beta-casein(Uniprot:P02666)
Beta-lactoglobulin(Uniprot:P02754)
Interleukin 17 receptor E like(Uniprot:G3N0X3)
Alpha-lactalbumin(Uniprot:P00711)
Ig-like domain-containing protein(Uniprot:G5E513)
Tetranectin(Uniprot:Q2KIS7)
Cathepsin D(Uniprot:F1MMR6)
Glycosylation-dependent cell adhesion molecule 1(Uniprot:P80195)
Tyrosine-protein kinase(Uniprot:A0A3Q1MYA3)
Lactotransferrin(Uniprot:P24627)
Complement C3(Uniprot:Q2UVX4)
Sulfhydryl oxidase(Uniprot:A0A3Q1N9Y5)
Polymeric immunoglobulin receptor(Uniprot:P81265)
Xanthine dehydrogenase/oxidase(Uniprot:P80457)
E1A binding protein p400(Uniprot:F1MLB1)
CIDEA protein(Uniprot:A4FUX1)
Serotransferrin(Uniprot:Q29443)
-5.0e-05
-2.5e-05
0.0e+00
2.5e-05
5.0e-05
Top 20 Lipids
Octadecanoic acid
9-octadecenoic acid
Hexadecanoic acid
Tetradecanoic acid
Octadecadienoic acid-(9Z,12Z)
Decanoic acid
FA 12:0 (lauric acid)
Tridecanoic acid
Pentadecanoic acid
Tetradecenoic acid-9Z
9Z-hexadecenoic acid
Octanoic acid
Heptadecanoic acid
Cholesterol
Octadecatrienoic acid-(9Z,12Z,15Z)
Docosahexaenoic acid-(4Z,7Z,10Z,13Z,16Z,19Z)
Icosatetraenoic acid-(5Z,8Z,11Z,14Z)
Docosapentaenoic acid-(7Z,10Z,13Z,16Z,19Z)
Icosapentaenoic acid-(5Z,8Z,11Z,14Z,17Z)
Docosapentaenoic acid-(4Z,7Z,10Z,13Z,16Z)
-2e-05
0e+00
2e-05
Top 20 Peptides
RPKHPIKHQGLPQEVLNENLL
RELEELNVPGE
TKVIPYVRYL
IPYVRYL
SDIPNPIGSENSEKTTMPLW
RPKHPIKHQGLPQEVLNENLLR
SSRQPQSQNPKLPLSILKEKHL
ALLDPSFFAKESVKDAAGGPGAPADPGRPTGYSGSSKALVSTLVPLA
AGEIQNKALLDPSFFAKESVKDAAGGPGAPADPGRPTGYSGSSKALVSTLVPLA
VIPYVRYL
LYQEPVLGPVRGPFPIIV
HKEMPFPKYPVEPFTESQSL
RPKHPIKHQGLPQEVLNENLLRF
HIQKEDVPSERYLGYLEQLLRLK
APSFSDIPNPIGSENSEKTTMPLW
EPVLGPVRGPFPIIV
ILNKPEDETHLEAQPTDASAQFIR
TKLTEEEKNRLNFLK
APFPEVFGKEKV
ALLDPSFFAKESVKDAAGGPGAPADPGRPTGYSGSSKALVSTLVPL
-2.5e-07
0.0e+00
2.5e-07

**Supplementary Figure 21. The top 20 features with the highest absolute weights of Factor 1 in each omics layer.**

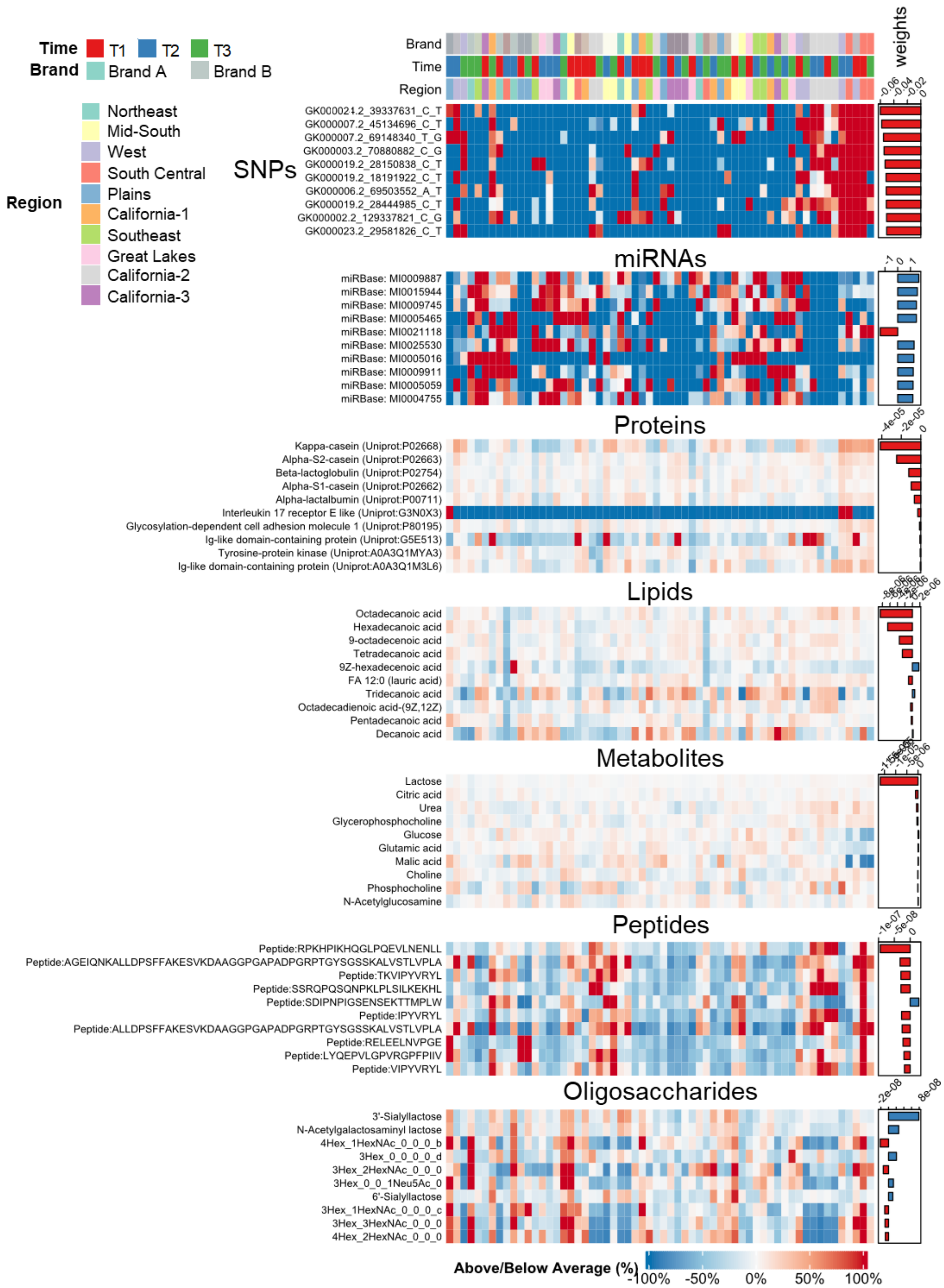

**Supplementary Figure 22. The omics profile subset with the top 10 absolute**

**weights of Factor 2 in each omics layer.**

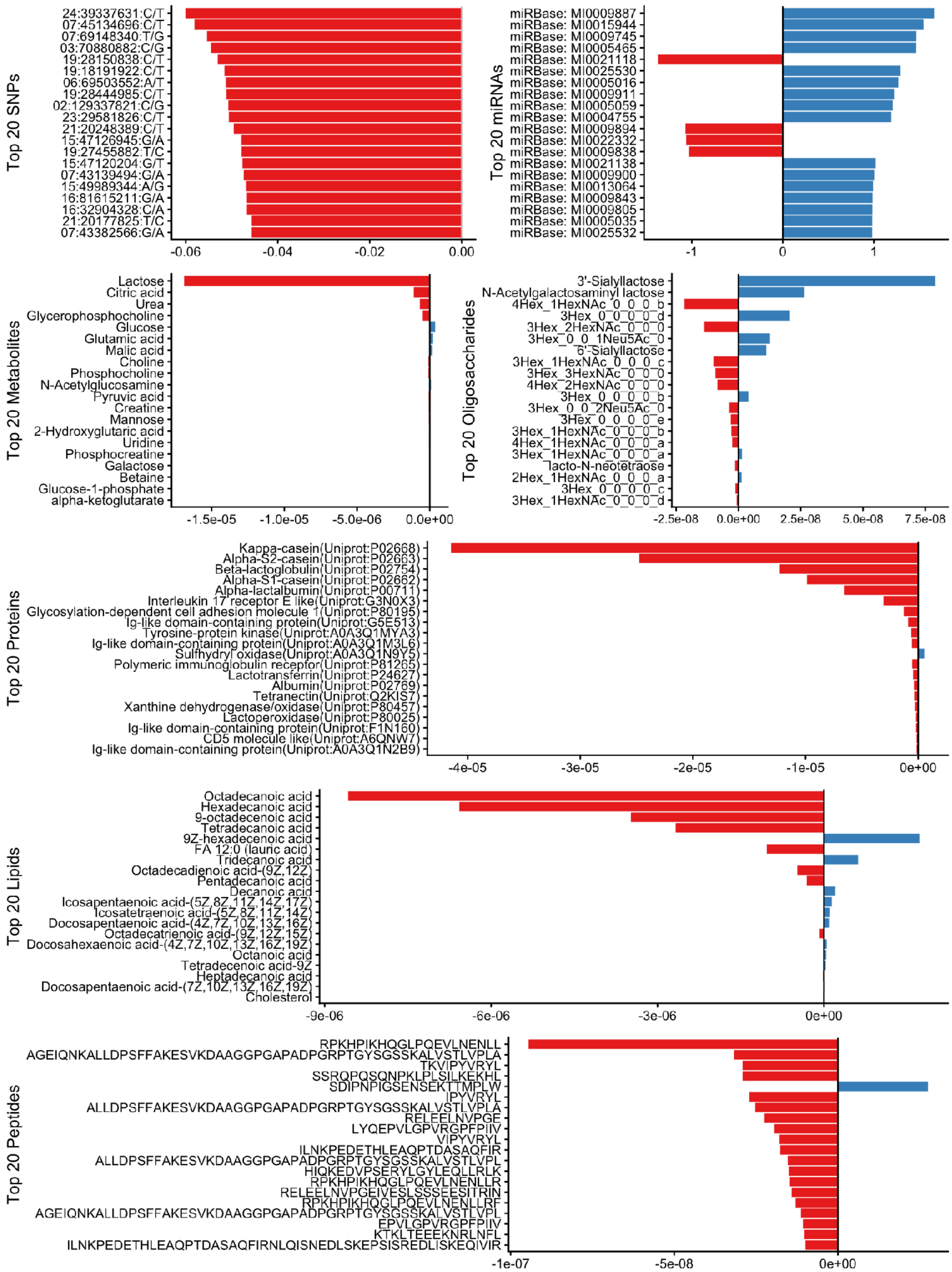

**Supplementary Figure 23. The top 20 features with the highest absolute weights of Factor 2 in each omics layer.**

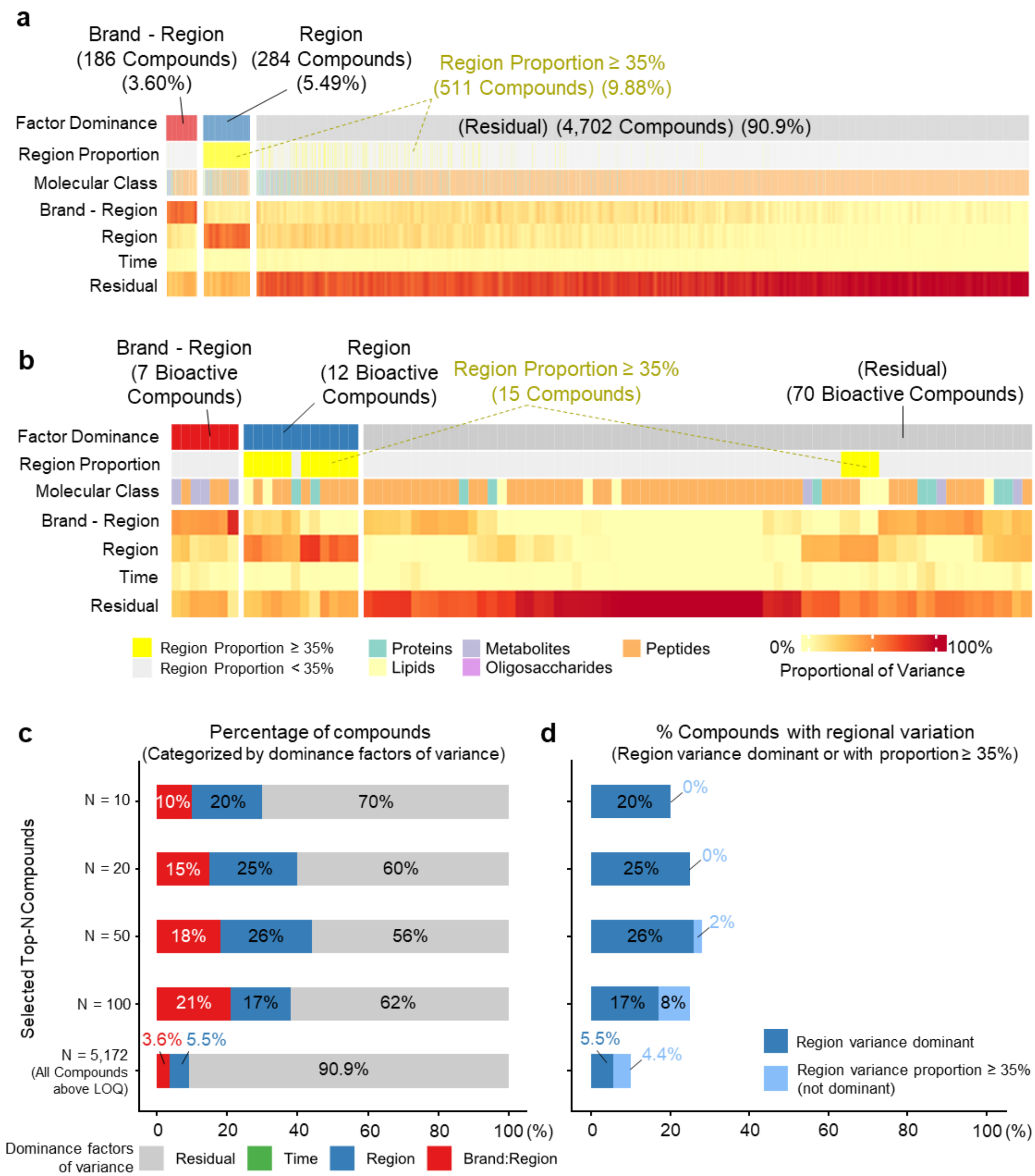

**Supplementary Figure 24. The variance decomposition analysis results. a.** The explained variance proportion contributed by region, time, brand (different brands for each region), and residual for each compound, **b.** The explained variance proportion contributed by region, time, brand (different brands for each region), and residual for each bioactive compound, **c.** The percentage of the compounds with different dominance factors of variance, and **d.** The percentage of the compounds with regional variation (Region mainly contributed to the variance or contributed to the variance with proportion no less than 35%).

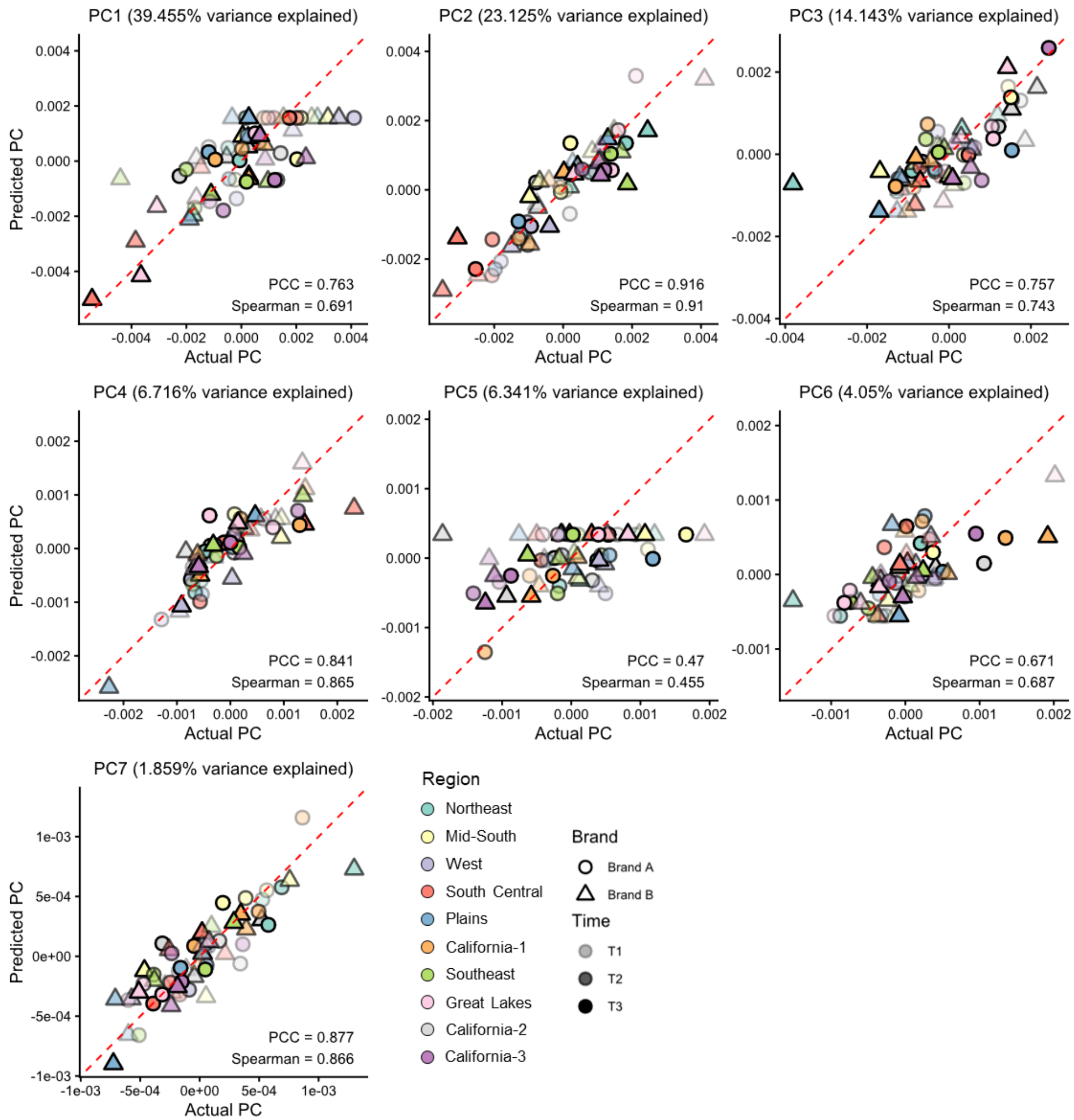

**Supplementary Figure 25. Observed versus predicted scores for the first 7 compound principal components using the selected 61 SNPs (full dataset is used for training the linear model).**

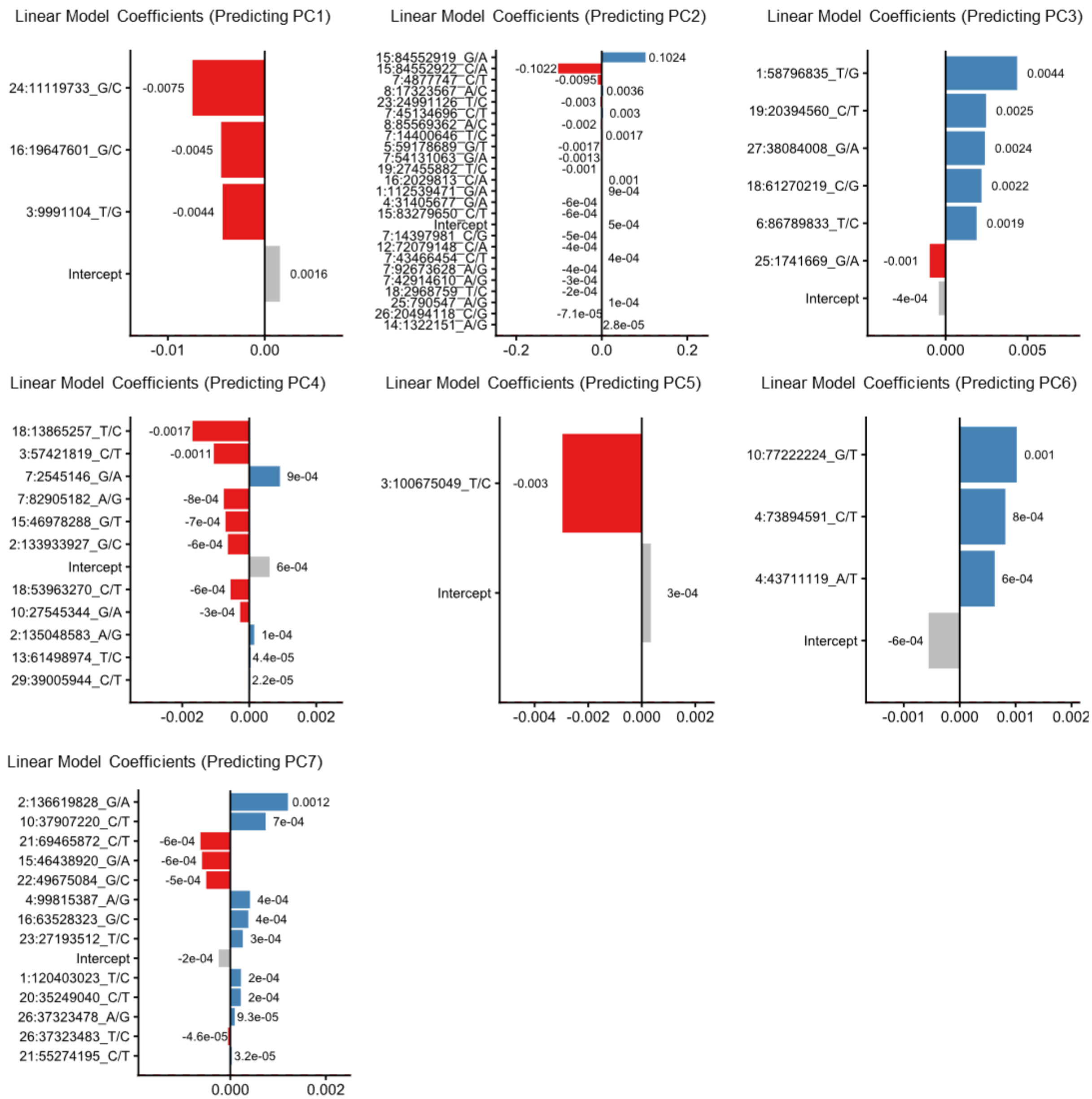

**Supplementary Figure 26. Coefficients of the linear models used to predict the first 7 compound principal components from SNP allele frequencies.** Models are shown for the final SNP set (61 SNPs selected).

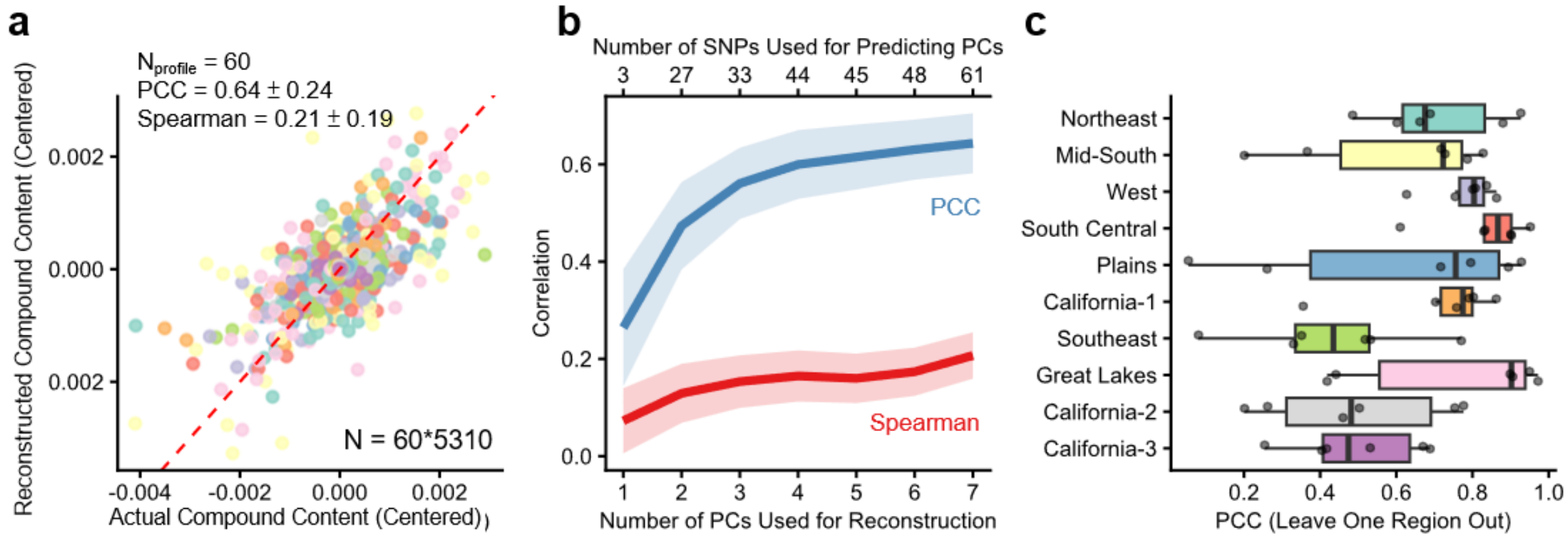

**Supplementary Figure 27. SNPs determine milk compound profiles. a.** Observed and predicted centered compound contents after subtracting the average concentration for each compound. Predictions were based on 61 selected SNPs and reconstructed from the predicted first seven principal components of the compound profiles using leave-one-region-out cross-validation. **b.** Prediction performance in leave-one-region-out cross-validation, reported as Pearson and Spearman correlations between observed and predicted compound profiles. **c.** Region-specific cross-validation performance.

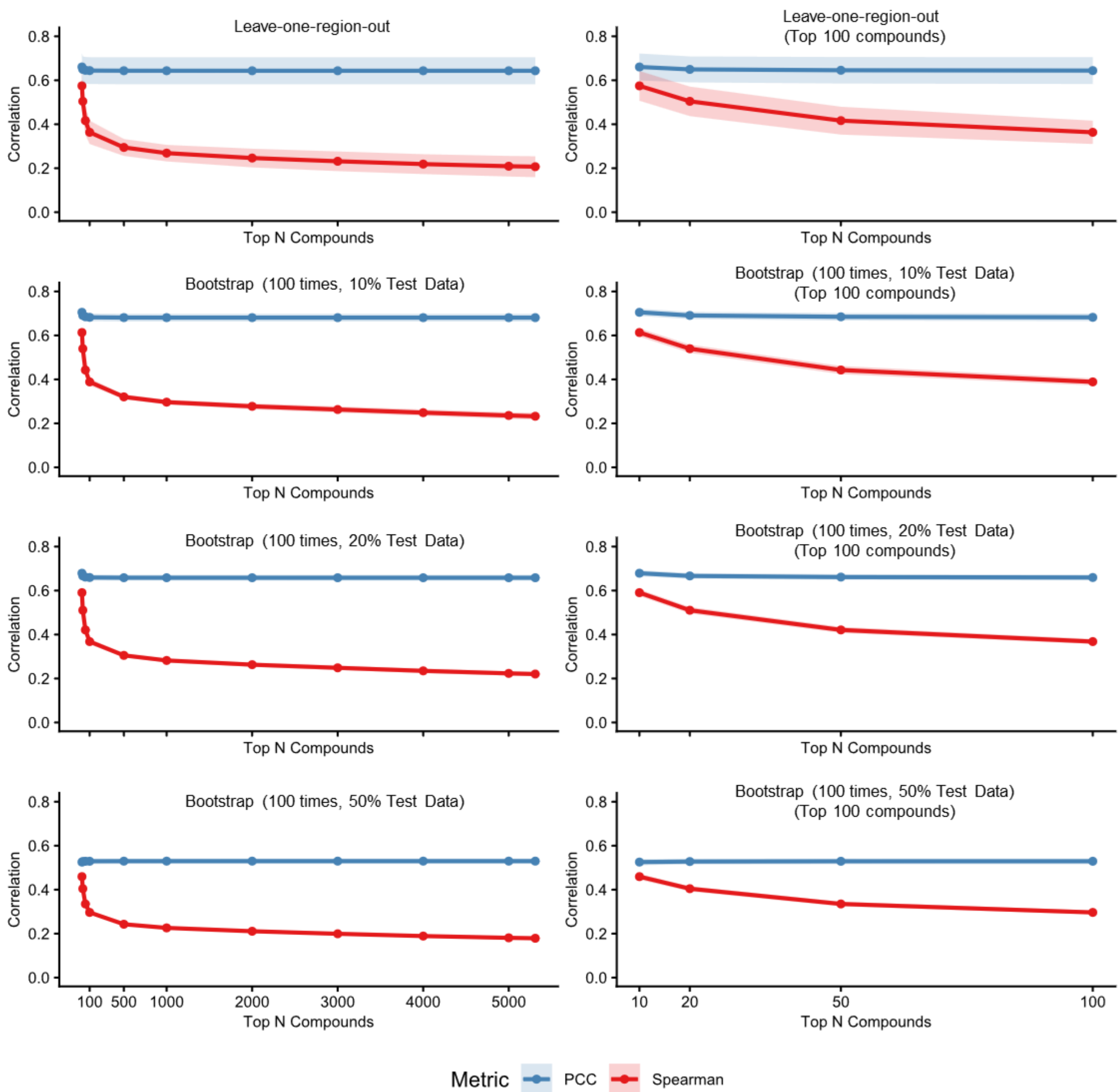

**Supplementary Figure 28. Reconstruction performance of the compound profile subset with top N highest concentration compounds based on seven predicted PCs.**

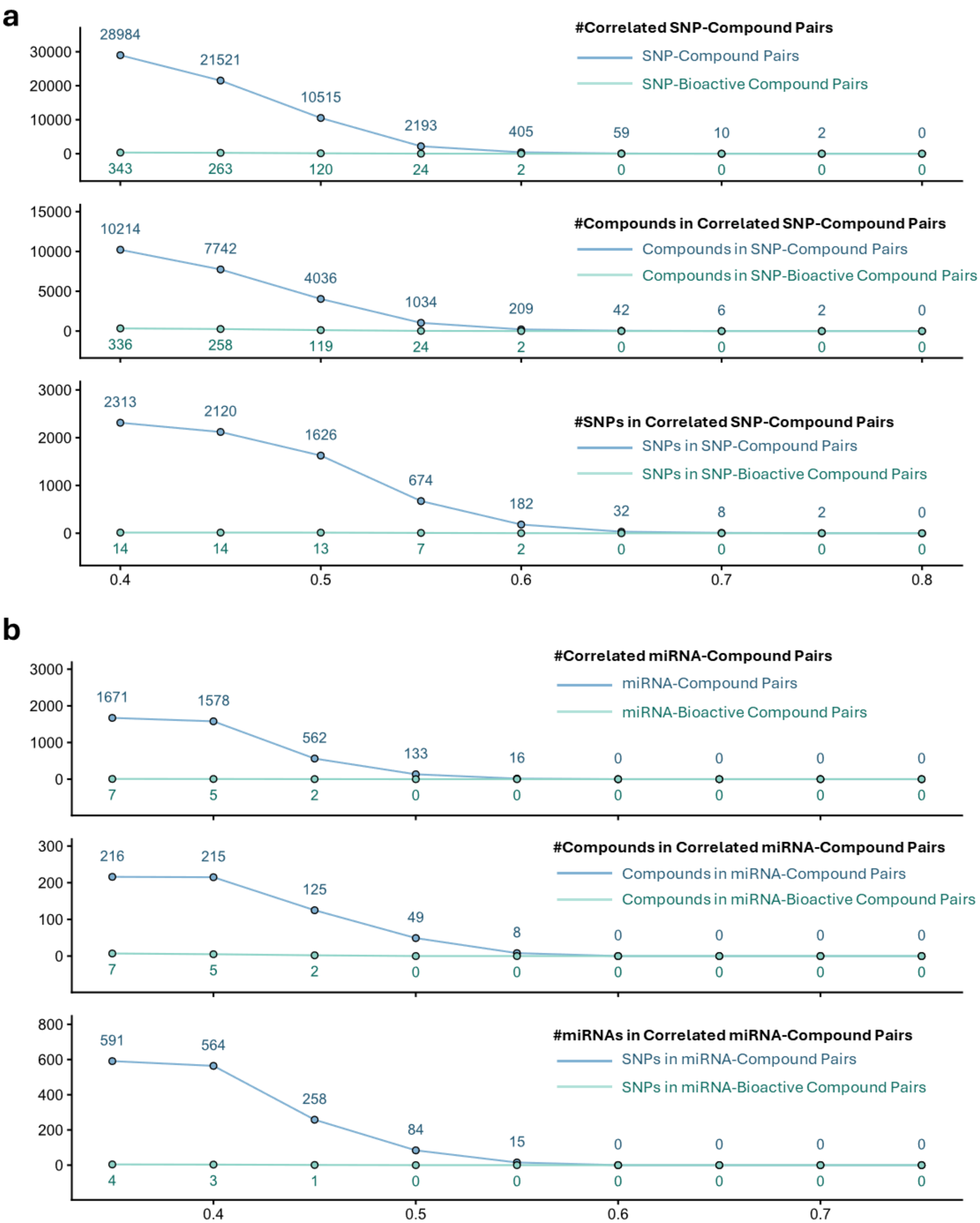
a
#Correlated SNP-Compound Pairs
SNP-Compound Pairs
SNP-Bioactive Compound Pairs
28984 21521 10515 2193 405 59 10 2 0
343 263 120 24 2 0 0 0 0
#Compounds in Correlated SNP-Compound Pairs
Compounds in SNP-Compound Pairs
Compounds in SNP-Bioactive Compound Pairs
10214 7742 4036 1034 209 42 6 2 0
336 258 119 24 2 0 0 0 0
#SNPs in Correlated SNP-Compound Pairs
SNPs in SNP-Compound Pairs
SNPs in SNP-Bioactive Compound Pairs
2313 2120 1626 674 182 32 8 2 0
14 14 13 7 2 0 0 0 0
0.4 0.5 0.6 0.7 0.8
b
#Correlated miRNA-Compound Pairs
miRNA-Compound Pairs
miRNA-Bioactive Compound Pairs
1671 1578 562 133 16 0 0 0 0
7 5 2 0 0 0 0 0 0
#Compounds in Correlated miRNA-Compound Pairs
Compounds in miRNA-Compound Pairs
Compounds in miRNA-Bioactive Compound Pairs
216 215 125 49 8 0 0 0 0
7 5 2 0 0 0 0 0 0
#miRNAs in Correlated miRNA-Compound Pairs
SNPs in miRNA-Compound Pairs
SNPs in miRNA-Bioactive Compound Pairs
591 564 258 84 15 0 0 0 0
4 3 1 0 0 0 0 0 0
0.4 0.5 0.6 0.7

**Supplementary Figure 29. Summary of significant SNP–compound and miRNA–compound associations. a.** Number of significant SNP–compound pairs and the corresponding numbers of unique SNPs and compounds. **b.** Number of significant miRNA–compound pairs and the corresponding numbers of unique miRNAs and compounds.

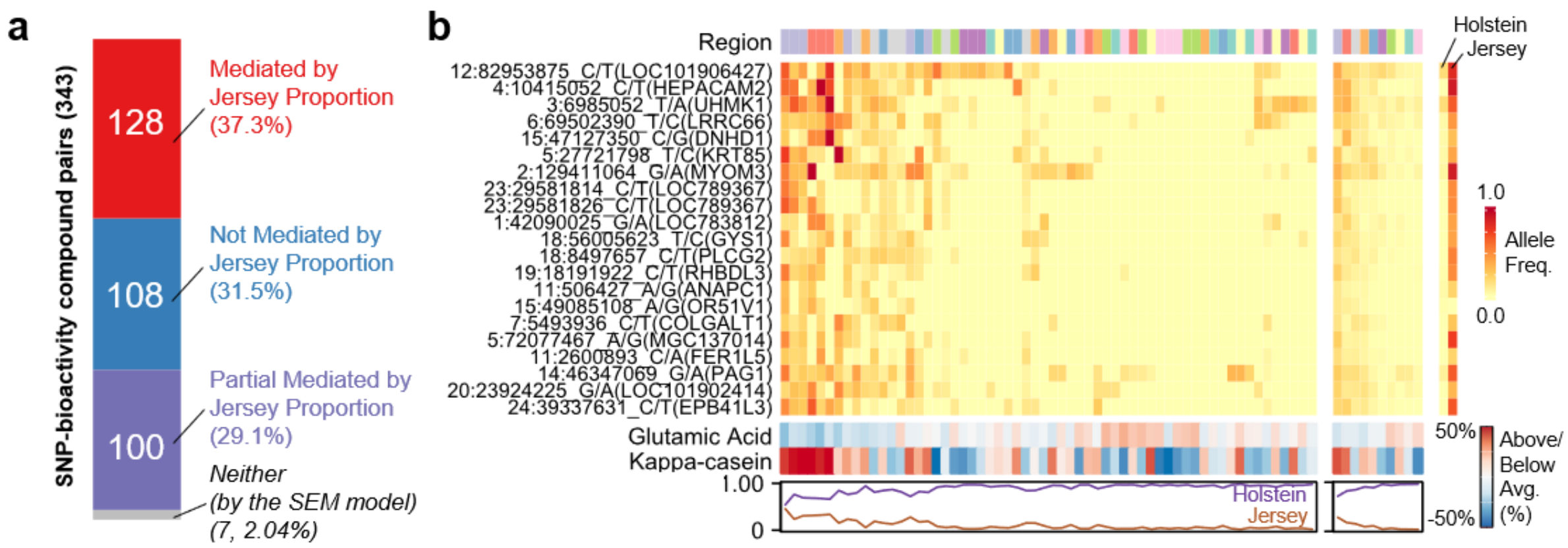

**Supplementary Figure 30. Mediation analysis results of SNP-bioactive compound pairs. a.** Classification of SNP-bioactive compound pairs by mediation through estimated Jersey breed proportion. **b.** SNP-bioactive compound pairs mediated through Jersey proportion, showing SNP allele-frequency differences between Holstein and Jersey reference profiles and associations between bioactive compounds and Jersey proportion.